\documentclass[sn-mathphys,Numbered]{sn-jnl}

\usepackage[utf8]{inputenc}
\usepackage{lineno}
\usepackage{amsmath, amssymb, amsfonts, amsthm}
\usepackage{mathrsfs} 
\usepackage{textgreek} 
\usepackage[version=4]{mhchem} 

\usepackage{graphicx}
\graphicspath{{figures/}}

\usepackage{multirow, hhline, colortbl}
\usepackage{booktabs} 
\usepackage{longtable} 

\usepackage{float}

\usepackage{algorithm}
\usepackage{algorithmicx}
\usepackage{algpseudocode} 
\usepackage{listings} 

\usepackage{lscape} 
\usepackage[title]{appendix} 

\usepackage{xr} 

\usepackage{hyperref} 
\hypersetup{
    colorlinks=true,
    linkcolor=blue,
    citecolor=blue,
    urlcolor=blue
}
\usepackage{xr-hyper} 
\usepackage{xcolor}
\usepackage{anyfontsize}

\usepackage{manyfoot}

\begin{document}

\title[QBIOL: A quantum bioelectrochemical software based on point stochastic processes]{QBIOL: A quantum bioelectrochemical software based on point stochastic processes}


\author*[1,2]{\fnm{Simon} \sur{Grall}}\email{sgrall@laas.fr}
\author[2]{\fnm{Ignacio} \sur{Madrid}}\email{madrid@sat.t.u-tokyo.ac.jp}
\author[2]{\fnm{Aramis} \sur{Dufour}}\email{aramis.dufour@polytechnique.org}
\author[2]{\fnm{Helen} \sur{Sands}}\email{helen.sands@polytechnique.org}
\author[4]{\fnm{Masaki} \sur{Kato}}\email{kato\_m@eis.hokudai.ac.jp}
\author[3]{\fnm{Akira} \sur{Fujiwara}}\email{akira.fujiwara@ntt.com}
\author[2]{\fnm{Soo Hyeon} \sur{Kim}}\email{shkim@iis.u-tokyo.ac.jp}
\author[5]{\fnm{Arnaud} \sur{Chovin}}\email{arnaud.chovin@u-paris.fr}
\author[5]{\fnm{Christophe} \sur{Demaille}}\email{christophe.demaille@u-paris.fr}
\author*[1,2]{\fnm{Nicolas} \sur{Clement}}\email{nclement@iis.u-tokyo.ac.jp}


\affil[1]{\orgdiv{LAAS}, \orgname{CNRS}, \orgaddress{\street{7, av. du colonel Roche}, \city{Toulouse}, \postcode{31031}, \country{France}}}

\affil[2]{\orgdiv{LIMMS}, \orgname{CNRS}, \orgaddress{\street{4-6-1 Komaba}, \city{Meguro-ku, Tokyo}, \postcode{153-8505}, \country{Japan}}}

\affil[3]{\orgdiv{ Basic Research Laboratories}, \orgname{NTT}, \orgaddress{\street{3-1 Morinosato Wakamiya}, \city{Atsugi-shi, Kanagawa}, \postcode{243-0198}, \country{Japan}}}

\affil[4]{\orgdiv{Graduate School of Chemical Sciences and Engineering}, \orgname{Hokkaido University}, \orgaddress{\street{Kita 13 Nishi 8}, \city{Kita-ku, Sapporo}, \postcode{060-8628}, \country{Japan}}}

\affil[5]{\orgdiv{Laboratoire d'Electrochimie Moléculaire}, \orgname{Université Paris Cité - CNRS}, \orgaddress{\street{15, rue Jean-Antoine de Baïf}, \city{Paris}, \postcode{75013}, \country{France}}}

\abstract{Bioelectrochemistry is crucial for understanding biological functions and driving applications in synthetic biology, healthcare, and catalysis. However, current simulation methods fail to capture both the stochastic nature of molecular motion and electron transfer across the relevant picosecond-to-minute timescales. We present QBIOL, a web-accessible software that integrates molecular dynamics, applied mathematics, GPU programming, and quantum charge transport to address this challenge. QBIOL enables quantitative stochastic electron transfer simulations and has the potential to reproduce numerically any (bio) electrochemical experiments. We illustrate this potential by comparing our simulations with experimental data on the current generated by electrode-attached redox-labeled DNA, or by nanoconfined redox species, in response to a variety of electrical excitation waveforms, configurations of interest in biosensing and catalysis. The adaptable architecture of QBIOL extends to the development of devices for quantum and molecular technologies, positioning our software as a powerful tool for enabling new research in this rapidly evolving field.}




\maketitle
\clearpage

\section{Introduction}\label{intro}

Over the last decades, more and more industries have moved toward simulations as a way to reduce risks and costs, with primary examples in aeronautics, semiconductors and health care.
Molecular systems attract heavy scientific interests due to their potential applications in bio-electrochemical systems, molecular electronics and quantum devices \cite{chen_molecular_2021, wu_device_2023,reynaud_rectifying_2020,wasielewski_exploiting_2020}.
The case of bio-electrochemistry is of particular interest as an attractive approach to understand nanoscale biological systems or use bio-mimetic approaches with low energy electron transfer toward catalytic or green energy applications \cite{cramer_architecture_2000, bocanegra_dna_2021, yao_evolution_2016, kumar_ins_2017, ramezani_building_2020, shi_small_2022}. 
It also has a dual role to play in the frame of the ``More than Moore'' and ``Beyond CMOS'' paradigms \cite{arden_more_2005}, contributing to the diversification of both the nature and scale of nanoelectronic devices. Redox molecules can be seen as nano-objects exhibiting exceptional reproducibility and compact size, they are already employed in biosensing applications \cite{li_redox-labelled_2022,arroyo-curras_subsecond-resolved_2018,dauphin-ducharme_high-precision_2019,nakatsuka_aptamerfield-effect_2018,verrinder_comparison_2023}, and as they can be integrated into electrochemical and biological systems, are envisioned for DNA data storage \cite{slinker_multiplexed_2010,nguyen_scaling_2021,yu_high-throughput_2024} or large scales integrated nanoelectrochemical sensing \cite{moazzenzade_stochastic_2020,bao_digital_2021}.
However, such nano-bio-systems exhibit a complexity (moving molecules, stochastic electron transfer) that is not captured by commercial finite element modeling (FEM) solutions \cite{COMSOL, ANSYS, DIGIELCH} nor by molecular dynamics (MD) and coarse-grained approaches alone.

MD provides invaluable insights on molecular movements, averaged electron transfer rates and molecular structure configurations, but does not include intermolecular or metal-molecule electron transfers especially at timescales that would allow a direct comparison with experimental measurements for most applications \cite{de_la_lande_progress_2015,kmiecik_coarse-grained_2016,schlick_biomolecular_2021,may_charge_2004,gray_long-range_2005}. Coarse-grained simulations helped greatly in bridging the gap between angtrom- and micrometer-scales descriptions of molecular movements \cite{kmiecik_coarse-grained_2016,schlick_biomolecular_2021, ouldridge_dna_2010, rovigatti_accurate_2014,poppleton_oxdna_2023}, even managing to bridge coarse-grained and FEM \cite{rudd_coarse-grained_2005}. However, they do not provide themselves a description of electron transfers resulting in current signals comparable with experiments. Molecular electronics achieves good agreement between simulations, models and molecular junction experiments \cite{capozzi_single-molecule_2015, gehring_single-molecule_2019}, based on averaged positions for the calculations of electronic rates, because molecular motions (or vibrational modes) are typically much faster than electron transfer rates. However, this assumption does not always hold for biomolecular motions, which occur across a wide range of timescale and over large distances \cite{yuan_controlling_2015,wang_dynamic_2022, zhang_molecular_2024}.
As for FEM, it models with great success macroscopic behaviors in terms of concentration profiles and average values \cite{dickinson_comsol_2014}, including currents, but is not able to take into account the stochasticity in electron transfer coupled to molecular movements arising at the nanoscale. As an example, for the past 20 years electrochemical DNA (E-DNA) sensors' measurements have led to the belief that molecular diffusion of tethered DNA was anomalously slow \cite{anne_dynamics_2006,anne_electron_2008,huang_random_2013,dauphin-ducharme_chain_2018}. The early developments of QBIOL helped to show that this was incorrect, and that the electrochemical response of these systems could be fully explained by a  reduced electron transfer rate, itself due the low probability of presence ($\rho$) of the redox head at the electrode surface, associated with low reorganization energy ($\lambda$) \cite{zheng_electrochemical_2023, madrid_ballistic_2023, zheng_activationless_2024} and complex energetics related to hydrophilic/hydrophobic interactions at interfaces \cite{martin_dipolar_2015}. This eventually led to the discovery of ballistic Brownian motion for nanoconfined DNA \cite{madrid_ballistic_2023}.
These results, along with remaining open questions and the general trend of performance demands driving development and device miniaturization for greater sensitivity and richer data, motivated the full-scope development of QBIOL presented in this work, addressing at the same time the need to transition from averaged to stochastic event representations of electrochemical reactions \cite{amatore_electrochemistry_2003,cutress_how_2011,singh_stochastic_2016,ren_stochasticity_2021}.

Since the rise of single-molecule electrochemistry \cite{bard_electrochemical_1996}, electrochemical systems started to be studied and modeled with a stochastic approach \cite{cutress_how_2011,cutress_electrochemical_2011,huang_random_2013,katelhon_noise_2013}, with such nanosystems already becoming more and more widespread. Recent works have shown that stochasticity could be used at the advantage of the experimentalist to obtain as much if not more information than through traditional macroscale measurements \cite{singh_stochastic_2016,ren_stochasticity_2021,grall_electrochemical_2023}. However, stochastic simulations remained behind both in terms of timescales and variety of the simulated entities, often limited to restrained timescales and free particles \cite{amatore_electrochemistry_2003,white_random_2005,white_electrochemistry_2008,krause_brownian_2014,samin_one-dimensional_2016}, with orders of magnitude discrepancies for experiments attempting to describe redox biomolecules \cite{huang_random_2013}. \\
Here we show that our simulator QBIOL enables the reproduction of macroscale electrochemical currents of various electrochemical addressing techniques on free particle as well as MD-simulated biomolecules from single-electron counting. The accuracy and timescales available allows direct comparison with experimental data. We choose to focus here on (bio-) electrochemical applications, and also show a working extension for quantum dots.\\

\begin{figure*}[h]
\centering
\includegraphics[clip, trim= 0.3cm 2.5cm 0.2cm 2cm, width = \textwidth]{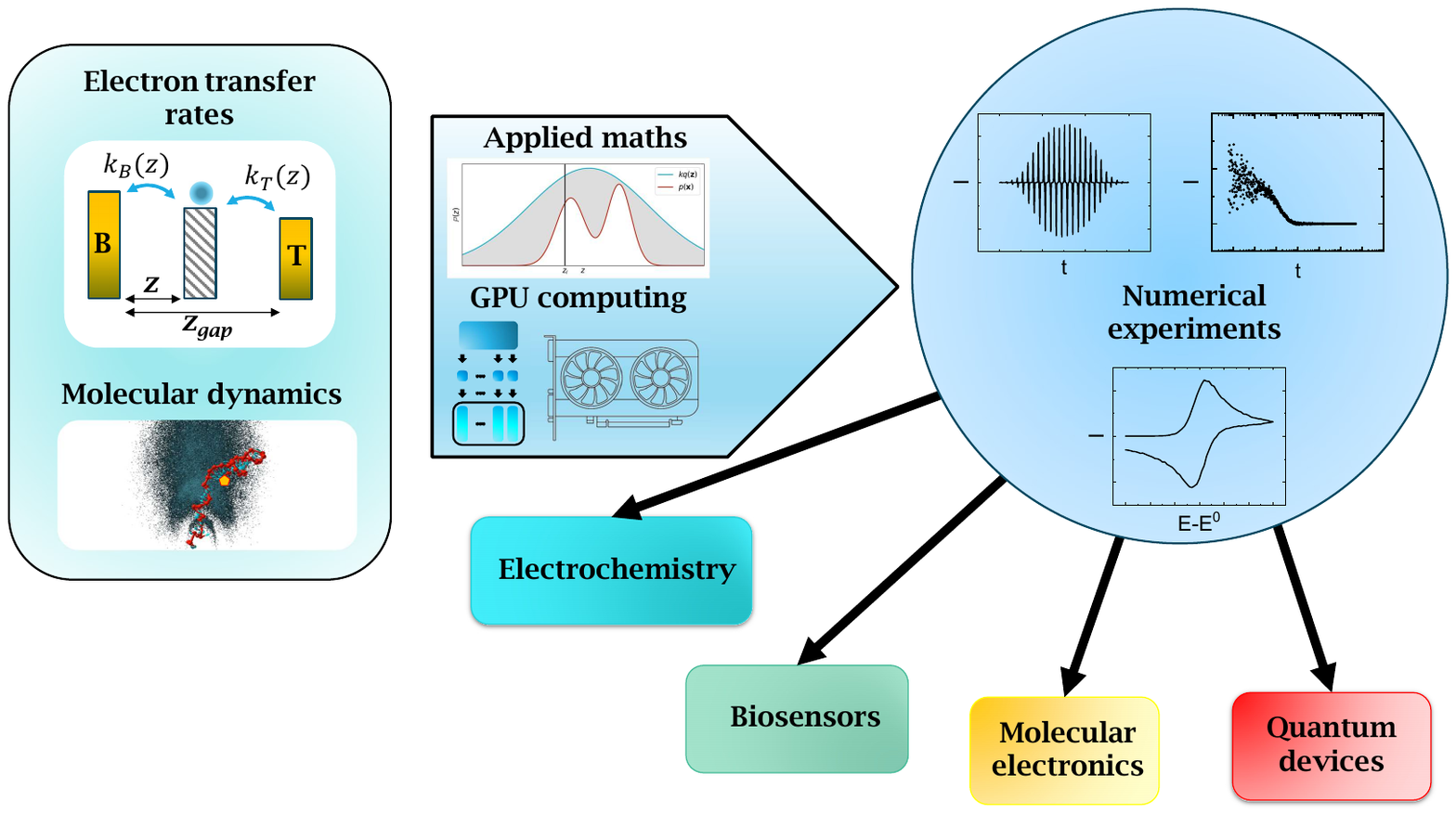}
\caption{\textbf{QBIOL overview.} QBIOL allows to count electrons individually based on a given rate equation. It takes into account the position in space of the electroactive center or molecule by leveraging molecular dynamic solutions such as oxDNA, with the possibility to simulate nanoconfined environments. The progress in GPU computing combined with applied maths engineering allowed numerical experiments such as cyclic voltammetry, AC voltammetry or chronoamperometry with realistic timescales. The ability to simulate realistically complex systems in confined environments has applications in biosensors development, quantum devices simulation and molecular electronics.}
\label{QBIOL_title}
\end{figure*}

\section{Results}
\subsection{QBIOL workflow and targets}

QBIOL is a web-accessible electron transfer simulation software that allows to obtain the current, its fluctuations and probability of presence for any molecule/system in liquid or solid state that can be simulated with molecular dynamics (MD) (Fig. \ref{QBIOL_title}). We introduce several innovations over previous electrochemistry-based approaches, incorporating accurate electron transfer rates informed by the latest advancements in the rapidly evolving field of bioelectrochemistry \cite{zheng_electrochemical_2023,madrid_ballistic_2023, zheng_activationless_2024}. This allows to count electrons one by one in complex moving systems, reproducing familiar macroscale measurements directly comparable with experimental data while keeping the full information on the current's fluctuations (Fig. \ref{QBIOL_title}).\\

The default electron transfer rates used are based on the Marcus-Hush (MH) theory \cite{marcus_theory_1965,hush_homogeneous_1968}. QBIOL uses existing MD libraries, such as oxDNA \cite{ouldridge_dna_2010, rovigatti_accurate_2014,poppleton_oxdna_2023}, to obtain the position of the electroactive part of the molecule of interest over time. Position-dependent electronic couplings and reorganization energies are computed from the MH formalism and used to generate time-dependent probabilities of electron transfer, which, combined with random number generation algorithms, are used to obtain the stochastic simulation \cite{blackman_scrambled_2021} at the picosecond scale. Estimating the evolution of electrochemical currents over minutes with one picosecond resolution is beyond the capabilities of current computers using a conventional ``naive'' approach, largely because of the computational cost of integrating MH rates and the trillions of stochastic processes to simulate per molecule. To overcome this limitation, we employ rejection sampling, GPU parallelization and precomputed rates, enabling the efficient calculation of electrochemical currents over extended timescales. In addition, we overcome the ``Markovian limitation'', which prevents massive parallel computation, by noticing that redox molecules typically have a small number of charge states (typically two). Details are provided in supplementary information (SI).\\
For the sake of illustration, a ferrocene molecule (Fc) is considered here as the electroactive part (or as a single molecule) unless explicitly mentioned. QBIOL uses the position of Fc to compute a probability of electron transfer to/from an electrode with a time resolution of $dt_{MD} = 9.09\times 10^{-13}$ s ($\approx 1$ ps) and up to several minutes. In practice, a user only has to input experimental conditions, such as the number of electrodes, type of experiment,  applied voltages, dimension of the confinement (if any), sequence of the molecule (if DNA is simulated), ionic strength, etc., to obtain a current versus voltage and/or versus time curve, called here ``numerical experiment''. Details about the time management, molecular dynamics and algorithms are available in SI.\\

The performances of QBIOL in terms of timescales, variety of electrochemical voltammetry experiments and accuracy of electron transfer rates are presented in Table \ref{stochastic_table} with comparisons to previous work on stochastic electrochemical simulations.
To illustrate QBIOL's capabilities and performance, we have focused on a few relevant application examples to highlight its significance for instrumentalists, biophysicists, sensor design, and its utility in addressing questions such as signal-to-noise ratio across various electrochemists' communities. Finally, the above-mentioned algorithms and computing effort can also be leveraged to model accurately quantum systems, as illustrated further in this paper.\\

\subsection{Performance evaluation: Bioelectrochemical experiments}
We present hereafter a range of electrochemical experiments available in QBIOL, representative of its reliability and versatility for various cases of interest for the community of electrochemists and bioelectrochemists.

\subsubsection{Cyclic voltammetry}
\begin{figure*}[h]%
\centering
\includegraphics[clip, trim= 5cm 0cm 4cm 1cm, width = 1\textwidth]{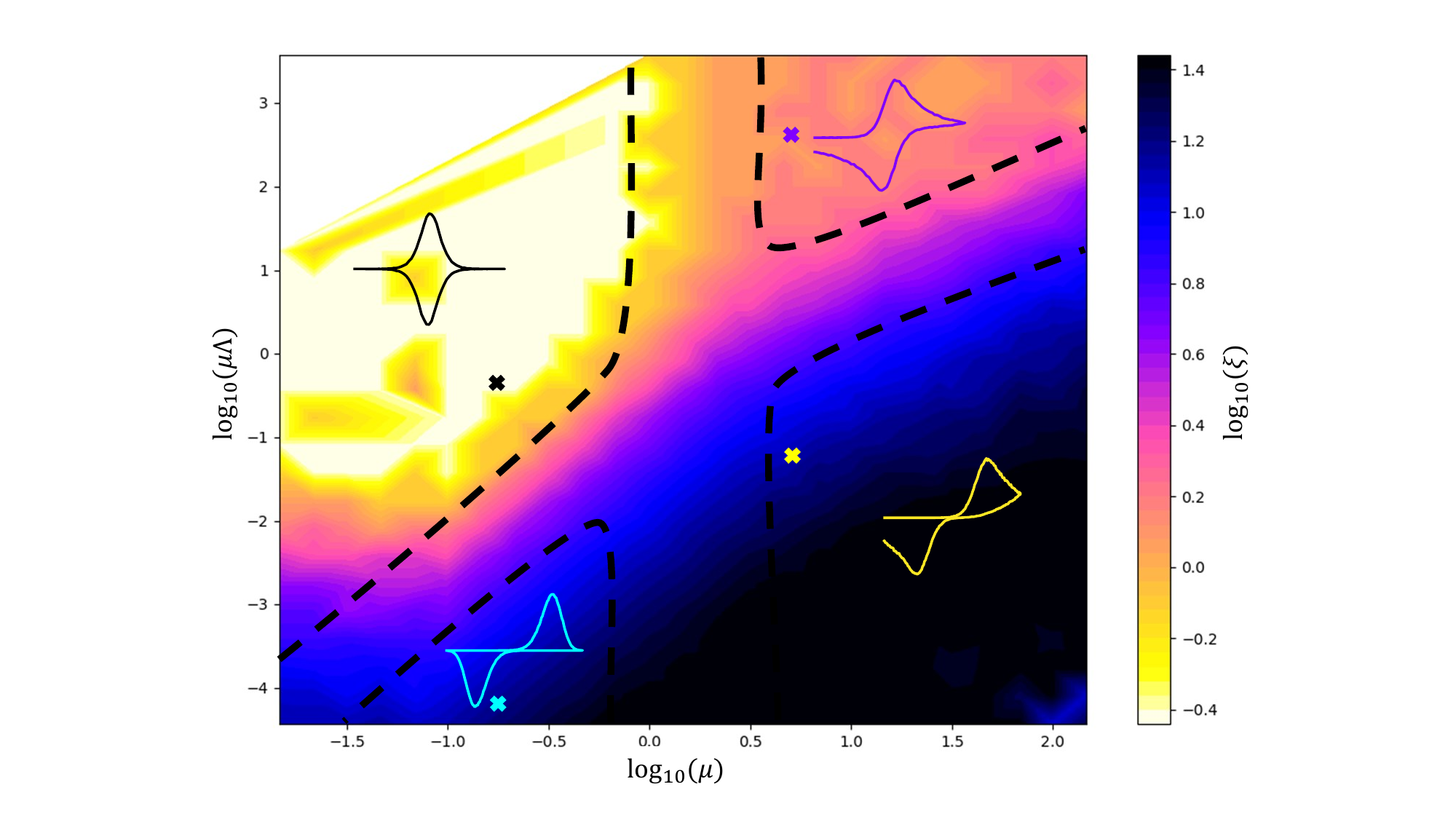}
\caption{\textbf{QBIOL-generated CV oxidation peak's voltage heatmap.} This heatmap generated from QBIOL cyclic voltammogram (CV) data represents the dimensionless oxidation peak's voltage $\xi$ as a function of the dimensionless parameters \textmu \textLambda~($\propto k_s$) and \textmu~($\propto \nu$) (other maps for current peak and current peak/sweep rate available in \ref{QBIOL_heatmaps_SI}). The variations of \textmu~and \textmu \textLambda~control the behavior of the CVs, with the crossed coordinates marking the shown CVs. The black dashes are obtained from the TLC analysis discussed section \ref{TCL_section}. ($z_{gap}$ = 60 nm)}
\label{QBIOL_map_CV}
\end{figure*}

Cyclic voltammetry (CV) has been the primary tool for electrochemists for nearly a century, allowing the extensive characterization of redox reactions as well as quantitative sensing through electroactive molecule measurements. If other more advanced addressing techniques are today used to circumvent some of its limitations, CV measurements remain a cornerstone of electrochemistry.\\
Fig. \ref{QBIOL_map_CV} shows, as a benchmark, a comparison of the cyclic voltammetry behavior of freely moving particles within a nanogap ranging from $\nu = 0.01$ V/s to $10^{10}$ V/s, with one of two walls of the gap acting as an electrode, either as derived from the analytical thin layer cell (TLC) model or from QBIOL simulations. Note that QBIOL simulations with the two-walls acting as electrodes, i.e. forming  an electrochemical nanogap, are also available.\\
The thin layer cell model \cite{laviron_general_1979} is developed in details in section \ref{TCL_section}, but briefly, it predicts the shape of CVs depending on the confinement length $z_{gap}$, the diffusion coefficient $D$, the MH electron transfer rates and the voltage sweep rate $\nu$. In particular, it can compute the dimensionless potential shift $\xi =  \frac{q}{k_BT}(E_p-E^0)$ (with $q$ the elementary charge, $k_B$ the Boltzmann constant, $T$ the temperature, $E^0$ the standard electrochemical potential and $E_p$ the potential of the oxidation peak) for every $(\mu, \mu \Lambda)$ coordinate couple, with $\mu = z_{gap}\sqrt{\frac{q\nu}{k_BTD}}$ and $\mu\Lambda = k_sz_{gap}/D$ ($k_s$ being here the relevant electron transfer rate), that can be thought as dimensionless sweep rate and electron transfer rate, respectively (details section \ref{TCL_section}).\\
As shown in Fig. \ref{QBIOL_map_CV}, QBIOL is in very good agreement with the model (see also Fig. \ref{TCL_vs_QBIOL}) except for $\log(\mu) < 10^{-1}$ and $\log(\mu \Lambda)< 10^{-2.5}$, at the bottom left of Fig. \ref{QBIOL_map_CV}. This corresponds to very small sweep rates combined with very slow electron transfer rates, in which cases very few events occur at the electrode, exposing the limit of electron counting in QBIOL when simply too few events occur over the simulated times to reproduce the macroscopically expected values.\\

Since the position and rates applied to the electroactive particle are known at each moment, QBIOL is able to reproduce concentration profiles. If simple cases such as a free particle confined between two electrodes are now trivial \cite{white_electrochemistry_2008, zevenbergen_fast_2009}, QBIOL offers a unique insight for complex systems, such as end-attached Fc DNA strands (Fig. \ref{PoxPred_dt20ds}).

\subsubsection{Chronoamperometry}
\begin{figure*}[h]%
\centering
\includegraphics[width=\textwidth,clip, trim = 0cm 5cm 0.5cm 4cm ]{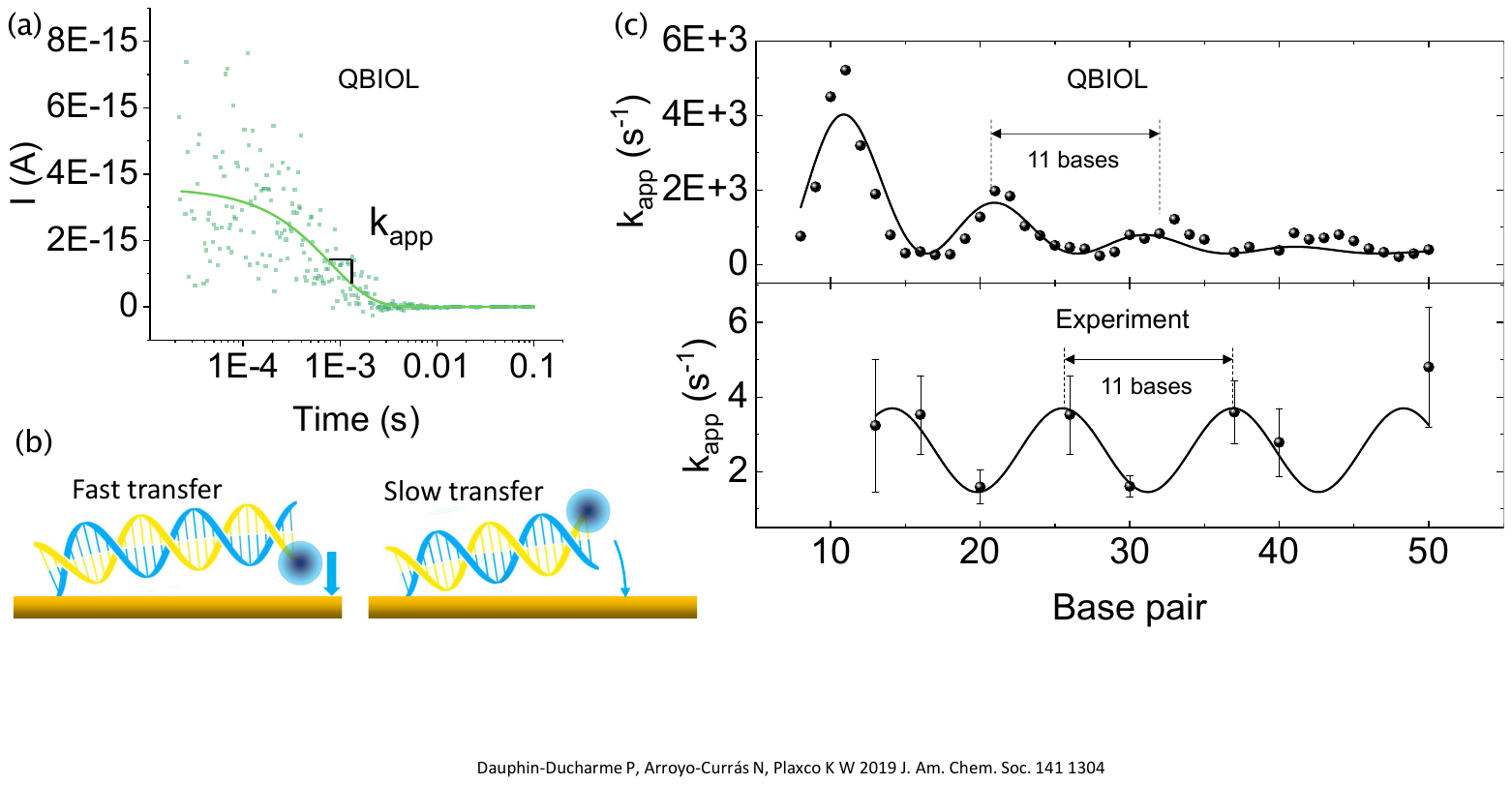}
\caption{\textbf{E-DNA chronoamperometry: comparison between QBIOL and experiment.} (a) Chronoamperometry of a dt20ds DNA. (E-E$^0$): $-50\rightarrow 0$ mV, with a single exponential decay fit. (b) Schematics showing the rate change to be expected from the rotation of the DNA helix with the number of base pair. (c) Measurement of $k_{app}$ for double-stranded DNA between $N_{bp}=$ 8 and 50 base pair. A periodic pattern is visible in both experimental (reproduced from \cite{dauphin-ducharme_high-precision_2019}) and QBIOL data. On the latter, the additional decay observed is due to shorter strands spending more time close to the electrode. Simulation parameters: MH rates, $\rho H^2/\hbar = 4\times 10^{-6}$ eV, $z= 11.5$ \AA. }\label{plaxco_rotation}
\end{figure*}

Chronoamperometry techniques are widely used to obtain information on the rate of electron transfer, with applications in biology, biosensors and electrochemistry. QBIOL implements chronoamperometry as described in section \ref{chronoamperometry_method}, with an example of chronoamperometry on a redox-labeled tethered DNA double strand T20 Fig. \ref{plaxco_rotation} (a). Plaxco et al. made experiments on tethered DNA strands of various lengths and showed an oscillating pattern \cite{dauphin-ducharme_high-precision_2019}, attributed to the DNA helix rotating the electrochemical head periodically closer to the electrode (Fig. \ref{plaxco_rotation} (b)). Chronoamperometry experiments on QBIOL were carried out on double strand DNA from 8 to 50 base pairs. The apparent rate ($k_{app}$) was obtained as the decay rate on the current versus time curves (Fig. \ref{plaxco_rotation} (a)). We observe the same periodicity  than in \cite{dauphin-ducharme_high-precision_2019} ($\approx$ 11 base pair) for $k_{app}$ with respect to the number of base pair (Fig. \ref{plaxco_rotation} (b) and (c)).\\
The apparent decay trend of $k_{app}$ on the QBIOL curve, though absent on the experimental data, was previously observed \cite{uzawa_mechanistic_2010,dauphin-ducharme_chain_2018,zheng_electrochemical_2023} and, simply explained, attributed to shorter strands spending more time close to the electrode. In other words, $k_{app} \propto \rho \propto 1/N_{bp}$, with $N_{bp}$ the length of the DNA strand in number of bases. Considering the error bars and the difficulty of the experiment, we predict that reiterating the experiment designed by \cite{dauphin-ducharme_high-precision_2019} with more intermediate bases and statistical reproductions should render the decay visible experimentally.\\
We observe a difference in phase that can be explained by a difference of linker, by the location of the electrochemical tag on the DNA head and by the type of electrochemical tag used (methylene blue in the experiment, Fc in the simulation). The difference in amplitude of $k_{app.}$ is due to the use of a Fc molecule in the simulation, with a single step electron transfer. Indeed, a full study to assess reorganization energy for the methylene blue used in the original article, which involves the exchange of a proton, remains to be done. The methylene blue shows an apparent electron transfer rate orders of magnitude lower than Fc \cite{uzawa_mechanistic_2010} explaining the difference here observed. This also shows that QBIOL is a powerful tool to explore and optimize the ideal number of bases for a given E-DNA sensor.

\subsubsection{SWV}

\begin{figure*}[h]%
\centering
\includegraphics[width=\textwidth,clip, trim = 1.5cm 5.7cm 2cm 5.5cm ]{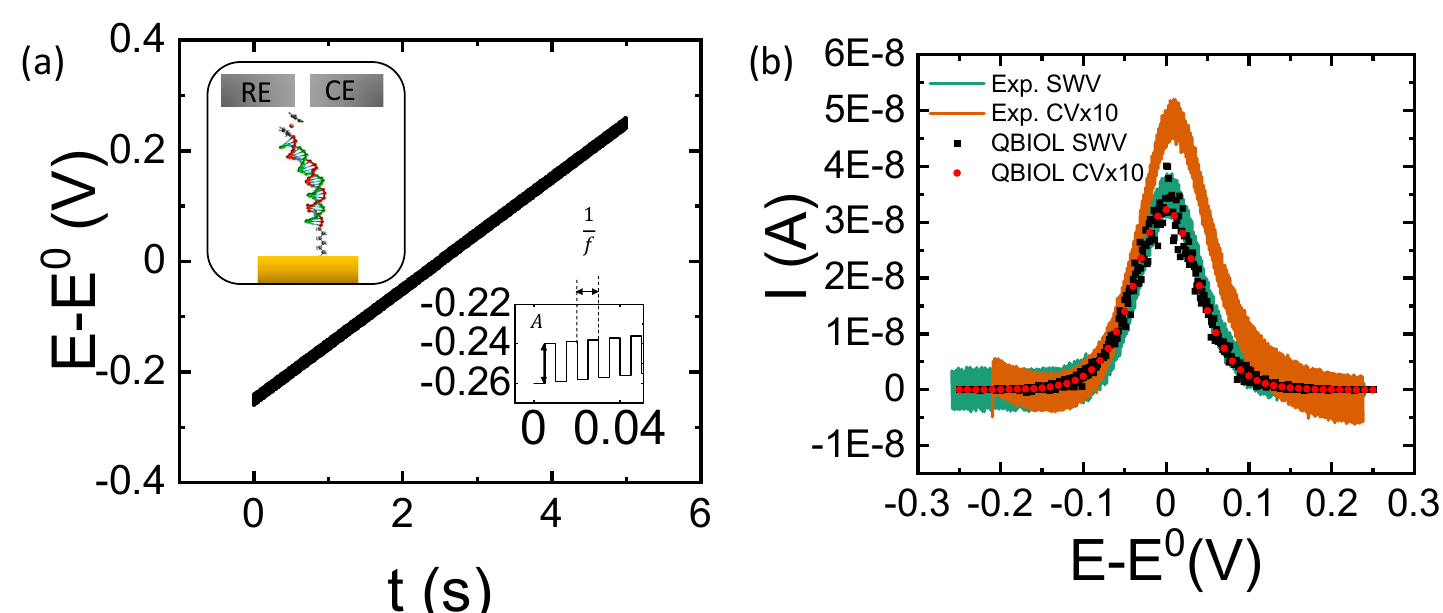}
\caption{\textbf{Square-wave voltammetry: comparison between QBIOL and experiment.} (a) Square wave voltammetry (SWV) voltage profile, with an enlarged version showing SWV parameters. Inset: schematics of the Fc-DNA-alkyl molecule tethered on a gold electrode. (b) Experimental SWV and CV (after baselines removal) compared to QBIOL numerical experiment with the same parameters. The CV currents are scaled 10 times for an easier comparison. Inset: raw SWV and CV data. SWV parameters: $A$ = 20 mV, $f$ = 100 Hz, $\nu = 0.1$~V/s. CV parameters: $\nu = 0.1$~V/s. Simulation parameters: MH rates, $\rho H^2/\hbar = 4\times 10^{-6}$ eV, $z= 4.2$ \AA.}\label{SWV}
\end{figure*}

Square wave voltammetry (SWV) (Fig. \ref{SWV} (a)) has been widely used for its ability to resolve a signal where CVs were too noisy or impeded by parasitic capacitive currents. However, the complex nature of the current answer to SWV waveforms prevents in practice the quantitative interpretation of experimental results, in spite of existing models \cite{osteryoung_square_1985,lovric_square-wave_1988,chen_electrochemical_2013}. As a result, empirical calibrations are typically used \cite{trasobares_estimation_2017, arroyo-curras_subsecond-resolved_2018,torbensen_immuno-based_2019}, using CVs as a base to take advantage of the increased sensitivity of SWV. QBIOL provides here a simulation platform to estimate both CVs and SWVs from complex systems, such as tethered electrochemical probes \cite{trasobares_estimation_2017,verrinder_comparison_2023}.\\

As an application example, we study a case where one wants to calibrate the SWV signal with CV measurements. We have performed experiments on a self-assembled monolayer (SAM) of dT35 double stranded DNA, as illustrated on Fig. \ref{SWV} (a), diluted with hexanethiol alkyl chains  (details in SI). Fig. \ref{SWV} (b) shows the currents recorded for both the CV and the SWV for such a SAM, where the amplitude of the CV peak is roughly 10 times smaller than that of the SWV. QBIOL simulations of both experiments with the same parameters are also shown on Fig. \ref{SWV} (b), with $N=2.76\times 10^{10}$ molecules (determined from the CV) and $z = 4.2$ \AA ~for the $e^{-\beta z}$ factor in the electrochemical rate expression (with $\beta$ the tunneling decay coefficient), coherent with the length of the hexanethiol used here. QBIOL shows a 1/10 ratio between CV and SWV peaks, similar to the experimental data, and also yields very close currents compared to the experimental values. These results show that QBIOL can be used to realistically simulate SWV and obtain quantitative information from SWV measurements.

\subsubsection{AC voltammetry}
\begin{figure*}[!h]%
\centering
\includegraphics[width=\textwidth,clip, trim = 5cm 3cm 4cm 3.1cm ]{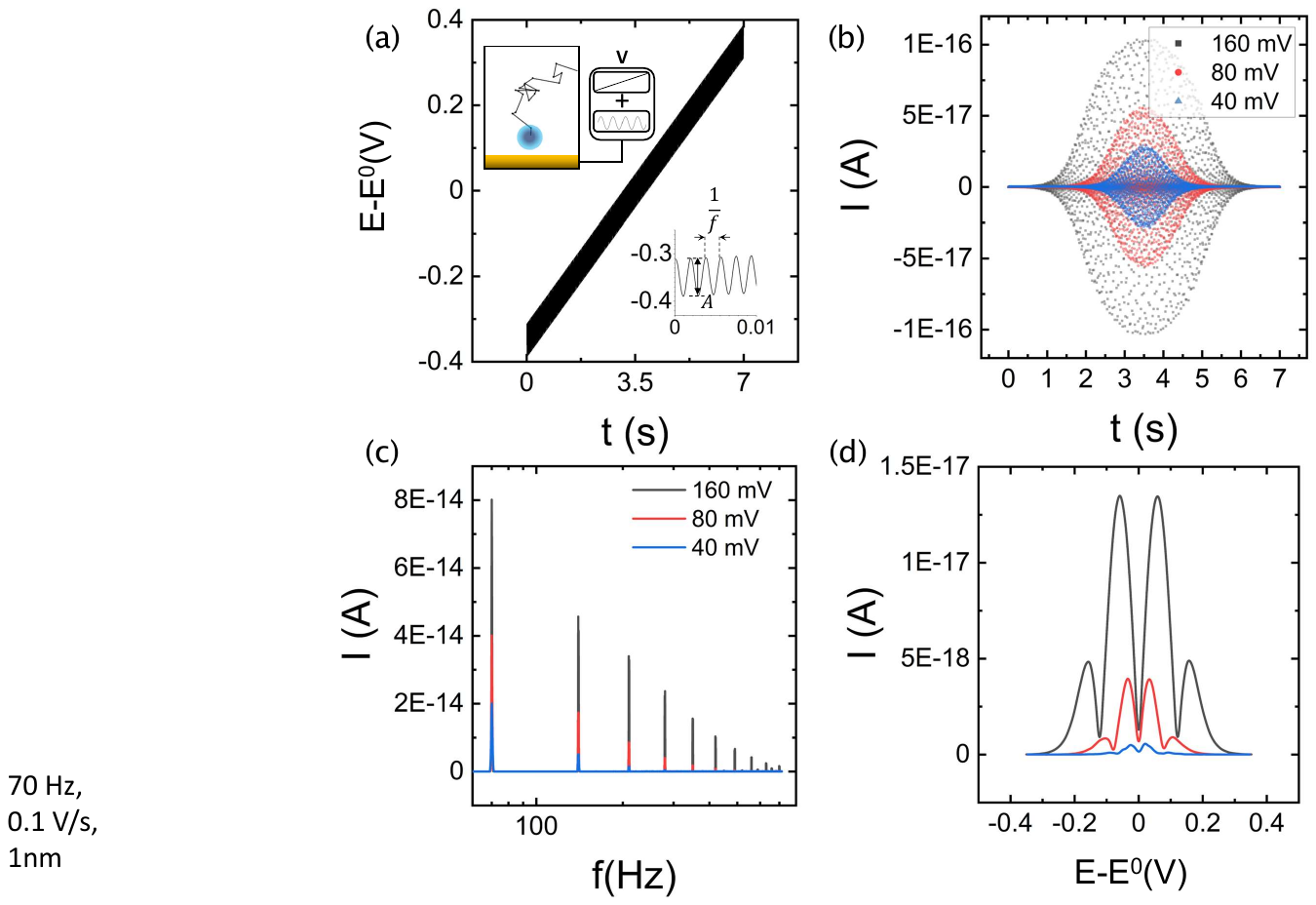}
\caption{\textbf{AC voltammetry QBIOL simulations} AC voltammetry with $f = 70$~Hz, $A = 40$, 80 or 160 mV, $\nu$ = 0.1 V/s, $z_{gap}$ = 60 nm. (a) Applied voltage (zoom in inset as well as schematics of the experiment). (b) Current versus time. (c) Fourier transform of the current, clearly showing well defined high harmonics for higher excitation amplitudes. (d) 4th harmonic envelope versus voltage for different excitation amplitudes.}\label{ACvolta}
\end{figure*}

Alternative current voltammetry (AC voltammetry) superimposes a linear sweep with a sinusoidal wave to study the frequency response of the interrogated electrochemical system. From this one can also address the frequency isolation techniques \cite{adamson_probing_2017, baranska_practical_2024} as well as impedance spectroscopy \cite{macdonald_reflections_2006, randviir_electrochemical_2013,wang_electrochemical_2021}, where the excitation frequency is swept. It allows to discriminate charge transfer rates of different origins and separate the different contribution to the system's impedance, with various applications in biology, batteries, fuel cells, corrosion, nanopores and porous materials. Most notably, high-harmonics are of particular interest as they almost completely remove the contribution of the double layer \cite{baranska_practical_2024}.  AC voltammetry is available in QBIOL as illustrated Fig. \ref{ACvolta}, where we simulate a Fc free particle in a 60 nm gap, reproducing the increase of readability for high harmonics with the increase of the excitation amplitude, as shown experimentally here \cite{baranska_practical_2024}.\\
Recent developments emphasize the importance of AC voltammetry techniques, as they allow to recover superior information compared to DC techniques, whether it is through advanced analysis of frequency-isolated currents \cite{baranska_practical_2024} or, as discussed further in this paper, by high-frequency capacitive amplification \cite{grall_attoampere_2021}.\\

\subsubsection{Electrochemical shot-noise}
\begin{figure}[h!]%
\centering
\includegraphics[width=0.7\textwidth,clip, trim = 5cm 5cm 5cm 5cm ]{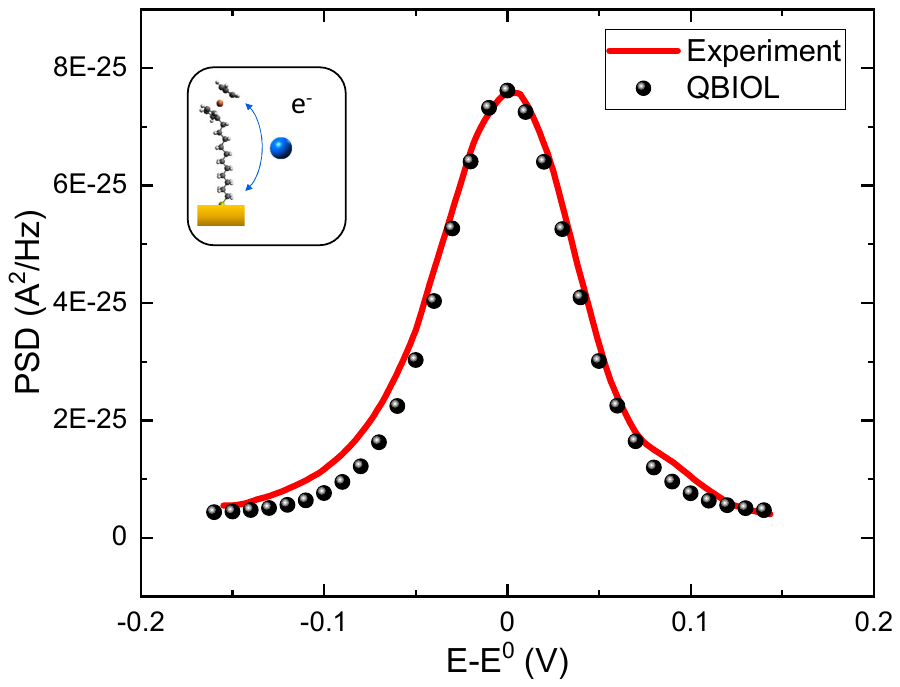}
\caption{\textbf{Voltnoisogram: comparison between QBIOL and experiment.} Comparison between experimentally recorded PSD (at 10 Hz) and QBIOL at low frequency. Inset: schematic of the experiment. Simulation parameters: BV rates, $dt_{sample}$ = 3.3e-5 s, $k_0 = 4\times 10^7$ s$^{-1}$, $z_{gap} = 1$~\AA, prefactor = 1.85e-4, $N=7.5\times 10^{10}$ molecules, noise background = $4\times 10^{-26}$ A$^2$/Hz. Experiment: ferrocene undecanethiol SAM diluted with undecanethiol alkyl chains with in total $\approx 7.5\times 10^{10}$ Fc molecules addressed in 0.5M NaCO$_4$ on microelectrodes (0.78 mm$^2$). Original data and detailed protocol in \cite{grall_electrochemical_2023}.}
\label{PSD_V}
\end{figure}

Experimental evidences were recently brought for the measurement of electrochemical shot-noise on a single-electrode device taking advantage of the stability and homogeneity of electrochemical self-assembled monolayers (SAMs) \cite{grall_electrochemical_2023}. As QBIOL inherently accounts for stochastic processes, it is particularly well-suited to simulate such noise experiments. The reproduction of this experiment in QBIOL, done by measuring the current over time at different voltages, is shown Figure \ref{PSD_V}, showing the low-frequency noise versus voltage. The data in QBIOL are obtained using a prefactor on the rates to account for the undecanethiol spacer, corresponding to an equivalent distance between Fc and the surface of the electrode of $\approx 9.1$~\AA, in reasonable agreement with literature \cite{smalley_kinetics_1995}. The experimentally measured noise baseline is not generated in QBIOL and added from its measured experimental values to QBIOL's noise results for comparison with the experiment.\\

The results are in good agreement with experimental data reproduced from \cite{grall_electrochemical_2023}. Interestingly the discrepancies on the width at half maximum between the model proposed in the original article and experimental data, attributed to interactions with neighboring molecules and electronic coupling variations, is here absent. This tends to indicate that a refining of the analytical model for electrochemical molecules is required, which we plan to address in a dedicated work.

\subsubsection{Designing biosensors}
If previous experiments showed archetypal DNA sequences, QBIOL is able to simulate any sequence such as biologically relevant aptamer sequences and be used as an engineering tool. Electrochemical DNA (E-DNA) probes consist in an end-attached electrochemical marker on a DNA strand or aptamer strand and have various biosensing applications \cite{li_redox-labelled_2022,arroyo-curras_subsecond-resolved_2018,dauphin-ducharme_high-precision_2019,nakatsuka_aptamerfield-effect_2018,verrinder_comparison_2023}. They have been used for the past decade as incredibly versatile biosensors, for food, metal, environment or cancer marker detection \cite{drummond_electrochemical_2003,diculescu_applications_2016,trotter_review_2020}.\\
We show here a case study for the design of such a biosensor based on a SYL3C aptamer\footnote{SYL3C sequence: 5'-Fc-CAC TAC AGA GGT TGC GTC TGT CCC ACG TTG TCA TGG GGG GTT GGC CTG-(CH$_2$)$_3$-SH-3'} for the EpCAM marker on cancer cell membrane \cite{li_redox-labelled_2022}. 
The aptamers are grafted on a gold electrode with nano pillars to suspend the cell, leaving the aptamer at a proper distance to both reach the surface of the cell and this electrode. The detection is made by monitoring the electrochemical current amplitude during cyclic voltammograms, as it would decrease when the aptamer matches the EpCAM receptor on the cell membrane and thus spend less time near the surface, decreasing the net current at the electrode. A crucial parameter in the development of the sensor was: what should be the height of the nano pillars to ensure that the aptamer can move freely without being compressed against the electrode, while still reaching the surface of the cell? If this was solved experimentally by trial and error, QBIOL now offers a tool to probe this question ahead of such experiments.\\

\begin{figure*}[h]%
\centering
\includegraphics[width=\textwidth,clip, trim = 1cm 2.5cm 3cm 3cm ]{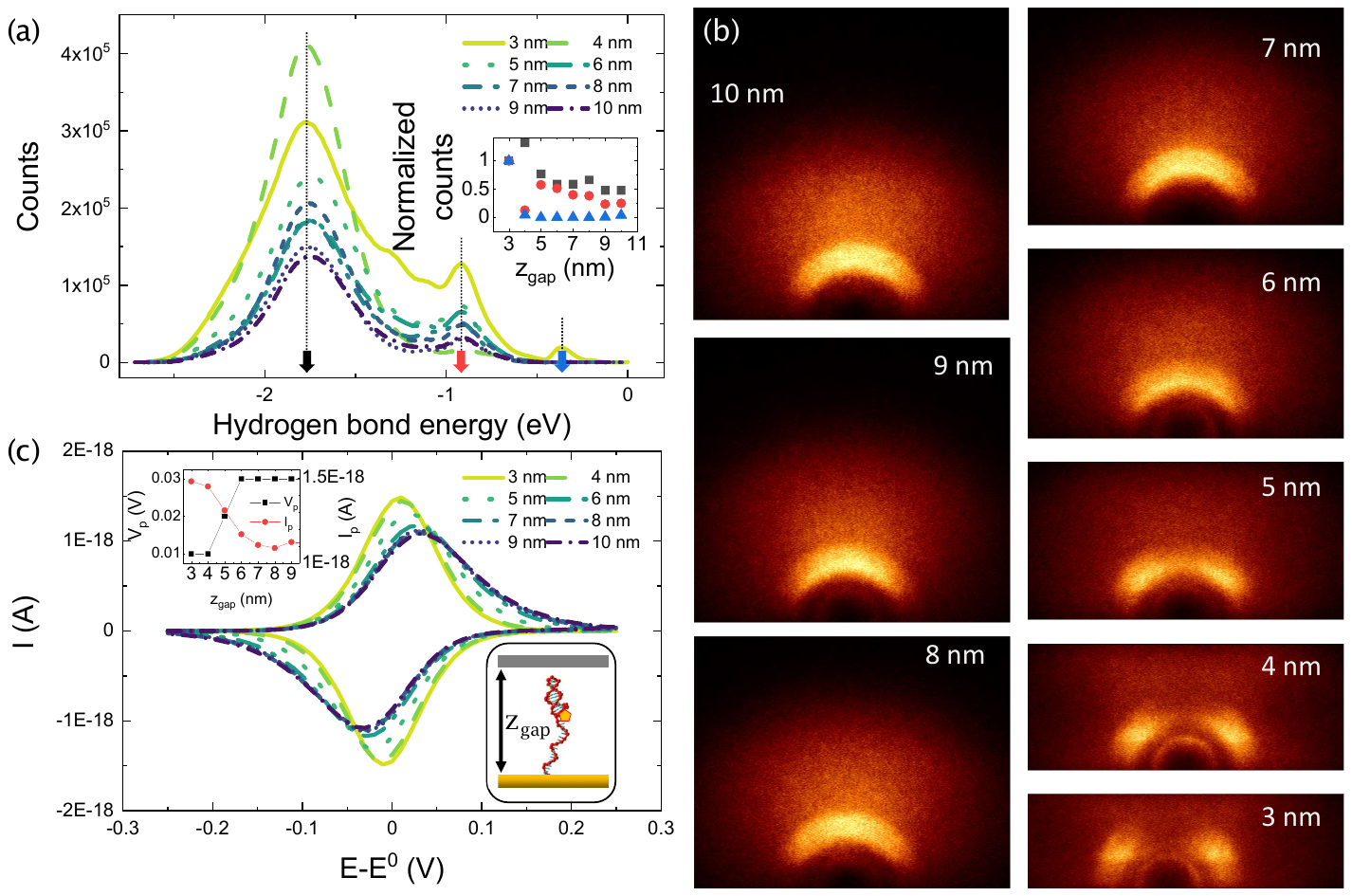}
\caption{\textbf{QBIOL as a tool for biosensors' design.} (a) Total hydrogen bond energy of SYL3C for different $z_{gap}$. Inset: peak height at the energies indicated by arrows versus $z_{gap}$. (b) Probability of presence of the Fc head within the gap. The height of images are to scale with the indicated gap. Scale bar: 1 nm. Simulation parameters: MD simulated time = 20 \textmu s, salt = 0.5 M, $T$ = 20\textdegree C. (c)~Corresponding QBIOL CVs with the following parameters: $\nu = 1$ V/s, 10 mV per step, MH rates, $\rho H^2 = 4\times 10^{-6}$ J, $z = 1$~nm in $e^{-\beta z}$. Insets: top left, voltage and current peak shift with $z_{gap}$, bottom right, a schematics of the simulated experiment.}
\label{biosensors}
\end{figure*}

Fig. \ref{biosensors} shows simulations of a single confined SYL3C strand with different $z_{gap}$. The total hydrogen energy bond Fig. \ref{biosensors} (a), important indication for the hairpins configurations of aptamers, and the probability of presence Fig. \ref{biosensors} (b) suggest that for a confinement down to 5 nm, the detection loop \cite{yang_predicting_2019,li_redox-labelled_2022} for EpCAM sensing is dominant most of the time. CVs are also simulated for the corresponding gaps, Fig. \ref{biosensors} (c). There is a clear transition from 4 nm downwards visible on hydrogen bonds, probability of presence and CVs, indicating a change in conformation. CVs remain the same from 10 nm gap, where the molecule is hardly confined, down to 5 nm where the peak clearly starts shifting toward a surface-only CV. This is due to higher probability of presence close to the electrode for more confined molecules.\\

It agrees with experimental findings where a 5 nm-high optimum gap was estimated between the surface of the cells and the electrode where SYL3C were grafted \cite{li_redox-labelled_2022}. The corresponding CVs Fig. \ref{biosensors} (c) show a shift of the current peak, due to different probabilities of presence at the interface for different confinements. The presence of polyethylene glycol oligomer that hinder the access to the surface in the real experiment is not included in QBIOL at this stage, but practically results in an impossibility to reach the electrode closer than $z \approx 1$~nm, which is translated into a reduction of the electron transfer rates by a factor $e^{\beta z}\approx 22000$. The simulation of a dense layer of laterally interacting Fc-DNA chains on an electrode is also possible (Fig. \ref{SAM_fig}), but since the Fc aptamer layer was here loosely packed, it was not especially relevant for this particular example. However, it can be an important feature for biosensor engineering.\\

If for this specific DNA sequence, the probability of presence near the interface plays the main role over the final current response, it is not true in general: for a given DNA sequence, the timescale of opening/closing of hairpins is for example not known in advance, and electron transfer rates can be of similar magnitudes. It is especially true for more complex waveforms applied on the electrode. Note that any DNA sequence can be used for simulation within QBIOL. Furthermore, QBIOL could be extended to simulate proteins by implementing a protein molecular dynamics library such as LAMMPS \cite{thompson_lammps_2022}.

\subsection{Pushing electrochemistry at the limit}

\begin{figure*}[h]%
\centering
\includegraphics[width=\textwidth,clip, trim = 3cm 5cm 2cm 5cm ]{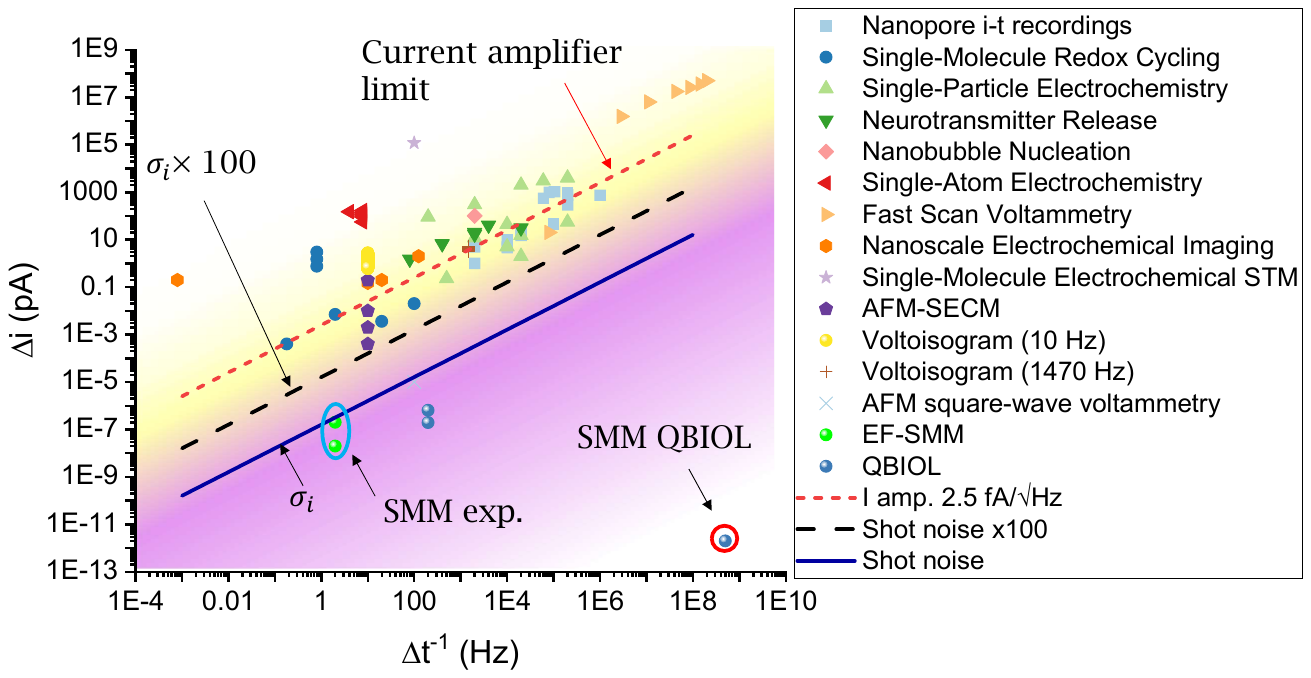}
\caption{\textbf{Electrochemical current limits.} Minimum faradaic current resolved $\Delta i$ versus its measurement bandwidth $\Delta t^{-1}$ for different electrochemical systems (adapted from \cite{gao_shot_2020}). See Tables \ref{ci_1} to \ref{ci_13} for numerical values and methods.}\label{Gao}
\end{figure*}

Electrochemistry at the nanoscale is often hindered by noise, and QBIOL is a natural tool to investigate such limits. A global perspective was initiated by Gao et al. \cite{gao_shot_2020} across a wide variety of electrochemical systems ranging from nanopores to nanoscale electrochemical imaging, including neurotransmitter release and single particle electrochemistry.  Fig. \ref{Gao} (a) represents the original data gathered by Gao et al., as well as QBIOL voltammetry simulations and other works investigated by the authors (details listed in SI). The original study suggested a measurement limit for electrochemical currents set from a shot noise definition based on the variation of the number of measured molecules changing charge state $N$ over a bandwidth $\Delta t^{-1}$ and on the ``Limit of Quantification'' \cite{macdougall_guidelines_1980} recommending a 10 times bigger measurement than the background noise. It translates practically in a limit current $\Delta i = 100 \sigma_i$, with $\sigma_i = \sigma_N * q/\Delta_t$ (black dashes on Fig. \ref{Gao}), $\sigma_i$ the current shot noise (blue line on Fig. \ref{Gao}) and $\sigma_N$ the variation of the number of elementary charges measured during $\Delta t$. If this definition of shot noise is debatable given the range of experiments considered, QBIOL simulations and literature data suggest that this limit is not absolute. In fact, many experiments were possibly constrained by their current amplifier (red small-dash line, Fig. \ref{Gao}) rather than true electrochemical noise.

In redox-cycling experiments, the electrochemical current and the background current (as compared in \cite{gao_shot_2020}) differ in nature. The electrochemical current results from redox cycles down to a single molecule \cite{kang_electrochemical_2013}, while the background current likely reflects the limit of the experiment's current amplifier. In contrast, voltnoisograms (Fig. \ref{PSD_V}) analyze a large number of molecules to determine their shared rate properties \cite{grall_electrochemical_2023}. A smaller number of molecules would proportionally reduce the current noise, highlighting the often-overlooked role of molecular count in measurements.\\

In addition to the considerations on the number of molecules, QBIOL simulations, free of any background noise, point toward lower current limits for CVs and voltnoisograms (blue spheres on Fig. \ref{Gao}). Experimentally, scanning microwave microscopy (SMM, blue circles on Fig. \ref{Gao}) also demonstrates that it is possible to break through the aforementioned limit thanks to its capacitive amplification and frequency isolation scheme \cite{grall_attoampere_2021}. QBIOL is also able to reproduce the SMM addressing scheme (see SI section \ref{Arb_section}) and highlights the potential improvements for SMM (red circle on Fig. \ref{Gao}), although the large difference in bandwidth is due to the absence of parasitic capacitances in QBIOL, experimentally ubiquitous and still challenging to avoid. Overall, this highlights the importance of alternative amplification schemes, such as employed in redox cycling and SMM, as well as the potential of frequency-isolation techniques to reach lower current limits, often synonym of a much smaller number of molecules, thus giving access to the stochastic properties of the system. Such advancements would benefit applications like single-entity measurements \cite{clarke_single_2024} or DNA synthesis, where precise, localized measurements are critical, for example, in DNA data storage technologies \cite{slinker_multiplexed_2010,nguyen_scaling_2021}.

\subsection{Toward quantum transport devices}
\begin{figure*}[h]%
\centering
\includegraphics[width=\textwidth,clip, trim = 2cm 3cm 3.5cm 3.5cm ]{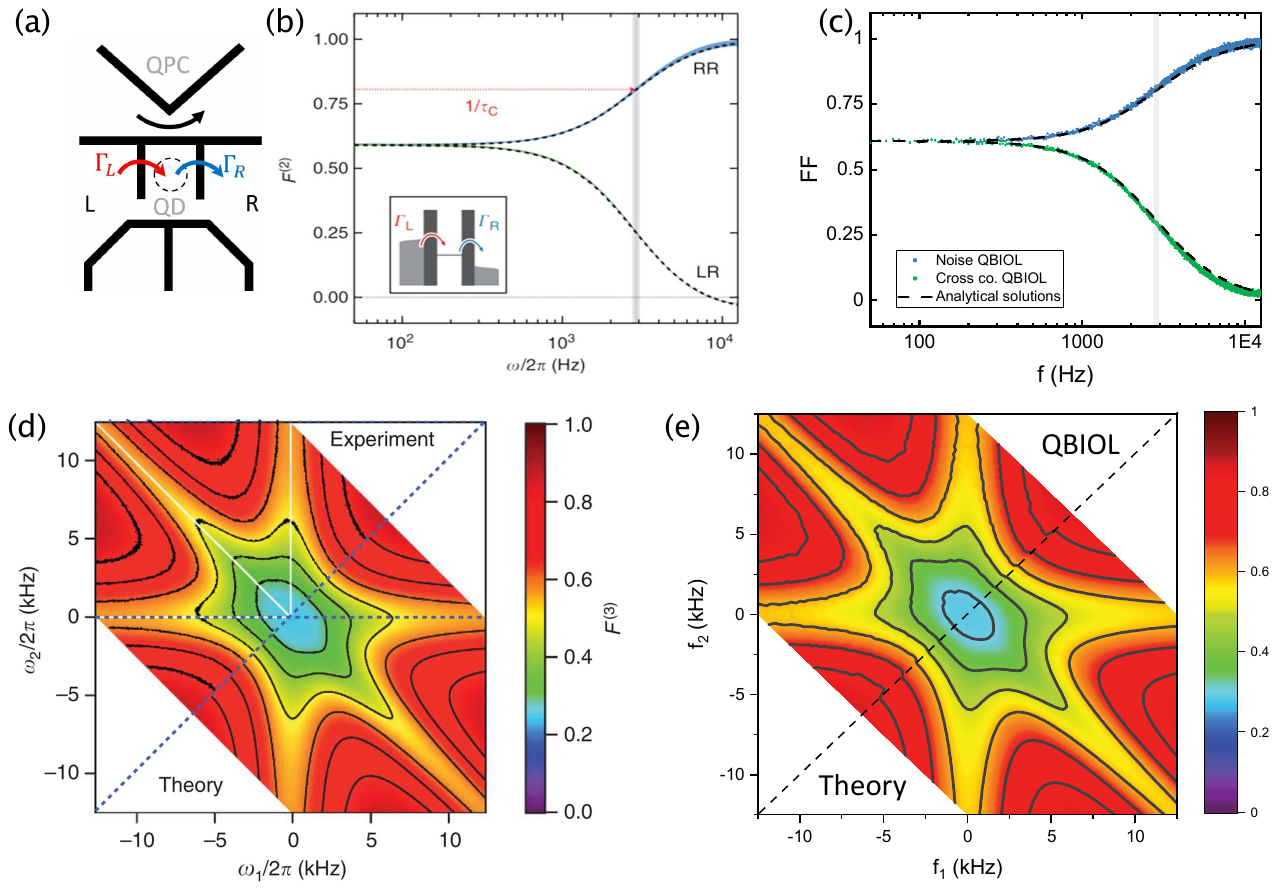}
\caption{\textbf{Solid-state electron counting: comparison between QBIOL and experiment.} Reproduction from QBIOL simulations of the results obtained by Ubbelohde et al. \cite{ubbelohde_measurement_2012}.\\ (a) Schematic representation of the experiment, where the state of the QPC allows to recover independently the electrons flowing in both leads. (b) Spectral noise and cross-correlation of the currents taken from \cite{ubbelohde_measurement_2012} and (c) the equivalent obtained from QBIOL, with analytical models derived in the original article in dashed lines. The average dwell time $1/(\Gamma_L + \Gamma_R)$ is well recovered (grayed zone) in QBIOL. (d) Third order fano factor (frequency dependent skewness) taken from \cite{ubbelohde_measurement_2012} and (e) the equivalent obtained from QBIOL. In both cases the bottom part is obtained from the analytical model and the top part is from experimental data in (d) and QBIOL data in (e). (QBIOL simulation parameters: $\Gamma_L = 13.23$ kHz, $\Gamma_R = 4.81$ kHz, binning window = 40 \textmu s, $dt_{MD} = 2$ \textmu s, $dt_{sample} = 2.77$ s $\times 1000$ threads $\approx$ 47 min simulated.) Original figures reproduced from \cite{ubbelohde_measurement_2012}. \label{ME}}
\end{figure*}

We so far focused on (bio-)electrochemical applications, but the architecture of QBIOL makes it suitable for any system where stochastic electron transfer occurs at the nanoscale. As an example, we chose to reproduce the following experiment carried out on a single-electron transistor. Ubbelohde et al. managed to measure precisely the electron statistics in a single-electron transistor (represented Fig. \ref{ME} (a)), using a quantum point contact (QPC) to monitor the presence or absence of electron within the quantum dot (QD) of the transistor\cite{ubbelohde_measurement_2012}. They manage not only to measure accurately the current, but also the higher-order moments of noise with precision and in agreement with analytical models. By using the rates reported in their work in QBIOL and adjusting time constants to match their experiment, we can reproduce remarkably well the experimentally measured electron distribution statistics, including the second moment (Fig. \ref{ME} (b)) and the third moment (Fig. \ref{ME} (c)) in the frequency domain. This shows that, provided the proper electron transfer rates, QBIOL is very versatile and can be structurally very easily adapted to quantum devices of interest for solid-states and molecular organic electronic communities.

\subsection{Web interface}
A web interface of QBIOL is available at qbiol.org. Up to date implemented features cover most of the examples shown in this work for anyone to test. In the future, QBIOL is meant to address most bioelectrochemical systems, in particular bioengineering applications, covering for example stochastic diffusion of particles in DNA synthesis \cite{nguyen_scaling_2021} or transistors aptasensors \cite{nakatsuka_aptamerfield-effect_2018}.

\section{Discussion}
QBIOL is able to tackle the challenge of electrochemical current simulations on large timescales thanks to its hybrid approach using coarse-grained MD and stochastic electron transfer simulations. QBIOL simulations are conducted in two stages. First, molecular dynamics (MD) libraries simulate the system over timescales representative of the target current measurement. Second, electron transfer probabilities are calculated over time using this spatial data, looping over it as necessary (see section \ref{long_experiments}). It effectively decouples electron transfer from structural conformation changes, providing a key simplification for computation speed. While it currently cannot model intermolecular electron transfers and shares some limitations with MD for systems with significant spatial evolution over long timescales, this approach allows to reach very long timescales compared to MD alone with a very good accuracy on the studied systems.\\
The electron transfer rates used in QBIOL reflect the state of the art in the biomolecular electron transfer rates understanding \cite{zheng_electrochemical_2023,madrid_ballistic_2023, zheng_activationless_2024}. Thanks to its flexible structure, QBIOL is able to take into account future refinements or user-input custom rates.\\
While MD-derived presence probabilities might seem sufficient to estimate current characteristics in simple cases, the full QBIOL approach is mandatory when molecular motion, conformation changes or electron transfers cover overlapping ranges. This can typically be the case for aptamers with hairpins forming at different time scales. Furthermore, a purely probabilistic approach would fail to capture the statistical distribution of transfer events, which provides critical insights beyond mean current values.\\

The QBIOL simulation time currently needed to reproduce Fig. \ref{plaxco_rotation} is about 4 days, including the MD and the electrochemistry. While this will be improved, it already advantageously compares to weeks of experimental work and therefore, shows that QBIOL can already be used as a screening technique.\\


QBIOL is the first numerical tool, accessible to all, aiming to generate any (bio)electrochemical experiment that can be directly compared to real experiments. Starting from stochastic processes (molecular and charge transfer) at its core while addressing computational challenges arising from the spread in timescales between molecular motion ($\approx$ ps) and electrical measurements ($\approx$ s), QBIOL proves to be the way to provide state-of-the-art quantitative results. It has already contributed to unveil the operational mechanism of E-DNA sensors \cite{zheng_electrochemical_2023}, the ballistic Brownian motion of DNA under nanoconfinement \cite{madrid_ballistic_2023} and the suppression of the reorganization energy for tethered DNA via the probability of presence at the metal interface \cite{ zheng_activationless_2024}. It allows direct comparison to a large variety of experiments, enabling anyone to unveil new science, engineer biosensors or simply for educational purposes.\\
We believe that with the continuous progress of molecular dynamic simulations, AI tools for macromolecular recognition and computing power of GPU, such numerical experiment will overcome in number real bioelectrochemical measurements in the next decade. It will also be transposed to other biosensing or imaging techniques, becoming a key tool for today and tomorrow's innovations, such as transistor aptasensors \cite{nakatsuka_aptamerfield-effect_2018} or large-arrays of nanoelectrodes for parallel electrochemical DNA synthesis \cite{nguyen_scaling_2021}.\\
%
%
%
%

\clearpage

\backmatter

\bmhead{Supplementary information}

Extensive supplementary information are available.

\bmhead{Acknowledgments}
The authors would like to thank Pierre Fauret and Christophe Zanon, engineers in the LAAS laboratory, for their major contribution in the making of the QBIOL web interface. The authors also thank the CNRS prematuration program QBIOL, EU-Attract Unicorn-Dx, MITI Pressure and ANR ESHOT for their financial support.

%
%
%

\bigskip\noindent

\bibliography{sn-bibliography}


\begin{thebibliography}{87}
\ifx \bisbn   \undefined \def \bisbn  #1{ISBN #1}\fi
\ifx \binits  \undefined \def \binits#1{#1}\fi
\ifx \bauthor  \undefined \def \bauthor#1{#1}\fi
\ifx \batitle  \undefined \def \batitle#1{#1}\fi
\ifx \bjtitle  \undefined \def \bjtitle#1{#1}\fi
\ifx \bvolume  \undefined \def \bvolume#1{\textbf{#1}}\fi
\ifx \byear  \undefined \def \byear#1{#1}\fi
\ifx \bissue  \undefined \def \bissue#1{#1}\fi
\ifx \bfpage  \undefined \def \bfpage#1{#1}\fi
\ifx \blpage  \undefined \def \blpage #1{#1}\fi
\ifx \burl  \undefined \def \burl#1{\textsf{#1}}\fi
\ifx \doiurl  \undefined \def \doiurl#1{\url{https://doi.org/#1}}\fi
\ifx \betal  \undefined \def \betal{\textit{et al.}}\fi
\ifx \binstitute  \undefined \def \binstitute#1{#1}\fi
\ifx \binstitutionaled  \undefined \def \binstitutionaled#1{#1}\fi
\ifx \bctitle  \undefined \def \bctitle#1{#1}\fi
\ifx \beditor  \undefined \def \beditor#1{#1}\fi
\ifx \bpublisher  \undefined \def \bpublisher#1{#1}\fi
\ifx \bbtitle  \undefined \def \bbtitle#1{#1}\fi
\ifx \bedition  \undefined \def \bedition#1{#1}\fi
\ifx \bseriesno  \undefined \def \bseriesno#1{#1}\fi
\ifx \blocation  \undefined \def \blocation#1{#1}\fi
\ifx \bsertitle  \undefined \def \bsertitle#1{#1}\fi
\ifx \bsnm \undefined \def \bsnm#1{#1}\fi
\ifx \bsuffix \undefined \def \bsuffix#1{#1}\fi
\ifx \bparticle \undefined \def \bparticle#1{#1}\fi
\ifx \barticle \undefined \def \barticle#1{#1}\fi
\bibcommenthead
\ifx \bconfdate \undefined \def \bconfdate #1{#1}\fi
\ifx \botherref \undefined \def \botherref #1{#1}\fi
\ifx \url \undefined \def \url#1{\textsf{#1}}\fi
\ifx \bchapter \undefined \def \bchapter#1{#1}\fi
\ifx \bbook \undefined \def \bbook#1{#1}\fi
\ifx \bcomment \undefined \def \bcomment#1{#1}\fi
\ifx \oauthor \undefined \def \oauthor#1{#1}\fi
\ifx \citeauthoryear \undefined \def \citeauthoryear#1{#1}\fi
\ifx \endbibitem  \undefined \def \endbibitem {}\fi
\ifx \bconflocation  \undefined \def \bconflocation#1{#1}\fi
\ifx \arxivurl  \undefined \def \arxivurl#1{\textsf{#1}}\fi
\csname PreBibitemsHook\endcsname

\bibitem[\protect\citeauthoryear{Chen and
  Fraser~Stoddart}{2021}]{chen_molecular_2021}
\begin{barticle}
\bauthor{\bsnm{Chen}, \binits{H.}},
\bauthor{\bsnm{Fraser~Stoddart}, \binits{J.}}:
\batitle{From molecular to supramolecular electronics}.
\bjtitle{Nature Reviews Materials}
\bvolume{6}(\bissue{9}),
\bfpage{804}--\blpage{828}
(\byear{2021})
\doiurl{10.1038/s41578-021-00302-2}
\end{barticle}
\endbibitem

\bibitem[\protect\citeauthoryear{Wu et~al.}{2023}]{wu_device_2023}
\begin{barticle}
\bauthor{\bsnm{Wu}, \binits{J.}},
\bauthor{\bsnm{Liu}, \binits{H.}},
\bauthor{\bsnm{Chen}, \binits{W.}},
\bauthor{\bsnm{Ma}, \binits{B.}},
\bauthor{\bsnm{Ju}, \binits{H.}}:
\batitle{Device integration of electrochemical biosensors}.
\bjtitle{Nature Reviews Bioengineering}
\bvolume{1}(\bissue{5}),
\bfpage{346}--\blpage{360}
(\byear{2023})
\doiurl{10.1038/s44222-023-00032-w}
\end{barticle}
\endbibitem

\bibitem[\protect\citeauthoryear{Reynaud
  et~al.}{2020}]{reynaud_rectifying_2020}
\begin{barticle}
\bauthor{\bsnm{Reynaud}, \binits{C.A.}},
\bauthor{\bsnm{Duché}, \binits{D.}},
\bauthor{\bsnm{Simon}, \binits{J.-J.}},
\bauthor{\bsnm{Sanchez-Adaime}, \binits{E.}},
\bauthor{\bsnm{Margeat}, \binits{O.}},
\bauthor{\bsnm{Ackermann}, \binits{J.}},
\bauthor{\bsnm{Jangid}, \binits{V.}},
\bauthor{\bsnm{Lebouin}, \binits{C.}},
\bauthor{\bsnm{Brunel}, \binits{D.}},
\bauthor{\bsnm{Dumur}, \binits{F.}},
\bauthor{\bsnm{Gigmes}, \binits{D.}},
\bauthor{\bsnm{Berginc}, \binits{G.}},
\bauthor{\bsnm{Nijhuis}, \binits{C.A.}},
\bauthor{\bsnm{Escoubas}, \binits{L.}}:
\batitle{Rectifying antennas for energy harvesting from the microwaves to
  visible light: {A} review}.
\bjtitle{Progress in Quantum Electronics}
\bvolume{72},
\bfpage{100265}
(\byear{2020})
\doiurl{10.1016/j.pquantelec.2020.100265}
\end{barticle}
\endbibitem

\bibitem[\protect\citeauthoryear{Wasielewski
  et~al.}{2020}]{wasielewski_exploiting_2020}
\begin{barticle}
\bauthor{\bsnm{Wasielewski}, \binits{M.R.}},
\bauthor{\bsnm{Forbes}, \binits{M.D.E.}},
\bauthor{\bsnm{Frank}, \binits{N.L.}},
\bauthor{\bsnm{Kowalski}, \binits{K.}},
\bauthor{\bsnm{Scholes}, \binits{G.D.}},
\bauthor{\bsnm{Yuen-Zhou}, \binits{J.}},
\bauthor{\bsnm{Baldo}, \binits{M.A.}},
\bauthor{\bsnm{Freedman}, \binits{D.E.}},
\bauthor{\bsnm{Goldsmith}, \binits{R.H.}},
\bauthor{\bsnm{Goodson}, \binits{T.}},
\bauthor{\bsnm{Kirk}, \binits{M.L.}},
\bauthor{\bsnm{McCusker}, \binits{J.K.}},
\bauthor{\bsnm{Ogilvie}, \binits{J.P.}},
\bauthor{\bsnm{Shultz}, \binits{D.A.}},
\bauthor{\bsnm{Stoll}, \binits{S.}},
\bauthor{\bsnm{Whaley}, \binits{K.B.}}:
\batitle{Exploiting chemistry and molecular systems for quantum information
  science}.
\bjtitle{Nature Reviews Chemistry}
\bvolume{4}(\bissue{9}),
\bfpage{490}--\blpage{504}
(\byear{2020})
\doiurl{10.1038/s41570-020-0200-5}
\end{barticle}
\endbibitem

\bibitem[\protect\citeauthoryear{Cramer
  et~al.}{2000}]{cramer_architecture_2000}
\begin{barticle}
\bauthor{\bsnm{Cramer}, \binits{P.}},
\bauthor{\bsnm{Bushnell}, \binits{D.A.}},
\bauthor{\bsnm{Fu}, \binits{J.}},
\bauthor{\bsnm{Gnatt}, \binits{A.L.}},
\bauthor{\bsnm{Maier-Davis}, \binits{B.}},
\bauthor{\bsnm{Thompson}, \binits{N.E.}},
\bauthor{\bsnm{Burgess}, \binits{R.R.}},
\bauthor{\bsnm{Edwards}, \binits{A.M.}},
\bauthor{\bsnm{David}, \binits{P.R.}},
\bauthor{\bsnm{Kornberg}, \binits{R.D.}}:
\batitle{Architecture of {RNA} {Polymerase} {II} and {Implications} for the
  {Transcription} {Mechanism}}.
\bjtitle{Science}
\bvolume{288}(\bissue{5466}),
\bfpage{640}--\blpage{649}
(\byear{2000})
\doiurl{10.1126/science.288.5466.640}
\end{barticle}
\endbibitem

\bibitem[\protect\citeauthoryear{Bocanegra et~al.}{2021}]{bocanegra_dna_2021}
\begin{barticle}
\bauthor{\bsnm{Bocanegra}, \binits{R.}},
\bauthor{\bsnm{Ismael~Plaza}, \binits{G.A.}},
\bauthor{\bsnm{Pulido}, \binits{C.R.}},
\bauthor{\bsnm{Ibarra}, \binits{B.}}:
\batitle{{DNA} replication machinery: {Insights} from in vitro single-molecule
  approaches}.
\bjtitle{Computational and Structural Biotechnology Journal}
\bvolume{19},
\bfpage{2057}--\blpage{2069}
(\byear{2021})
\doiurl{10.1016/j.csbj.2021.04.013}
\end{barticle}
\endbibitem

\bibitem[\protect\citeauthoryear{Yao and
  O’Donnell}{2016}]{yao_evolution_2016}
\begin{barticle}
\bauthor{\bsnm{Yao}, \binits{N.Y.}},
\bauthor{\bsnm{O’Donnell}, \binits{M.E.}}:
\batitle{Evolution of replication machines}.
\bjtitle{Critical Reviews in Biochemistry and Molecular Biology}
\bvolume{51}(\bissue{3}),
\bfpage{135}--\blpage{149}
(\byear{2016})
\doiurl{10.3109/10409238.2015.1125845}
\end{barticle}
\endbibitem

\bibitem[\protect\citeauthoryear{Kumar et~al.}{2017}]{kumar_ins_2017}
\begin{barticle}
\bauthor{\bsnm{Kumar}, \binits{A.}},
\bauthor{\bsnm{Hsu}, \binits{L.H.-H.}},
\bauthor{\bsnm{Kavanagh}, \binits{P.}},
\bauthor{\bsnm{Barrière}, \binits{F.}},
\bauthor{\bsnm{Lens}, \binits{P.N.L.}},
\bauthor{\bsnm{Lapinsonnière}, \binits{L.}},
\bauthor{\bsnm{Lienhard~V}, \binits{J.H.}},
\bauthor{\bsnm{Schröder}, \binits{U.}},
\bauthor{\bsnm{Jiang}, \binits{X.}},
\bauthor{\bsnm{Leech}, \binits{D.}}:
\batitle{The ins and outs of microorganism–electrode electron transfer
  reactions}.
\bjtitle{Nature Reviews Chemistry}
\bvolume{1}(\bissue{3}),
\bfpage{0024}
(\byear{2017})
\doiurl{10.1038/s41570-017-0024}
\end{barticle}
\endbibitem

\bibitem[\protect\citeauthoryear{Ramezani and
  Dietz}{2020}]{ramezani_building_2020}
\begin{barticle}
\bauthor{\bsnm{Ramezani}, \binits{H.}},
\bauthor{\bsnm{Dietz}, \binits{H.}}:
\batitle{Building machines with {DNA} molecules}.
\bjtitle{Nature Reviews Genetics}
\bvolume{21}(\bissue{1}),
\bfpage{5}--\blpage{26}
(\byear{2020})
\doiurl{10.1038/s41576-019-0175-6}
\end{barticle}
\endbibitem

\bibitem[\protect\citeauthoryear{Shi et~al.}{2022}]{shi_small_2022}
\begin{barticle}
\bauthor{\bsnm{Shi}, \binits{Y.}},
\bauthor{\bsnm{Chang}, \binits{Y.}},
\bauthor{\bsnm{Lu}, \binits{K.}},
\bauthor{\bsnm{Chen}, \binits{Z.}},
\bauthor{\bsnm{Zhang}, \binits{J.}},
\bauthor{\bsnm{Yan}, \binits{Y.}},
\bauthor{\bsnm{Qiu}, \binits{D.}},
\bauthor{\bsnm{Liu}, \binits{Y.}},
\bauthor{\bsnm{Adil}, \binits{M.A.}},
\bauthor{\bsnm{Ma}, \binits{W.}},
\bauthor{\bsnm{Hao}, \binits{X.}},
\bauthor{\bsnm{Zhu}, \binits{L.}},
\bauthor{\bsnm{Wei}, \binits{Z.}}:
\batitle{Small reorganization energy acceptors enable low energy losses in
  non-fullerene organic solar cells}.
\bjtitle{Nature Communications}
\bvolume{13}(\bissue{1}),
\bfpage{3256}
(\byear{2022})
\doiurl{10.1038/s41467-022-30927-y}
\end{barticle}
\endbibitem

\bibitem[\protect\citeauthoryear{Arden et~al.}{2005}]{arden_more_2005}
\begin{bchapter}
\bauthor{\bsnm{Arden}, \binits{W.}},
\bauthor{\bsnm{Brillouet}, \binits{M.}},
\bauthor{\bsnm{Cogez}},
\bauthor{\bsnm{Graef}, \binits{M.}},
\bauthor{\bsnm{Huizing}, \binits{B.}},
\bauthor{\bsnm{Mahnkopf}, \binits{R.}}:
\bctitle{More {Than} {Moore} {White} {Paper}}.
In: \bbtitle{2005 {International} {Technology} {Roadmap} for {Semiconductors}},
pp. \bfpage{1}--\blpage{47}
(\byear{2005})
\end{bchapter}
\endbibitem

\bibitem[\protect\citeauthoryear{Li et~al.}{2022}]{li_redox-labelled_2022}
\begin{barticle}
\bauthor{\bsnm{Li}, \binits{S.}},
\bauthor{\bsnm{Coffinier}, \binits{Y.}},
\bauthor{\bsnm{Lagadec}, \binits{C.}},
\bauthor{\bsnm{Cleri}, \binits{F.}},
\bauthor{\bsnm{Nishiguchi}, \binits{K.}},
\bauthor{\bsnm{Fujiwara}, \binits{A.}},
\bauthor{\bsnm{Fujii}, \binits{T.}},
\bauthor{\bsnm{Kim}, \binits{S.-H.}},
\bauthor{\bsnm{Clément}, \binits{N.}}:
\batitle{Redox-labelled electrochemical aptasensors with nanosupported cancer
  cells}.
\bjtitle{Biosensors and Bioelectronics}
\bvolume{216},
\bfpage{114643}
(\byear{2022})
\doiurl{10.1016/j.bios.2022.114643}
\end{barticle}
\endbibitem

\bibitem[\protect\citeauthoryear{Arroyo-Currás
  et~al.}{2018}]{arroyo-curras_subsecond-resolved_2018}
\begin{barticle}
\bauthor{\bsnm{Arroyo-Currás}, \binits{N.}},
\bauthor{\bsnm{Dauphin-Ducharme}, \binits{P.}},
\bauthor{\bsnm{Ortega}, \binits{G.}},
\bauthor{\bsnm{Ploense}, \binits{K.L.}},
\bauthor{\bsnm{Kippin}, \binits{T.E.}},
\bauthor{\bsnm{Plaxco}, \binits{K.W.}}:
\batitle{Subsecond-{Resolved} {Molecular} {Measurements} in the {Living} {Body}
  {Using} {Chronoamperometrically} {Interrogated} {Aptamer}-{Based} {Sensors}}.
\bjtitle{ACS Sensors}
\bvolume{3}(\bissue{2}),
\bfpage{360}--\blpage{366}
(\byear{2018})
\doiurl{10.1021/acssensors.7b00787}
\end{barticle}
\endbibitem

\bibitem[\protect\citeauthoryear{Dauphin-Ducharme
  et~al.}{2019}]{dauphin-ducharme_high-precision_2019}
\begin{barticle}
\bauthor{\bsnm{Dauphin-Ducharme}, \binits{P.}},
\bauthor{\bsnm{Arroyo-Currás}, \binits{N.}},
\bauthor{\bsnm{Plaxco}, \binits{K.W.}}:
\batitle{High-{Precision} {Electrochemical} {Measurements} of the {Guanine}-,
  {Mismatch}-, and {Length}-{Dependence} of {Electron} {Transfer} from
  {Electrode}-{Bound} {DNA} {Are} {Consistent} with a {Contact}-{Mediated}
  {Mechanism}}.
\bjtitle{Journal of the American Chemical Society}
\bvolume{141}(\bissue{3}),
\bfpage{1304}--\blpage{1311}
(\byear{2019})
\doiurl{10.1021/jacs.8b11341}
\end{barticle}
\endbibitem

\bibitem[\protect\citeauthoryear{Nakatsuka
  et~al.}{2018}]{nakatsuka_aptamerfield-effect_2018}
\begin{barticle}
\bauthor{\bsnm{Nakatsuka}, \binits{N.}},
\bauthor{\bsnm{Yang}, \binits{K.-A.}},
\bauthor{\bsnm{Abendroth}, \binits{J.M.}},
\bauthor{\bsnm{Cheung}, \binits{K.M.}},
\bauthor{\bsnm{Xu}, \binits{X.}},
\bauthor{\bsnm{Yang}, \binits{H.}},
\bauthor{\bsnm{Zhao}, \binits{C.}},
\bauthor{\bsnm{Zhu}, \binits{B.}},
\bauthor{\bsnm{Rim}, \binits{Y.S.}},
\bauthor{\bsnm{Yang}, \binits{Y.}},
\bauthor{\bsnm{Weiss}, \binits{P.S.}},
\bauthor{\bsnm{Stojanović}, \binits{M.N.}},
\bauthor{\bsnm{Andrews}, \binits{A.M.}}:
\batitle{Aptamer–field-effect transistors overcome {Debye} length limitations
  for small-molecule sensing}.
\bjtitle{Science}
\bvolume{362}(\bissue{6412}),
\bfpage{319}--\blpage{324}
(\byear{2018})
\doiurl{10.1126/science.aao6750}
\end{barticle}
\endbibitem

\bibitem[\protect\citeauthoryear{Verrinder
  et~al.}{2023}]{verrinder_comparison_2023}
\begin{botherref}
\oauthor{\bsnm{Verrinder}, \binits{E.}},
\oauthor{\bsnm{Leung}, \binits{K.K.}},
\oauthor{\bsnm{Erdal}, \binits{M.K.}},
\oauthor{\bsnm{Sepunaru}, \binits{L.}},
\oauthor{\bsnm{Plaxco}, \binits{K.W.}}:
Comparison of voltammetric methods used in the interrogation of electrochemical
  aptamer-based sensors.
Sensors \& Diagnostics,
10--1039300083
(2023)
\doiurl{10.1039/D3SD00083D}
\end{botherref}
\endbibitem

\bibitem[\protect\citeauthoryear{Slinker
  et~al.}{2010}]{slinker_multiplexed_2010}
\begin{barticle}
\bauthor{\bsnm{Slinker}, \binits{J.D.}},
\bauthor{\bsnm{Muren}, \binits{N.B.}},
\bauthor{\bsnm{Gorodetsky}, \binits{A.A.}},
\bauthor{\bsnm{Barton}, \binits{J.K.}}:
\batitle{Multiplexed {DNA}-{Modified} {Electrodes}}.
\bjtitle{Journal of the American Chemical Society}
\bvolume{132}(\bissue{8}),
\bfpage{2769}--\blpage{2774}
(\byear{2010})
\doiurl{10.1021/ja909915m}
\end{barticle}
\endbibitem

\bibitem[\protect\citeauthoryear{Nguyen et~al.}{2021}]{nguyen_scaling_2021}
\begin{barticle}
\bauthor{\bsnm{Nguyen}, \binits{B.H.}},
\bauthor{\bsnm{Takahashi}, \binits{C.N.}},
\bauthor{\bsnm{Gupta}, \binits{G.}},
\bauthor{\bsnm{Smith}, \binits{J.A.}},
\bauthor{\bsnm{Rouse}, \binits{R.}},
\bauthor{\bsnm{Berndt}, \binits{P.}},
\bauthor{\bsnm{Yekhanin}, \binits{S.}},
\bauthor{\bsnm{Ward}, \binits{D.P.}},
\bauthor{\bsnm{Ang}, \binits{S.D.}},
\bauthor{\bsnm{Garvan}, \binits{P.}},
\bauthor{\bsnm{Parker}, \binits{H.-Y.}},
\bauthor{\bsnm{Carlson}, \binits{R.}},
\bauthor{\bsnm{Carmean}, \binits{D.}},
\bauthor{\bsnm{Ceze}, \binits{L.}},
\bauthor{\bsnm{Strauss}, \binits{K.}}:
\batitle{Scaling {DNA} data storage with nanoscale electrode wells}.
\bjtitle{Science Advances}
\bvolume{7}(\bissue{48}),
\bfpage{6714}
(\byear{2021})
\doiurl{10.1126/sciadv.abi6714}
\end{barticle}
\endbibitem

\bibitem[\protect\citeauthoryear{Yu et~al.}{2024}]{yu_high-throughput_2024}
\begin{barticle}
\bauthor{\bsnm{Yu}, \binits{M.}},
\bauthor{\bsnm{Tang}, \binits{X.}},
\bauthor{\bsnm{Li}, \binits{Z.}},
\bauthor{\bsnm{Wang}, \binits{W.}},
\bauthor{\bsnm{Wang}, \binits{S.}},
\bauthor{\bsnm{Li}, \binits{M.}},
\bauthor{\bsnm{Yu}, \binits{Q.}},
\bauthor{\bsnm{Xie}, \binits{S.}},
\bauthor{\bsnm{Zuo}, \binits{X.}},
\bauthor{\bsnm{Chen}, \binits{C.}}:
\batitle{High-throughput {DNA} synthesis for data storage}.
\bjtitle{Chemical Society Reviews}
\bvolume{53}(\bissue{9}),
\bfpage{4463}--\blpage{4489}
(\byear{2024})
\doiurl{10.1039/D3CS00469D}
\end{barticle}
\endbibitem

\bibitem[\protect\citeauthoryear{Moazzenzade
  et~al.}{2020}]{moazzenzade_stochastic_2020}
\begin{barticle}
\bauthor{\bsnm{Moazzenzade}, \binits{T.}},
\bauthor{\bsnm{Huskens}, \binits{J.}},
\bauthor{\bsnm{Lemay}, \binits{S.G.}}:
\batitle{Stochastic electrochemistry at ultralow concentrations: the case for
  digital sensors}.
\bjtitle{The Analyst}
\bvolume{145}(\bissue{3}),
\bfpage{750}--\blpage{758}
(\byear{2020})
\doiurl{10.1039/C9AN01832H}
\end{barticle}
\endbibitem

\bibitem[\protect\citeauthoryear{Bao et~al.}{2021}]{bao_digital_2021}
\begin{barticle}
\bauthor{\bsnm{Bao}, \binits{B.}},
\bauthor{\bsnm{Rivkin}, \binits{B.}},
\bauthor{\bsnm{Akbar}, \binits{F.}},
\bauthor{\bsnm{Karnaushenko}, \binits{D.D.}},
\bauthor{\bsnm{Bandari}, \binits{V.K.}},
\bauthor{\bsnm{Teuerle}, \binits{L.}},
\bauthor{\bsnm{Becker}, \binits{C.}},
\bauthor{\bsnm{Baunack}, \binits{S.}},
\bauthor{\bsnm{Karnaushenko}, \binits{D.}},
\bauthor{\bsnm{Schmidt}, \binits{O.G.}}:
\batitle{Digital {Electrochemistry} for {On}‐{Chip} {Heterogeneous}
  {Material} {Integration}}.
\bjtitle{Advanced Materials}
\bvolume{33}(\bissue{26}),
\bfpage{2101272}
(\byear{2021})
\doiurl{10.1002/adma.202101272}
\end{barticle}
\endbibitem

\bibitem[\protect\citeauthoryear{}{2024}]{COMSOL}
\begin{botherref}
COMSOL electrochemistry module.
\url{https://www.comsol.com/electrochemistry-module}.
Accessed: 01/10/2024
(2024)
\end{botherref}
\endbibitem

\bibitem[\protect\citeauthoryear{}{2024}]{ANSYS}
\begin{botherref}
Ansys battery application.
\url{https://www.ansys.com/applications/battery/battery-cell-and-electrode}.
Accessed: 01/10/2024
(2024)
\end{botherref}
\endbibitem

\bibitem[\protect\citeauthoryear{}{2024}]{DIGIELCH}
\begin{botherref}
Digielch electrochemical simulation software.
\url{https://www.gamry.com/cells-and-accessories/digielch-electrochemical-simulation-software/}.
Accessed: 01/10/2024
(2024)
\end{botherref}
\endbibitem

\bibitem[\protect\citeauthoryear{De~La~Lande
  et~al.}{2015}]{de_la_lande_progress_2015}
\begin{barticle}
\bauthor{\bsnm{De~La~Lande}, \binits{A.}},
\bauthor{\bsnm{Gillet}, \binits{N.}},
\bauthor{\bsnm{Chen}, \binits{S.}},
\bauthor{\bsnm{Salahub}, \binits{D.R.}}:
\batitle{Progress and challenges in simulating and understanding electron
  transfer in proteins}.
\bjtitle{Archives of Biochemistry and Biophysics}
\bvolume{582},
\bfpage{28}--\blpage{41}
(\byear{2015})
\doiurl{10.1016/j.abb.2015.06.016}
\end{barticle}
\endbibitem

\bibitem[\protect\citeauthoryear{Kmiecik
  et~al.}{2016}]{kmiecik_coarse-grained_2016}
\begin{barticle}
\bauthor{\bsnm{Kmiecik}, \binits{S.}},
\bauthor{\bsnm{Gront}, \binits{D.}},
\bauthor{\bsnm{Kolinski}, \binits{M.}},
\bauthor{\bsnm{Wieteska}, \binits{L.}},
\bauthor{\bsnm{Dawid}, \binits{A.E.}},
\bauthor{\bsnm{Kolinski}, \binits{A.}}:
\batitle{Coarse-{Grained} {Protein} {Models} and {Their} {Applications}}.
\bjtitle{Chemical Reviews}
\bvolume{116}(\bissue{14}),
\bfpage{7898}--\blpage{7936}
(\byear{2016})
\doiurl{10.1021/acs.chemrev.6b00163}
\end{barticle}
\endbibitem

\bibitem[\protect\citeauthoryear{Schlick
  et~al.}{2021}]{schlick_biomolecular_2021}
\begin{barticle}
\bauthor{\bsnm{Schlick}, \binits{T.}},
\bauthor{\bsnm{Portillo-Ledesma}, \binits{S.}},
\bauthor{\bsnm{Myers}, \binits{C.G.}},
\bauthor{\bsnm{Beljak}, \binits{L.}},
\bauthor{\bsnm{Chen}, \binits{J.}},
\bauthor{\bsnm{Dakhel}, \binits{S.}},
\bauthor{\bsnm{Darling}, \binits{D.}},
\bauthor{\bsnm{Ghosh}, \binits{S.}},
\bauthor{\bsnm{Hall}, \binits{J.}},
\bauthor{\bsnm{Jan}, \binits{M.}},
\bauthor{\bsnm{Liang}, \binits{E.}},
\bauthor{\bsnm{Saju}, \binits{S.}},
\bauthor{\bsnm{Vohr}, \binits{M.}},
\bauthor{\bsnm{Wu}, \binits{C.}},
\bauthor{\bsnm{Xu}, \binits{Y.}},
\bauthor{\bsnm{Xue}, \binits{E.}}:
\batitle{Biomolecular {Modeling} and {Simulation}: {A} {Prospering}
  {Multidisciplinary} {Field}}.
\bjtitle{Annual Review of Biophysics}
\bvolume{50}(\bissue{1}),
\bfpage{267}--\blpage{301}
(\byear{2021})
\doiurl{10.1146/annurev-biophys-091720-102019}
\end{barticle}
\endbibitem

\bibitem[\protect\citeauthoryear{May and Kühn}{2004}]{may_charge_2004}
\begin{bbook}
\bauthor{\bsnm{May}, \binits{V.}},
\bauthor{\bsnm{Kühn}, \binits{O.}}:
\bbtitle{Charge and Energy Transfer Dynamics in Molecular Systems},
\bedition{2., rev. and enlarged ed., 2. reprint} edn.
\bpublisher{Wiley-VCH},
\blocation{Weinheim}
(\byear{2004})
\end{bbook}
\endbibitem

\bibitem[\protect\citeauthoryear{Gray and Winkler}{2005}]{gray_long-range_2005}
\begin{barticle}
\bauthor{\bsnm{Gray}, \binits{H.B.}},
\bauthor{\bsnm{Winkler}, \binits{J.R.}}:
\batitle{Long-range electron transfer}.
\bjtitle{Proceedings of the National Academy of Sciences}
\bvolume{102}(\bissue{10}),
\bfpage{3534}--\blpage{3539}
(\byear{2005})
\doiurl{10.1073/pnas.0408029102}
\end{barticle}
\endbibitem

\bibitem[\protect\citeauthoryear{Ouldridge et~al.}{2010}]{ouldridge_dna_2010}
\begin{barticle}
\bauthor{\bsnm{Ouldridge}, \binits{T.E.}},
\bauthor{\bsnm{Louis}, \binits{A.A.}},
\bauthor{\bsnm{Doye}, \binits{J.P.K.}}:
\batitle{{DNA} {Nanotweezers} {Studied} with a {Coarse}-{Grained} {Model} of
  {DNA}}.
\bjtitle{Physical Review Letters}
\bvolume{104}(\bissue{17}),
\bfpage{178101}
(\byear{2010})
\doiurl{10.1103/PhysRevLett.104.178101}
\end{barticle}
\endbibitem

\bibitem[\protect\citeauthoryear{Rovigatti
  et~al.}{2014}]{rovigatti_accurate_2014}
\begin{barticle}
\bauthor{\bsnm{Rovigatti}, \binits{L.}},
\bauthor{\bsnm{Bomboi}, \binits{F.}},
\bauthor{\bsnm{Sciortino}, \binits{F.}}:
\batitle{Accurate phase diagram of tetravalent {DNA} nanostars}.
\bjtitle{The Journal of Chemical Physics}
\bvolume{140}(\bissue{15}),
\bfpage{154903}
(\byear{2014})
\doiurl{10.1063/1.4870467}
\end{barticle}
\endbibitem

\bibitem[\protect\citeauthoryear{Poppleton et~al.}{2023}]{poppleton_oxdna_2023}
\begin{barticle}
\bauthor{\bsnm{Poppleton}, \binits{E.}},
\bauthor{\bsnm{Matthies}, \binits{M.}},
\bauthor{\bsnm{Mandal}, \binits{D.}},
\bauthor{\bsnm{Romano}, \binits{F.}},
\bauthor{\bsnm{Šulc}, \binits{P.}},
\bauthor{\bsnm{Rovigatti}, \binits{L.}}:
\batitle{{oxDNA}: coarse-grained simulations of nucleic acids madesimple}.
\bjtitle{Journal of Open Source Software}
\bvolume{8}(\bissue{81}),
\bfpage{4693}
(\byear{2023})
\doiurl{10.21105/joss.04693}
\end{barticle}
\endbibitem

\bibitem[\protect\citeauthoryear{Rudd and
  Broughton}{2005}]{rudd_coarse-grained_2005}
\begin{barticle}
\bauthor{\bsnm{Rudd}, \binits{R.E.}},
\bauthor{\bsnm{Broughton}, \binits{J.Q.}}:
\batitle{Coarse-grained molecular dynamics: {Nonlinear} finite elements and
  finite temperature}.
\bjtitle{Physical Review B}
\bvolume{72}(\bissue{14}),
\bfpage{144104}
(\byear{2005})
\doiurl{10.1103/PhysRevB.72.144104}
\end{barticle}
\endbibitem

\bibitem[\protect\citeauthoryear{Capozzi
  et~al.}{2015}]{capozzi_single-molecule_2015}
\begin{barticle}
\bauthor{\bsnm{Capozzi}, \binits{B.}},
\bauthor{\bsnm{Xia}, \binits{J.}},
\bauthor{\bsnm{Adak}, \binits{O.}},
\bauthor{\bsnm{Dell}, \binits{E.J.}},
\bauthor{\bsnm{Liu}, \binits{Z.-F.}},
\bauthor{\bsnm{Taylor}, \binits{J.C.}},
\bauthor{\bsnm{Neaton}, \binits{J.B.}},
\bauthor{\bsnm{Campos}, \binits{L.M.}},
\bauthor{\bsnm{Venkataraman}, \binits{L.}}:
\batitle{Single-molecule diodes with high rectification ratios through
  environmental control}.
\bjtitle{Nature Nanotechnology}
\bvolume{10}(\bissue{6}),
\bfpage{522}--\blpage{527}
(\byear{2015})
\doiurl{10.1038/nnano.2015.97}
\end{barticle}
\endbibitem

\bibitem[\protect\citeauthoryear{Gehring
  et~al.}{2019}]{gehring_single-molecule_2019}
\begin{barticle}
\bauthor{\bsnm{Gehring}, \binits{P.}},
\bauthor{\bsnm{Thijssen}, \binits{J.M.}},
\bauthor{\bsnm{Van Der~Zant}, \binits{H.S.J.}}:
\batitle{Single-molecule quantum-transport phenomena in break junctions}.
\bjtitle{Nature Reviews Physics}
\bvolume{1}(\bissue{6}),
\bfpage{381}--\blpage{396}
(\byear{2019})
\doiurl{10.1038/s42254-019-0055-1}
\end{barticle}
\endbibitem

\bibitem[\protect\citeauthoryear{Yuan et~al.}{2015}]{yuan_controlling_2015}
\begin{barticle}
\bauthor{\bsnm{Yuan}, \binits{L.}},
\bauthor{\bsnm{Nerngchamnong}, \binits{N.}},
\bauthor{\bsnm{Cao}, \binits{L.}},
\bauthor{\bsnm{Hamoudi}, \binits{H.}},
\bauthor{\bsnm{Barco}, \binits{E.}},
\bauthor{\bsnm{Roemer}, \binits{M.}},
\bauthor{\bsnm{Sriramula}, \binits{R.K.}},
\bauthor{\bsnm{Thompson}, \binits{D.}},
\bauthor{\bsnm{Nijhuis}, \binits{C.A.}}:
\batitle{Controlling the direction of rectification in a molecular diode}.
\bjtitle{Nature Communications}
\bvolume{6}(\bissue{1}),
\bfpage{6324}
(\byear{2015})
\doiurl{10.1038/ncomms7324}
\end{barticle}
\endbibitem

\bibitem[\protect\citeauthoryear{Wang et~al.}{2022}]{wang_dynamic_2022}
\begin{barticle}
\bauthor{\bsnm{Wang}, \binits{Y.}},
\bauthor{\bsnm{Zhang}, \binits{Q.}},
\bauthor{\bsnm{Astier}, \binits{H.P.A.G.}},
\bauthor{\bsnm{Nickle}, \binits{C.}},
\bauthor{\bsnm{Soni}, \binits{S.}},
\bauthor{\bsnm{Alami}, \binits{F.A.}},
\bauthor{\bsnm{Borrini}, \binits{A.}},
\bauthor{\bsnm{Zhang}, \binits{Z.}},
\bauthor{\bsnm{Honnigfort}, \binits{C.}},
\bauthor{\bsnm{Braunschweig}, \binits{B.}},
\bauthor{\bsnm{Leoncini}, \binits{A.}},
\bauthor{\bsnm{Qi}, \binits{D.-C.}},
\bauthor{\bsnm{Han}, \binits{Y.}},
\bauthor{\bsnm{Del~Barco}, \binits{E.}},
\bauthor{\bsnm{Thompson}, \binits{D.}},
\bauthor{\bsnm{Nijhuis}, \binits{C.A.}}:
\batitle{Dynamic molecular switches with hysteretic negative differential
  conductance emulating synaptic behaviour}.
\bjtitle{Nature Materials}
\bvolume{21}(\bissue{12}),
\bfpage{1403}--\blpage{1411}
(\byear{2022})
\doiurl{10.1038/s41563-022-01402-2}
\end{barticle}
\endbibitem

\bibitem[\protect\citeauthoryear{Zhang et~al.}{2024}]{zhang_molecular_2024}
\begin{barticle}
\bauthor{\bsnm{Zhang}, \binits{Q.}},
\bauthor{\bsnm{Wang}, \binits{Y.}},
\bauthor{\bsnm{Nickle}, \binits{C.}},
\bauthor{\bsnm{Zhang}, \binits{Z.}},
\bauthor{\bsnm{Leoncini}, \binits{A.}},
\bauthor{\bsnm{Qi}, \binits{D.-C.}},
\bauthor{\bsnm{Sotthewes}, \binits{K.}},
\bauthor{\bsnm{Borrini}, \binits{A.}},
\bauthor{\bsnm{Zandvliet}, \binits{H.J.W.}},
\bauthor{\bsnm{Del~Barco}, \binits{E.}},
\bauthor{\bsnm{Thompson}, \binits{D.}},
\bauthor{\bsnm{Nijhuis}, \binits{C.A.}}:
\batitle{Molecular switching by proton-coupled electron transport drives giant
  negative differential resistance}.
\bjtitle{Nature Communications}
\bvolume{15}(\bissue{1}),
\bfpage{8300}
(\byear{2024})
\doiurl{10.1038/s41467-024-52496-y}
\end{barticle}
\endbibitem

\bibitem[\protect\citeauthoryear{Dickinson
  et~al.}{2014}]{dickinson_comsol_2014}
\begin{barticle}
\bauthor{\bsnm{Dickinson}, \binits{E.J.F.}},
\bauthor{\bsnm{Ekström}, \binits{H.}},
\bauthor{\bsnm{Fontes}, \binits{E.}}:
\batitle{{COMSOL} {Multiphysics}®: {Finite} element software for
  electrochemical analysis. {A} mini-review}.
\bjtitle{Electrochemistry Communications}
\bvolume{40},
\bfpage{71}--\blpage{74}
(\byear{2014})
\doiurl{10.1016/j.elecom.2013.12.020}
\end{barticle}
\endbibitem

\bibitem[\protect\citeauthoryear{Anne and Demaille}{2006}]{anne_dynamics_2006}
\begin{barticle}
\bauthor{\bsnm{Anne}, \binits{A.}},
\bauthor{\bsnm{Demaille}, \binits{C.}}:
\batitle{Dynamics of {Electron} {Transport} by {Elastic} {Bending} of {Short}
  {DNA} {Duplexes}. {Experimental} {Study} and {Quantitative} {Modeling} of the
  {Cyclic} {Voltammetric} {Behavior} of 3‘-{Ferrocenyl} {DNA} {End}-{Grafted}
  on {Gold}}.
\bjtitle{Journal of the American Chemical Society}
\bvolume{128}(\bissue{2}),
\bfpage{542}--\blpage{557}
(\byear{2006})
\doiurl{10.1021/ja055112a}
\end{barticle}
\endbibitem

\bibitem[\protect\citeauthoryear{Anne and Demaille}{2008}]{anne_electron_2008}
\begin{barticle}
\bauthor{\bsnm{Anne}, \binits{A.}},
\bauthor{\bsnm{Demaille}, \binits{C.}}:
\batitle{Electron {Transport} by {Molecular} {Motion} of redox-{DNA} {Strands}:
  {Unexpectedly} {Slow} {Rotational} {Dynamics} of 20-mer ds-{DNA} {Chains}
  {End}-{Grafted} onto {Surfaces} via {C}$_{\textrm{6}}$ {Linkers}}.
\bjtitle{Journal of the American Chemical Society}
\bvolume{130}(\bissue{30}),
\bfpage{9812}--\blpage{9823}
(\byear{2008})
\doiurl{10.1021/ja801074m}
\end{barticle}
\endbibitem

\bibitem[\protect\citeauthoryear{Huang and White}{2013}]{huang_random_2013}
\begin{barticle}
\bauthor{\bsnm{Huang}, \binits{K.-C.}},
\bauthor{\bsnm{White}, \binits{R.J.}}:
\batitle{Random {Walk} on a {Leash}: {A} {Simple} {Single}-{Molecule}
  {Diffusion} {Model} for {Surface}-{Tethered} {Redox} {Molecules} with
  {Flexible} {Linkers}}.
\bjtitle{Journal of the American Chemical Society}
\bvolume{135}(\bissue{34}),
\bfpage{12808}--\blpage{12817}
(\byear{2013})
\doiurl{10.1021/ja4060788}
\end{barticle}
\endbibitem

\bibitem[\protect\citeauthoryear{Dauphin-Ducharme
  et~al.}{2018}]{dauphin-ducharme_chain_2018}
\begin{barticle}
\bauthor{\bsnm{Dauphin-Ducharme}, \binits{P.}},
\bauthor{\bsnm{Arroyo-Currás}, \binits{N.}},
\bauthor{\bsnm{Adhikari}, \binits{R.}},
\bauthor{\bsnm{Somerson}, \binits{J.}},
\bauthor{\bsnm{Ortega}, \binits{G.}},
\bauthor{\bsnm{Makarov}, \binits{D.E.}},
\bauthor{\bsnm{Plaxco}, \binits{K.W.}}:
\batitle{Chain {Dynamics} {Limit} {Electron} {Transfer} from
  {Electrode}-{Bound}, {Single}-{Stranded} {Oligonucleotides}}.
\bjtitle{The Journal of Physical Chemistry C}
\bvolume{122}(\bissue{37}),
\bfpage{21441}--\blpage{21448}
(\byear{2018})
\doiurl{10.1021/acs.jpcc.8b06111}
\end{barticle}
\endbibitem

\bibitem[\protect\citeauthoryear{Zheng
  et~al.}{2023}]{zheng_electrochemical_2023}
\begin{barticle}
\bauthor{\bsnm{Zheng}, \binits{Z.}},
\bauthor{\bsnm{Kim}, \binits{S.H.}},
\bauthor{\bsnm{Chovin}, \binits{A.}},
\bauthor{\bsnm{Clement}, \binits{N.}},
\bauthor{\bsnm{Demaille}, \binits{C.}}:
\batitle{Electrochemical response of surface-attached redox {DNA} governed by
  low activation energy electron transfer kinetics}.
\bjtitle{Chemical Science}
\bvolume{14}(\bissue{13}),
\bfpage{3652}--\blpage{3660}
(\byear{2023})
\doiurl{10.1039/D3SC00320E}
\end{barticle}
\endbibitem

\bibitem[\protect\citeauthoryear{Madrid et~al.}{2023}]{madrid_ballistic_2023}
\begin{barticle}
\bauthor{\bsnm{Madrid}, \binits{I.}},
\bauthor{\bsnm{Zheng}, \binits{Z.}},
\bauthor{\bsnm{Gerbelot}, \binits{C.}},
\bauthor{\bsnm{Fujiwara}, \binits{A.}},
\bauthor{\bsnm{Li}, \binits{S.}},
\bauthor{\bsnm{Grall}, \binits{S.}},
\bauthor{\bsnm{Nishiguchi}, \binits{K.}},
\bauthor{\bsnm{Kim}, \binits{S.H.}},
\bauthor{\bsnm{Chovin}, \binits{A.}},
\bauthor{\bsnm{Demaille}, \binits{C.}},
\bauthor{\bsnm{Clement}, \binits{N.}}:
\batitle{Ballistic {Brownian} {Motion} of {Nanoconfined} {DNA}}.
\bjtitle{ACS Nano}
\bvolume{17}(\bissue{17}),
\bfpage{17031}--\blpage{17040}
(\byear{2023})
\doiurl{10.1021/acsnano.3c04349}
\end{barticle}
\endbibitem

\bibitem[\protect\citeauthoryear{Zheng
  et~al.}{2024}]{zheng_activationless_2024}
\begin{barticle}
\bauthor{\bsnm{Zheng}, \binits{Z.}},
\bauthor{\bsnm{Grall}, \binits{S.}},
\bauthor{\bsnm{Kim}, \binits{S.H.}},
\bauthor{\bsnm{Chovin}, \binits{A.}},
\bauthor{\bsnm{Clement}, \binits{N.}},
\bauthor{\bsnm{Demaille}, \binits{C.}}:
\batitle{Activationless {Electron} {Transfer} of redox-{DNA} in
  {Electrochemical} {Nanogaps}}.
\bjtitle{Journal of the American Chemical Society}
\bvolume{146}(\bissue{9}),
\bfpage{6094}--\blpage{6103}
(\byear{2024})
\doiurl{10.1021/jacs.3c13532}
\end{barticle}
\endbibitem

\bibitem[\protect\citeauthoryear{Martin and
  Matyushov}{2015}]{martin_dipolar_2015}
\begin{barticle}
\bauthor{\bsnm{Martin}, \binits{D.R.}},
\bauthor{\bsnm{Matyushov}, \binits{D.V.}}:
\batitle{Dipolar {Nanodomains} in {Protein} {Hydration} {Shells}}.
\bjtitle{The Journal of Physical Chemistry Letters}
\bvolume{6}(\bissue{3}),
\bfpage{407}--\blpage{412}
(\byear{2015})
\doiurl{10.1021/jz5025433}
\end{barticle}
\endbibitem

\bibitem[\protect\citeauthoryear{Amatore
  et~al.}{2003}]{amatore_electrochemistry_2003}
\begin{barticle}
\bauthor{\bsnm{Amatore}, \binits{C.}},
\bauthor{\bsnm{Grün}, \binits{F.}},
\bauthor{\bsnm{Maisonhaute}, \binits{E.}}:
\batitle{Electrochemistry within a {Limited} {Number} of {Molecules}:
  {Delineating} the {Fringe} {Between} {Stochastic} and {Statistical}
  {Behavior}}.
\bjtitle{Angewandte Chemie International Edition}
\bvolume{42}(\bissue{40}),
\bfpage{4944}--\blpage{4947}
(\byear{2003})
\doiurl{10.1002/anie.200352353}
\end{barticle}
\endbibitem

\bibitem[\protect\citeauthoryear{Cutress and Compton}{2011}]{cutress_how_2011}
\begin{barticle}
\bauthor{\bsnm{Cutress}, \binits{I.J.}},
\bauthor{\bsnm{Compton}, \binits{R.G.}}:
\batitle{How many molecules are required to measure a cyclic voltammogram?}
\bjtitle{Chemical Physics Letters}
\bvolume{508}(\bissue{4-6}),
\bfpage{306}--\blpage{313}
(\byear{2011})
\doiurl{10.1016/j.cplett.2011.04.036} .
\bcomment{Number: 4-6}
\end{barticle}
\endbibitem

\bibitem[\protect\citeauthoryear{Singh and Lemay}{2016}]{singh_stochastic_2016}
\begin{barticle}
\bauthor{\bsnm{Singh}, \binits{P.S.}},
\bauthor{\bsnm{Lemay}, \binits{S.G.}}:
\batitle{Stochastic {Processes} in {Electrochemistry}}.
\bjtitle{Analytical Chemistry}
\bvolume{88}(\bissue{10}),
\bfpage{5017}--\blpage{5027}
(\byear{2016})
\doiurl{10.1021/acs.analchem.6b00683}
\end{barticle}
\endbibitem

\bibitem[\protect\citeauthoryear{Ren and
  Edwards}{2021}]{ren_stochasticity_2021}
\begin{barticle}
\bauthor{\bsnm{Ren}, \binits{H.}},
\bauthor{\bsnm{Edwards}, \binits{M.A.}}:
\batitle{Stochasticity in single-entity electrochemistry}.
\bjtitle{Current Opinion in Electrochemistry}
\bvolume{25},
\bfpage{100632}
(\byear{2021})
\doiurl{10.1016/j.coelec.2020.08.014}
\end{barticle}
\endbibitem

\bibitem[\protect\citeauthoryear{Bard and
  Fan}{1996}]{bard_electrochemical_1996}
\begin{barticle}
\bauthor{\bsnm{Bard}, \binits{A.J.}},
\bauthor{\bsnm{Fan}, \binits{F.-R.F.}}:
\batitle{Electrochemical {Detection} of {Single} {Molecules}}.
\bjtitle{Accounts of Chemical Research}
\bvolume{29},
\bfpage{572}--\blpage{578}
(\byear{1996})
\doiurl{10.1126/science.267.5199.871}
\end{barticle}
\endbibitem

\bibitem[\protect\citeauthoryear{Cutress
  et~al.}{2011}]{cutress_electrochemical_2011}
\begin{barticle}
\bauthor{\bsnm{Cutress}, \binits{I.J.}},
\bauthor{\bsnm{Dickinson}, \binits{E.J.F.}},
\bauthor{\bsnm{Compton}, \binits{R.G.}}:
\batitle{Electrochemical random-walk theory}.
\bjtitle{Journal of Electroanalytical Chemistry}
\bvolume{655}(\bissue{1}),
\bfpage{1}--\blpage{8}
(\byear{2011})
\doiurl{10.1016/j.jelechem.2011.02.023}
\end{barticle}
\endbibitem

\bibitem[\protect\citeauthoryear{Kätelhön et~al.}{2013}]{katelhon_noise_2013}
\begin{barticle}
\bauthor{\bsnm{Kätelhön}, \binits{E.}},
\bauthor{\bsnm{Krause}, \binits{K.J.}},
\bauthor{\bsnm{Singh}, \binits{P.S.}},
\bauthor{\bsnm{Lemay}, \binits{S.G.}},
\bauthor{\bsnm{Wolfrum}, \binits{B.}}:
\batitle{Noise {Characteristics} of {Nanoscaled} {Redox}-{Cycling} {Sensors}:
  {Investigations} {Based} on {Random} {Walks}}.
\bjtitle{Journal of the American Chemical Society}
\bvolume{135}(\bissue{24}),
\bfpage{8874}--\blpage{8881}
(\byear{2013})
\doiurl{10.1021/ja3121313}
\end{barticle}
\endbibitem

\bibitem[\protect\citeauthoryear{Grall
  et~al.}{2023}]{grall_electrochemical_2023}
\begin{barticle}
\bauthor{\bsnm{Grall}, \binits{S.}},
\bauthor{\bsnm{Li}, \binits{S.}},
\bauthor{\bsnm{Jalabert}, \binits{L.}},
\bauthor{\bsnm{Kim}, \binits{S.H.}},
\bauthor{\bsnm{Chovin}, \binits{A.}},
\bauthor{\bsnm{Demaille}, \binits{C.}},
\bauthor{\bsnm{Clément}, \binits{N.}}:
\batitle{Electrochemical {Shot} {Noise} of a {Redox} {Monolayer}}.
\bjtitle{Physical Review Letters}
\bvolume{130}(\bissue{21}),
\bfpage{218001}
(\byear{2023})
\doiurl{10.1103/PhysRevLett.130.218001}
\end{barticle}
\endbibitem

\bibitem[\protect\citeauthoryear{White and White}{2005}]{white_random_2005}
\begin{barticle}
\bauthor{\bsnm{White}, \binits{R.J.}},
\bauthor{\bsnm{White}, \binits{H.S.}}:
\batitle{A {Random} {Walk} through {Electron}-{Transfer} {Kinetics}}.
\bjtitle{Analytical Chemistry}
\bvolume{77}(\bissue{11}),
\bfpage{214}--\blpage{220}
(\byear{2005})
\doiurl{10.1021/ac053391e}
\end{barticle}
\endbibitem

\bibitem[\protect\citeauthoryear{White and
  White}{2008}]{white_electrochemistry_2008}
\begin{barticle}
\bauthor{\bsnm{White}, \binits{R.J.}},
\bauthor{\bsnm{White}, \binits{H.S.}}:
\batitle{Electrochemistry in {Nanometer}-{Wide} {Electrochemical} {Cells}}.
\bjtitle{Langmuir}
\bvolume{24}(\bissue{6}),
\bfpage{2850}--\blpage{2855}
(\byear{2008})
\doiurl{10.1021/la7031779}
\end{barticle}
\endbibitem

\bibitem[\protect\citeauthoryear{Krause et~al.}{2014}]{krause_brownian_2014}
\begin{barticle}
\bauthor{\bsnm{Krause}, \binits{K.J.}},
\bauthor{\bsnm{Mathwig}, \binits{K.}},
\bauthor{\bsnm{Wolfrum}, \binits{B.}},
\bauthor{\bsnm{Lemay}, \binits{S.G.}}:
\batitle{Brownian motion in electrochemical nanodevices}.
\bjtitle{The European Physical Journal Special Topics}
\bvolume{223}(\bissue{14}),
\bfpage{3165}--\blpage{3178}
(\byear{2014})
\doiurl{10.1140/epjst/e2014-02325-5}
\end{barticle}
\endbibitem

\bibitem[\protect\citeauthoryear{Samin}{2016}]{samin_one-dimensional_2016}
\begin{barticle}
\bauthor{\bsnm{Samin}, \binits{A.J.}}:
\batitle{A one-dimensional stochastic approach to the study of cyclic
  voltammetry with adsorption effects}.
\bjtitle{AIP Advances}
\bvolume{6}(\bissue{5}),
\bfpage{055101}
(\byear{2016})
\doiurl{10.1063/1.4948698}
\end{barticle}
\endbibitem

\bibitem[\protect\citeauthoryear{Marcus}{1965}]{marcus_theory_1965}
\begin{barticle}
\bauthor{\bsnm{Marcus}, \binits{R.A.}}:
\batitle{On the {Theory} of {Electron}‐{Transfer} {Reactions}. {VI}.
  {Unified} {Treatment} for {Homogeneous} and {Electrode} {Reactions}}.
\bjtitle{The Journal of Chemical Physics}
\bvolume{43}(\bissue{2}),
\bfpage{679}--\blpage{701}
(\byear{1965})
\doiurl{10.1063/1.1696792}
\end{barticle}
\endbibitem

\bibitem[\protect\citeauthoryear{Hush}{1968}]{hush_homogeneous_1968}
\begin{barticle}
\bauthor{\bsnm{Hush}, \binits{N.S.}}:
\batitle{Homogeneous and heterogeneous optical and thermal electron transfer}.
\bjtitle{Electrochimica Acta}
\bvolume{13}(\bissue{5}),
\bfpage{1005}--\blpage{1023}
(\byear{1968})
\doiurl{10.1016/0013-4686(68)80032-5}
\end{barticle}
\endbibitem

\bibitem[\protect\citeauthoryear{Blackman and
  Vigna}{2021}]{blackman_scrambled_2021}
\begin{barticle}
\bauthor{\bsnm{Blackman}, \binits{D.}},
\bauthor{\bsnm{Vigna}, \binits{S.}}:
\batitle{Scrambled {Linear} {Pseudorandom} {Number} {Generators}}.
\bjtitle{ACM Transactions on Mathematical Software}
\bvolume{47}(\bissue{4}),
\bfpage{1}--\blpage{32}
(\byear{2021})
\doiurl{10.1145/3460772}
\end{barticle}
\endbibitem

\bibitem[\protect\citeauthoryear{Laviron}{1979}]{laviron_general_1979}
\begin{botherref}
\oauthor{\bsnm{Laviron}, \binits{E.}}:
General expression of the linear potential sweep voltammogram in the case of
  diffusionless electrochemical systems.
Journal of Electroanalytical Chemistry and Interfacial Electrochemistry,
10
(1979)
\doiurl{10.1016/S0022-0728(79)80075-3}
\end{botherref}
\endbibitem

\bibitem[\protect\citeauthoryear{Zevenbergen
  et~al.}{2009}]{zevenbergen_fast_2009}
\begin{barticle}
\bauthor{\bsnm{Zevenbergen}, \binits{M.A.G.}},
\bauthor{\bsnm{Wolfrum}, \binits{B.L.}},
\bauthor{\bsnm{Goluch}, \binits{E.D.}},
\bauthor{\bsnm{Singh}, \binits{P.S.}},
\bauthor{\bsnm{Lemay}, \binits{S.G.}}:
\batitle{Fast {Electron}-{Transfer} {Kinetics} {Probed} in {Nanofluidic}
  {Channels}}.
\bjtitle{Journal of the American Chemical Society}
\bvolume{131}(\bissue{32}),
\bfpage{11471}--\blpage{11477}
(\byear{2009})
\doiurl{10.1021/ja902331u}
\end{barticle}
\endbibitem

\bibitem[\protect\citeauthoryear{Uzawa et~al.}{2010}]{uzawa_mechanistic_2010}
\begin{barticle}
\bauthor{\bsnm{Uzawa}, \binits{T.}},
\bauthor{\bsnm{Cheng}, \binits{R.R.}},
\bauthor{\bsnm{White}, \binits{R.J.}},
\bauthor{\bsnm{Makarov}, \binits{D.E.}},
\bauthor{\bsnm{Plaxco}, \binits{K.W.}}:
\batitle{A {Mechanistic} {Study} of {Electron} {Transfer} from the {Distal}
  {Termini} of {Electrode}-{Bound}, {Single}-{Stranded} {DNAs}}.
\bjtitle{Journal of the American Chemical Society}
\bvolume{132}(\bissue{45}),
\bfpage{16120}--\blpage{16126}
(\byear{2010})
\doiurl{10.1021/ja106345d}
\end{barticle}
\endbibitem

\bibitem[\protect\citeauthoryear{Osteryoung and
  Osteryoung}{1985}]{osteryoung_square_1985}
\begin{barticle}
\bauthor{\bsnm{Osteryoung}, \binits{J.G.}},
\bauthor{\bsnm{Osteryoung}, \binits{R.A.}}:
\batitle{Square wave voltammetry}.
\bjtitle{Analytical Chemistry}
\bvolume{57}(\bissue{1}),
\bfpage{101}--\blpage{110}
(\byear{1985})
\doiurl{10.1021/ac00279a004}
\end{barticle}
\endbibitem

\bibitem[\protect\citeauthoryear{Lovri\'c and
  Komorsky-Lovri\'c}{1988}]{lovric_square-wave_1988}
\begin{barticle}
\bauthor{\bsnm{Lovri\'c}, \binits{M.}},
\bauthor{\bsnm{Komorsky-Lovri\'c}, \binits{{\v{S}}.}}:
\batitle{Square-wave voltammetry of an adsorbed reactant}.
\bjtitle{Journal of Electroanalytical Chemistry and Interfacial
  Electrochemistry}
\bvolume{248}(\bissue{2}),
\bfpage{239}--\blpage{253}
(\byear{1988})
\doiurl{10.1016/0022-0728(88)85089-7}
\end{barticle}
\endbibitem

\bibitem[\protect\citeauthoryear{Chen and
  Shah}{2013}]{chen_electrochemical_2013}
\begin{barticle}
\bauthor{\bsnm{Chen}, \binits{A.}},
\bauthor{\bsnm{Shah}, \binits{B.}}:
\batitle{Electrochemical sensing and biosensing based on square wave
  voltammetry}.
\bjtitle{Analytical Methods}
\bvolume{5}(\bissue{9}),
\bfpage{2158}
(\byear{2013})
\doiurl{10.1039/c3ay40155c}
\end{barticle}
\endbibitem

\bibitem[\protect\citeauthoryear{Trasobares
  et~al.}{2017}]{trasobares_estimation_2017}
\begin{barticle}
\bauthor{\bsnm{Trasobares}, \binits{J.}},
\bauthor{\bsnm{Rech}, \binits{J.}},
\bauthor{\bsnm{Jonckheere}, \binits{T.}},
\bauthor{\bsnm{Martin}, \binits{T.}},
\bauthor{\bsnm{Aleveque}, \binits{O.}},
\bauthor{\bsnm{Levillain}, \binits{E.}},
\bauthor{\bsnm{Diez-Cabanes}, \binits{V.}},
\bauthor{\bsnm{Olivier}, \binits{Y.}},
\bauthor{\bsnm{Cornil}, \binits{J.}},
\bauthor{\bsnm{Nys}, \binits{J.P.}},
\bauthor{\bsnm{Sivakumarasamy}, \binits{R.}},
\bauthor{\bsnm{Smaali}, \binits{K.}},
\bauthor{\bsnm{Leclere}, \binits{P.}},
\bauthor{\bsnm{Fujiwara}, \binits{A.}},
\bauthor{\bsnm{Théron}, \binits{D.}},
\bauthor{\bsnm{Vuillaume}, \binits{D.}},
\bauthor{\bsnm{Clément}, \binits{N.}}:
\batitle{Estimation of $\pi-\pi$ electronic couplings from current
  measurements}.
\bjtitle{Nano Letters}
\bvolume{17}(\bissue{5}),
\bfpage{3215}--\blpage{3224}
(\byear{2017})
\doiurl{10.1021/acs.nanolett.7b00804}
\end{barticle}
\endbibitem

\bibitem[\protect\citeauthoryear{Torbensen
  et~al.}{2019}]{torbensen_immuno-based_2019}
\begin{barticle}
\bauthor{\bsnm{Torbensen}, \binits{K.}},
\bauthor{\bsnm{Patel}, \binits{A.N.}},
\bauthor{\bsnm{Anne}, \binits{A.}},
\bauthor{\bsnm{Chovin}, \binits{A.}},
\bauthor{\bsnm{Demaille}, \binits{C.}},
\bauthor{\bsnm{Bataille}, \binits{L.}},
\bauthor{\bsnm{Michon}, \binits{T.}},
\bauthor{\bsnm{Grelet}, \binits{E.}}:
\batitle{Immuno-{Based} {Molecular} {Scaffolding} of {Glucose} {Dehydrogenase}
  and {Ferrocene} {Mediator} on \textit{fd} {Viral} {Particles} {Yields}
  {Enhanced} {Bioelectrocatalysis}}.
\bjtitle{ACS Catalysis}
\bvolume{9}(\bissue{6}),
\bfpage{5783}--\blpage{5796}
(\byear{2019})
\doiurl{10.1021/acscatal.9b01263}
\end{barticle}
\endbibitem

\bibitem[\protect\citeauthoryear{Adamson et~al.}{2017}]{adamson_probing_2017}
\begin{barticle}
\bauthor{\bsnm{Adamson}, \binits{H.}},
\bauthor{\bsnm{Bond}, \binits{A.M.}},
\bauthor{\bsnm{Parkin}, \binits{A.}}:
\batitle{Probing biological redox chemistry with large amplitude {Fourier}
  transformed ac voltammetry}.
\bjtitle{Chemical Communications}
\bvolume{53}(\bissue{69}),
\bfpage{9519}--\blpage{9533}
(\byear{2017})
\doiurl{10.1039/C7CC03870D}
\end{barticle}
\endbibitem

\bibitem[\protect\citeauthoryear{Baranska
  et~al.}{2024}]{baranska_practical_2024}
\begin{barticle}
\bauthor{\bsnm{Baranska}, \binits{N.G.}},
\bauthor{\bsnm{Jones}, \binits{B.}},
\bauthor{\bsnm{Dowsett}, \binits{M.R.}},
\bauthor{\bsnm{Rhodes}, \binits{C.}},
\bauthor{\bsnm{Elton}, \binits{D.M.}},
\bauthor{\bsnm{Zhang}, \binits{J.}},
\bauthor{\bsnm{Bond}, \binits{A.M.}},
\bauthor{\bsnm{Gavaghan}, \binits{D.}},
\bauthor{\bsnm{Lloyd-Laney}, \binits{H.O.}},
\bauthor{\bsnm{Parkin}, \binits{A.}}:
\batitle{Practical {Guide} to {Large} {Amplitude} {Fourier}-{Transformed}
  {Alternating} {Current} {Voltammetry}-{What}, {How}, and {Why}}.
\bjtitle{ACS Measurement Science Au}
\bvolume{4}(\bissue{4}),
\bfpage{418}--\blpage{431}
(\byear{2024})
\doiurl{10.1021/acsmeasuresciau.4c00008}
\end{barticle}
\endbibitem

\bibitem[\protect\citeauthoryear{Macdonald}{2006}]{macdonald_reflections_2006}
\begin{barticle}
\bauthor{\bsnm{Macdonald}, \binits{D.D.}}:
\batitle{Reflections on the history of electrochemical impedance spectroscopy}.
\bjtitle{Electrochimica Acta}
\bvolume{51}(\bissue{8-9}),
\bfpage{1376}--\blpage{1388}
(\byear{2006})
\doiurl{10.1016/j.electacta.2005.02.107}
\end{barticle}
\endbibitem

\bibitem[\protect\citeauthoryear{Randviir and
  Banks}{2013}]{randviir_electrochemical_2013}
\begin{barticle}
\bauthor{\bsnm{Randviir}, \binits{E.P.}},
\bauthor{\bsnm{Banks}, \binits{C.E.}}:
\batitle{Electrochemical impedance spectroscopy: an overview of bioanalytical
  applications}.
\bjtitle{Analytical Methods}
\bvolume{5}(\bissue{5}),
\bfpage{1098}
(\byear{2013})
\doiurl{10.1039/c3ay26476a}
\end{barticle}
\endbibitem

\bibitem[\protect\citeauthoryear{Wang et~al.}{2021}]{wang_electrochemical_2021}
\begin{barticle}
\bauthor{\bsnm{Wang}, \binits{S.}},
\bauthor{\bsnm{Zhang}, \binits{J.}},
\bauthor{\bsnm{Gharbi}, \binits{O.}},
\bauthor{\bsnm{Vivier}, \binits{V.}},
\bauthor{\bsnm{Gao}, \binits{M.}},
\bauthor{\bsnm{Orazem}, \binits{M.E.}}:
\batitle{Electrochemical impedance spectroscopy}.
\bjtitle{Nature Reviews Methods Primers}
\bvolume{1}(\bissue{1}),
\bfpage{41}
(\byear{2021})
\doiurl{10.1038/s43586-021-00039-w}
\end{barticle}
\endbibitem

\bibitem[\protect\citeauthoryear{Grall et~al.}{2021}]{grall_attoampere_2021}
\begin{barticle}
\bauthor{\bsnm{Grall}, \binits{S.}},
\bauthor{\bsnm{Alić}, \binits{I.}},
\bauthor{\bsnm{Pavoni}, \binits{E.}},
\bauthor{\bsnm{Awadein}, \binits{M.}},
\bauthor{\bsnm{Fujii}, \binits{T.}},
\bauthor{\bsnm{Müllegger}, \binits{S.}},
\bauthor{\bsnm{Farina}, \binits{M.}},
\bauthor{\bsnm{Clément}, \binits{N.}},
\bauthor{\bsnm{Gramse}, \binits{G.}}:
\batitle{Attoampere {Nanoelectrochemistry}}.
\bjtitle{Small}
\bvolume{17}(\bissue{29}),
\bfpage{2101253}
(\byear{2021})
\doiurl{10.1002/smll.202101253}
\end{barticle}
\endbibitem

\bibitem[\protect\citeauthoryear{Smalley et~al.}{1995}]{smalley_kinetics_1995}
\begin{barticle}
\bauthor{\bsnm{Smalley}, \binits{J.F.}},
\bauthor{\bsnm{Feldberg}, \binits{S.W.}},
\bauthor{\bsnm{Chidsey}, \binits{C.E.D.}},
\bauthor{\bsnm{Linford}, \binits{M.R.}},
\bauthor{\bsnm{Newton}, \binits{M.D.}},
\bauthor{\bsnm{Liu}, \binits{Y.-P.}}:
\batitle{The {Kinetics} of {Electron} {Transfer} {Through}
  {Ferrocene}-{Terminated} {Alkanethiol} {Monolayers} on {Gold}}.
\bjtitle{The Journal of Physical Chemistry}
\bvolume{99}(\bissue{35}),
\bfpage{13141}--\blpage{13149}
(\byear{1995})
\doiurl{10.1021/j100035a016}
\end{barticle}
\endbibitem

\bibitem[\protect\citeauthoryear{Drummond
  et~al.}{2003}]{drummond_electrochemical_2003}
\begin{barticle}
\bauthor{\bsnm{Drummond}, \binits{T.G.}},
\bauthor{\bsnm{Hill}, \binits{M.G.}},
\bauthor{\bsnm{Barton}, \binits{J.K.}}:
\batitle{Electrochemical {DNA} sensors}.
\bjtitle{Nature Biotechnology}
\bvolume{21}(\bissue{10}),
\bfpage{1192}--\blpage{1199}
(\byear{2003})
\doiurl{10.1038/nbt873}
\end{barticle}
\endbibitem

\bibitem[\protect\citeauthoryear{Diculescu
  et~al.}{2016}]{diculescu_applications_2016}
\begin{barticle}
\bauthor{\bsnm{Diculescu}, \binits{V.C.}},
\bauthor{\bsnm{Chiorcea-Paquim}, \binits{A.-M.}},
\bauthor{\bsnm{Oliveira-Brett}, \binits{A.M.}}:
\batitle{Applications of a {DNA}-electrochemical biosensor}.
\bjtitle{TrAC Trends in Analytical Chemistry}
\bvolume{79},
\bfpage{23}--\blpage{36}
(\byear{2016})
\doiurl{10.1016/j.trac.2016.01.019}
\end{barticle}
\endbibitem

\bibitem[\protect\citeauthoryear{Trotter et~al.}{2020}]{trotter_review_2020}
\begin{barticle}
\bauthor{\bsnm{Trotter}, \binits{M.}},
\bauthor{\bsnm{Borst}, \binits{N.}},
\bauthor{\bsnm{Thewes}, \binits{R.}},
\bauthor{\bsnm{Von~Stetten}, \binits{F.}}:
\batitle{Review: {Electrochemical} {DNA} sensing – {Principles}, commercial
  systems, and applications}.
\bjtitle{Biosensors and Bioelectronics}
\bvolume{154},
\bfpage{112069}
(\byear{2020})
\doiurl{10.1016/j.bios.2020.112069}
\end{barticle}
\endbibitem

\bibitem[\protect\citeauthoryear{Yang et~al.}{2019}]{yang_predicting_2019}
\begin{barticle}
\bauthor{\bsnm{Yang}, \binits{H.-W.}},
\bauthor{\bsnm{Ju}, \binits{S.-P.}},
\bauthor{\bsnm{Lin}, \binits{Y.-S.}}:
\batitle{Predicting the {Most} {Stable} {Aptamer}/{Target} {Molecule} {Complex}
  {Configuration} {Using} a {Stochastic}-{Tunnelling} {Basin}-{Hopping}
  {Discrete} {Molecular} {Dynamics} {Method}: {A} {Novel} {Global} {Minimum}
  {Search} {Method} for a {Biomolecule} {Complex}}.
\bjtitle{Computational and Structural Biotechnology Journal}
\bvolume{17},
\bfpage{812}--\blpage{820}
(\byear{2019})
\doiurl{10.1016/j.csbj.2019.06.021}
\end{barticle}
\endbibitem

\bibitem[\protect\citeauthoryear{Thompson et~al.}{2022}]{thompson_lammps_2022}
\begin{barticle}
\bauthor{\bsnm{Thompson}, \binits{A.P.}},
\bauthor{\bsnm{Aktulga}, \binits{H.M.}},
\bauthor{\bsnm{Berger}, \binits{R.}},
\bauthor{\bsnm{Bolintineanu}, \binits{D.S.}},
\bauthor{\bsnm{Brown}, \binits{W.M.}},
\bauthor{\bsnm{Crozier}, \binits{P.S.}},
\bauthor{\bsnm{In~'T~Veld}, \binits{P.J.}},
\bauthor{\bsnm{Kohlmeyer}, \binits{A.}},
\bauthor{\bsnm{Moore}, \binits{S.G.}},
\bauthor{\bsnm{Nguyen}, \binits{T.D.}},
\bauthor{\bsnm{Shan}, \binits{R.}},
\bauthor{\bsnm{Stevens}, \binits{M.J.}},
\bauthor{\bsnm{Tranchida}, \binits{J.}},
\bauthor{\bsnm{Trott}, \binits{C.}},
\bauthor{\bsnm{Plimpton}, \binits{S.J.}}:
\batitle{{LAMMPS} - a flexible simulation tool for particle-based materials
  modeling at the atomic, meso, and continuum scales}.
\bjtitle{Computer Physics Communications}
\bvolume{271},
\bfpage{108171}
(\byear{2022})
\doiurl{10.1016/j.cpc.2021.108171}
\end{barticle}
\endbibitem

\bibitem[\protect\citeauthoryear{Gao et~al.}{2020}]{gao_shot_2020}
\begin{barticle}
\bauthor{\bsnm{Gao}, \binits{R.}},
\bauthor{\bsnm{Edwards}, \binits{M.A.}},
\bauthor{\bsnm{Harris}, \binits{J.M.}},
\bauthor{\bsnm{White}, \binits{H.S.}}:
\batitle{Shot noise sets the limit of quantification in electrochemical
  measurements}.
\bjtitle{Current Opinion in Electrochemistry}
\bvolume{22},
\bfpage{170}--\blpage{177}
(\byear{2020})
\doiurl{10.1016/j.coelec.2020.05.010}
\end{barticle}
\endbibitem

\bibitem[\protect\citeauthoryear{MacDougall
  et~al.}{1980}]{macdougall_guidelines_1980}
\begin{barticle}
\bauthor{\bsnm{MacDougall}, \binits{D.}},
\bauthor{\bsnm{Crummett}, \binits{W.B.}},
\bauthor{\bsnm{Et~Al.}, \binits{.}}:
\batitle{Guidelines for data acquisition and data quality evaluation in
  environmental chemistry}.
\bjtitle{Analytical Chemistry}
\bvolume{52}(\bissue{14}),
\bfpage{2242}--\blpage{2249}
(\byear{1980})
\doiurl{10.1021/ac50064a004}
\end{barticle}
\endbibitem

\bibitem[\protect\citeauthoryear{Kang et~al.}{2013}]{kang_electrochemical_2013}
\begin{barticle}
\bauthor{\bsnm{Kang}, \binits{S.}},
\bauthor{\bsnm{Nieuwenhuis}, \binits{A.F.}},
\bauthor{\bsnm{Mathwig}, \binits{K.}},
\bauthor{\bsnm{Mampallil}, \binits{D.}},
\bauthor{\bsnm{Lemay}, \binits{S.G.}}:
\batitle{Electrochemical {Single}-{Molecule} {Detection} in {Aqueous}
  {Solution} {Using} {Self}-{Aligned} {Nanogap} {Transducers}}.
\bjtitle{ACS Nano}
\bvolume{7}(\bissue{12}),
\bfpage{10931}--\blpage{10937}
(\byear{2013})
\doiurl{10.1021/nn404440v}
\end{barticle}
\endbibitem

\bibitem[\protect\citeauthoryear{Clarke et~al.}{2024}]{clarke_single_2024}
\begin{barticle}
\bauthor{\bsnm{Clarke}, \binits{T.B.}},
\bauthor{\bsnm{Krushinski}, \binits{L.E.}},
\bauthor{\bsnm{Vannoy}, \binits{K.J.}},
\bauthor{\bsnm{Colón-Quintana}, \binits{G.}},
\bauthor{\bsnm{Roy}, \binits{K.}},
\bauthor{\bsnm{Rana}, \binits{A.}},
\bauthor{\bsnm{Renault}, \binits{C.}},
\bauthor{\bsnm{Hill}, \binits{M.L.}},
\bauthor{\bsnm{Dick}, \binits{J.E.}}:
\batitle{Single {Entity} {Electrocatalysis}}.
\bjtitle{Chemical Reviews}
\bvolume{124}(\bissue{15}),
\bfpage{9015}--\blpage{9080}
(\byear{2024})
\doiurl{10.1021/acs.chemrev.3c00723}
\end{barticle}
\endbibitem

\bibitem[\protect\citeauthoryear{Ubbelohde
  et~al.}{2012}]{ubbelohde_measurement_2012}
\begin{barticle}
\bauthor{\bsnm{Ubbelohde}, \binits{N.}},
\bauthor{\bsnm{Fricke}, \binits{C.}},
\bauthor{\bsnm{Flindt}, \binits{C.}},
\bauthor{\bsnm{Hohls}, \binits{F.}},
\bauthor{\bsnm{Haug}, \binits{R.J.}}:
\batitle{Measurement of finite-frequency current statistics in a
  single-electron transistor}.
\bjtitle{Nature Communications}
\bvolume{3}(\bissue{1}),
\bfpage{612}
(\byear{2012})
\doiurl{10.1038/ncomms1620}
\end{barticle}
\endbibitem

\end{thebibliography}



\begin{thebibliography}{73}
\ifx \bisbn   \undefined \def \bisbn  #1{ISBN #1}\fi
\ifx \binits  \undefined \def \binits#1{#1}\fi
\ifx \bauthor  \undefined \def \bauthor#1{#1}\fi
\ifx \batitle  \undefined \def \batitle#1{#1}\fi
\ifx \bjtitle  \undefined \def \bjtitle#1{#1}\fi
\ifx \bvolume  \undefined \def \bvolume#1{\textbf{#1}}\fi
\ifx \byear  \undefined \def \byear#1{#1}\fi
\ifx \bissue  \undefined \def \bissue#1{#1}\fi
\ifx \bfpage  \undefined \def \bfpage#1{#1}\fi
\ifx \blpage  \undefined \def \blpage #1{#1}\fi
\ifx \burl  \undefined \def \burl#1{\textsf{#1}}\fi
\ifx \doiurl  \undefined \def \doiurl#1{\url{https://doi.org/#1}}\fi
\ifx \betal  \undefined \def \betal{\textit{et al.}}\fi
\ifx \binstitute  \undefined \def \binstitute#1{#1}\fi
\ifx \binstitutionaled  \undefined \def \binstitutionaled#1{#1}\fi
\ifx \bctitle  \undefined \def \bctitle#1{#1}\fi
\ifx \beditor  \undefined \def \beditor#1{#1}\fi
\ifx \bpublisher  \undefined \def \bpublisher#1{#1}\fi
\ifx \bbtitle  \undefined \def \bbtitle#1{#1}\fi
\ifx \bedition  \undefined \def \bedition#1{#1}\fi
\ifx \bseriesno  \undefined \def \bseriesno#1{#1}\fi
\ifx \blocation  \undefined \def \blocation#1{#1}\fi
\ifx \bsertitle  \undefined \def \bsertitle#1{#1}\fi
\ifx \bsnm \undefined \def \bsnm#1{#1}\fi
\ifx \bsuffix \undefined \def \bsuffix#1{#1}\fi
\ifx \bparticle \undefined \def \bparticle#1{#1}\fi
\ifx \barticle \undefined \def \barticle#1{#1}\fi
\bibcommenthead
\ifx \bconfdate \undefined \def \bconfdate #1{#1}\fi
\ifx \botherref \undefined \def \botherref #1{#1}\fi
\ifx \url \undefined \def \url#1{\textsf{#1}}\fi
\ifx \bchapter \undefined \def \bchapter#1{#1}\fi
\ifx \bbook \undefined \def \bbook#1{#1}\fi
\ifx \bcomment \undefined \def \bcomment#1{#1}\fi
\ifx \oauthor \undefined \def \oauthor#1{#1}\fi
\ifx \citeauthoryear \undefined \def \citeauthoryear#1{#1}\fi
\ifx \endbibitem  \undefined \def \endbibitem {}\fi
\ifx \bconflocation  \undefined \def \bconflocation#1{#1}\fi
\ifx \arxivurl  \undefined \def \arxivurl#1{\textsf{#1}}\fi
\csname PreBibitemsHook\endcsname

\bibitem[\protect\citeauthoryear{Marcus}{1965}]{marcus_theory_1965}
\begin{barticle}
\bauthor{\bsnm{Marcus}, \binits{R.A.}}:
\batitle{On the {Theory} of {Electron}‐{Transfer} {Reactions}. {VI}.
  {Unified} {Treatment} for {Homogeneous} and {Electrode} {Reactions}}.
\bjtitle{The Journal of Chemical Physics}
\bvolume{43}(\bissue{2}),
\bfpage{679}--\blpage{701}
(\byear{1965})
\doiurl{10.1063/1.1696792}
\end{barticle}
\endbibitem

\bibitem[\protect\citeauthoryear{Hush}{1968}]{hush_homogeneous_1968}
\begin{barticle}
\bauthor{\bsnm{Hush}, \binits{N.S.}}:
\batitle{Homogeneous and heterogeneous optical and thermal electron transfer}.
\bjtitle{Electrochimica Acta}
\bvolume{13}(\bissue{5}),
\bfpage{1005}--\blpage{1023}
(\byear{1968})
\doiurl{10.1016/0013-4686(68)80032-5}
\end{barticle}
\endbibitem

\bibitem[\protect\citeauthoryear{Zheng
  et~al.}{2023}]{zheng_electrochemical_2023}
\begin{barticle}
\bauthor{\bsnm{Zheng}, \binits{Z.}},
\bauthor{\bsnm{Kim}, \binits{S.H.}},
\bauthor{\bsnm{Chovin}, \binits{A.}},
\bauthor{\bsnm{Clement}, \binits{N.}},
\bauthor{\bsnm{Demaille}, \binits{C.}}:
\batitle{Electrochemical response of surface-attached redox {DNA} governed by
  low activation energy electron transfer kinetics}.
\bjtitle{Chemical Science}
\bvolume{14}(\bissue{13}),
\bfpage{3652}--\blpage{3660}
(\byear{2023})
\doiurl{10.1039/D3SC00320E}
\end{barticle}
\endbibitem

\bibitem[\protect\citeauthoryear{Madrid et~al.}{2023}]{madrid_ballistic_2023}
\begin{barticle}
\bauthor{\bsnm{Madrid}, \binits{I.}},
\bauthor{\bsnm{Zheng}, \binits{Z.}},
\bauthor{\bsnm{Gerbelot}, \binits{C.}},
\bauthor{\bsnm{Fujiwara}, \binits{A.}},
\bauthor{\bsnm{Li}, \binits{S.}},
\bauthor{\bsnm{Grall}, \binits{S.}},
\bauthor{\bsnm{Nishiguchi}, \binits{K.}},
\bauthor{\bsnm{Kim}, \binits{S.H.}},
\bauthor{\bsnm{Chovin}, \binits{A.}},
\bauthor{\bsnm{Demaille}, \binits{C.}},
\bauthor{\bsnm{Clement}, \binits{N.}}:
\batitle{Ballistic {Brownian} {Motion} of {Nanoconfined} {DNA}}.
\bjtitle{ACS Nano}
\bvolume{17}(\bissue{17}),
\bfpage{17031}--\blpage{17040}
(\byear{2023})
\doiurl{10.1021/acsnano.3c04349}
\end{barticle}
\endbibitem

\bibitem[\protect\citeauthoryear{Zheng
  et~al.}{2024}]{zheng_activationless_2024}
\begin{barticle}
\bauthor{\bsnm{Zheng}, \binits{Z.}},
\bauthor{\bsnm{Grall}, \binits{S.}},
\bauthor{\bsnm{Kim}, \binits{S.H.}},
\bauthor{\bsnm{Chovin}, \binits{A.}},
\bauthor{\bsnm{Clement}, \binits{N.}},
\bauthor{\bsnm{Demaille}, \binits{C.}}:
\batitle{Activationless {Electron} {Transfer} of redox-{DNA} in
  {Electrochemical} {Nanogaps}}.
\bjtitle{Journal of the American Chemical Society}
\bvolume{146}(\bissue{9}),
\bfpage{6094}--\blpage{6103}
(\byear{2024})
\doiurl{10.1021/jacs.3c13532}
\end{barticle}
\endbibitem

\bibitem[\protect\citeauthoryear{Fumagalli
  et~al.}{2018}]{fumagalli_anomalously_2018}
\begin{barticle}
\bauthor{\bsnm{Fumagalli}, \binits{L.}},
\bauthor{\bsnm{Esfandiar}, \binits{A.}},
\bauthor{\bsnm{Fabregas}, \binits{R.}},
\bauthor{\bsnm{Hu}, \binits{S.}},
\bauthor{\bsnm{Ares}, \binits{P.}},
\bauthor{\bsnm{Janardanan}, \binits{A.}},
\bauthor{\bsnm{Yang}, \binits{Q.}},
\bauthor{\bsnm{Radha}, \binits{B.}},
\bauthor{\bsnm{Taniguchi}, \binits{T.}},
\bauthor{\bsnm{Watanabe}, \binits{K.}},
\bauthor{\bsnm{Gomila}, \binits{G.}},
\bauthor{\bsnm{Novoselov}, \binits{K.S.}},
\bauthor{\bsnm{Geim}, \binits{A.K.}}:
\batitle{Anomalously low dielectric constant of confined water}.
\bjtitle{Science}
\bvolume{360}(\bissue{6395}),
\bfpage{1339}--\blpage{1342}
(\byear{2018})
\doiurl{10.1126/science.aat4191}
\end{barticle}
\endbibitem

\bibitem[\protect\citeauthoryear{Zevenbergen
  et~al.}{2009}]{zevenbergen_fast_2009}
\begin{barticle}
\bauthor{\bsnm{Zevenbergen}, \binits{M.A.G.}},
\bauthor{\bsnm{Wolfrum}, \binits{B.L.}},
\bauthor{\bsnm{Goluch}, \binits{E.D.}},
\bauthor{\bsnm{Singh}, \binits{P.S.}},
\bauthor{\bsnm{Lemay}, \binits{S.G.}}:
\batitle{Fast {Electron}-{Transfer} {Kinetics} {Probed} in {Nanofluidic}
  {Channels}}.
\bjtitle{Journal of the American Chemical Society}
\bvolume{131}(\bissue{32}),
\bfpage{11471}--\blpage{11477}
(\byear{2009})
\doiurl{10.1021/ja902331u}
\end{barticle}
\endbibitem

\bibitem[\protect\citeauthoryear{White and
  White}{2008}]{white_electrochemistry_2008}
\begin{barticle}
\bauthor{\bsnm{White}, \binits{R.J.}},
\bauthor{\bsnm{White}, \binits{H.S.}}:
\batitle{Electrochemistry in {Nanometer}-{Wide} {Electrochemical} {Cells}}.
\bjtitle{Langmuir}
\bvolume{24}(\bissue{6}),
\bfpage{2850}--\blpage{2855}
(\byear{2008})
\doiurl{10.1021/la7031779}
\end{barticle}
\endbibitem

\bibitem[\protect\citeauthoryear{Oldham and
  Zoski}{1988}]{oldham_comparison_1988}
\begin{barticle}
\bauthor{\bsnm{Oldham}, \binits{K.B.}},
\bauthor{\bsnm{Zoski}, \binits{C.G.}}:
\batitle{Comparison of voltammetric steady states at hemispherical and disc
  microelectrodes}.
\bjtitle{Journal of Electroanalytical Chemistry and Interfacial
  Electrochemistry}
\bvolume{256}(\bissue{1}),
\bfpage{11}--\blpage{19}
(\byear{1988})
\doiurl{10.1016/0022-0728(88)85002-2}
\end{barticle}
\endbibitem

\bibitem[\protect\citeauthoryear{Cutress and Compton}{2011}]{cutress_how_2011}
\begin{barticle}
\bauthor{\bsnm{Cutress}, \binits{I.J.}},
\bauthor{\bsnm{Compton}, \binits{R.G.}}:
\batitle{How many molecules are required to measure a cyclic voltammogram?}
\bjtitle{Chemical Physics Letters}
\bvolume{508}(\bissue{4-6}),
\bfpage{306}--\blpage{313}
(\byear{2011})
\doiurl{10.1016/j.cplett.2011.04.036} .
\bcomment{Number: 4-6}
\end{barticle}
\endbibitem

\bibitem[\protect\citeauthoryear{Cutress
  et~al.}{2011}]{cutress_electrochemical_2011}
\begin{barticle}
\bauthor{\bsnm{Cutress}, \binits{I.J.}},
\bauthor{\bsnm{Dickinson}, \binits{E.J.F.}},
\bauthor{\bsnm{Compton}, \binits{R.G.}}:
\batitle{Electrochemical random-walk theory}.
\bjtitle{Journal of Electroanalytical Chemistry}
\bvolume{655}(\bissue{1}),
\bfpage{1}--\blpage{8}
(\byear{2011})
\doiurl{10.1016/j.jelechem.2011.02.023}
\end{barticle}
\endbibitem

\bibitem[\protect\citeauthoryear{Huang and White}{2013}]{huang_random_2013}
\begin{barticle}
\bauthor{\bsnm{Huang}, \binits{K.-C.}},
\bauthor{\bsnm{White}, \binits{R.J.}}:
\batitle{Random {Walk} on a {Leash}: {A} {Simple} {Single}-{Molecule}
  {Diffusion} {Model} for {Surface}-{Tethered} {Redox} {Molecules} with
  {Flexible} {Linkers}}.
\bjtitle{Journal of the American Chemical Society}
\bvolume{135}(\bissue{34}),
\bfpage{12808}--\blpage{12817}
(\byear{2013})
\doiurl{10.1021/ja4060788}
\end{barticle}
\endbibitem

\bibitem[\protect\citeauthoryear{Kätelhön et~al.}{2013}]{katelhon_noise_2013}
\begin{barticle}
\bauthor{\bsnm{Kätelhön}, \binits{E.}},
\bauthor{\bsnm{Krause}, \binits{K.J.}},
\bauthor{\bsnm{Singh}, \binits{P.S.}},
\bauthor{\bsnm{Lemay}, \binits{S.G.}},
\bauthor{\bsnm{Wolfrum}, \binits{B.}}:
\batitle{Noise {Characteristics} of {Nanoscaled} {Redox}-{Cycling} {Sensors}:
  {Investigations} {Based} on {Random} {Walks}}.
\bjtitle{Journal of the American Chemical Society}
\bvolume{135}(\bissue{24}),
\bfpage{8874}--\blpage{8881}
(\byear{2013})
\doiurl{10.1021/ja3121313}
\end{barticle}
\endbibitem

\bibitem[\protect\citeauthoryear{Samin}{2016}]{samin_one-dimensional_2016}
\begin{barticle}
\bauthor{\bsnm{Samin}, \binits{A.J.}}:
\batitle{A one-dimensional stochastic approach to the study of cyclic
  voltammetry with adsorption effects}.
\bjtitle{AIP Advances}
\bvolume{6}(\bissue{5}),
\bfpage{055101}
(\byear{2016})
\doiurl{10.1063/1.4948698}
\end{barticle}
\endbibitem

\bibitem[\protect\citeauthoryear{Casella
  et~al.}{2004}]{casella_generalized_2004}
\begin{bchapter}
\bauthor{\bsnm{Casella}, \binits{G.}},
\bauthor{\bsnm{Robert}, \binits{C.P.}},
\bauthor{\bsnm{Wells}, \binits{M.T.}}:
\bctitle{Generalized {Accept}-{Reject} sampling schemes}.
In: \bbtitle{Institute of {Mathematical} {Statistics} {Lecture} {Notes} -
  {Monograph} {Series}},
pp. \bfpage{342}--\blpage{347}.
\bpublisher{Institute of Mathematical Statistics},
\blocation{Beachwood, Ohio, USA}
(\byear{2004}).
\doiurl{10.1214/lnms/1196285403} .
\burl{http://projecteuclid.org/euclid.lnms/1196285403}
\end{bchapter}
\endbibitem

\bibitem[\protect\citeauthoryear{Anderson and
  Reilley}{1965}]{anderson_thin-layer_1965}
\begin{barticle}
\bauthor{\bsnm{Anderson}, \binits{L.B.}},
\bauthor{\bsnm{Reilley}, \binits{C.N.}}:
\batitle{Thin-layer electrochemistry: steady-state methods of studying rate
  processes}.
\bjtitle{Journal of Electroanalytical Chemistry (1959)}
\bvolume{10}(\bissue{4}),
\bfpage{295}--\blpage{305}
(\byear{1965})
\doiurl{10.1016/0022-0728(65)85063-X}
\end{barticle}
\endbibitem

\bibitem[\protect\citeauthoryear{Awadein et~al.}{2022}]{awadein_nanoscale_2022}
\begin{botherref}
\oauthor{\bsnm{Awadein}, \binits{M.}},
\oauthor{\bsnm{Sparey}, \binits{M.}},
\oauthor{\bsnm{Grall}, \binits{S.}},
\oauthor{\bsnm{Kienberger}, \binits{F.}},
\oauthor{\bsnm{Clement}, \binits{N.}},
\oauthor{\bsnm{Gramse}, \binits{G.}}:
Nanoscale electrochemical charge transfer kinetics investigated by
  electrochemical scanning microwave microscopy.
Nanoscale Advances,
10--1039200671
(2022)
\doiurl{10.1039/D2NA00671E}
\end{botherref}
\endbibitem

\bibitem[\protect\citeauthoryear{Ouldridge et~al.}{2010}]{ouldridge_dna_2010}
\begin{barticle}
\bauthor{\bsnm{Ouldridge}, \binits{T.E.}},
\bauthor{\bsnm{Louis}, \binits{A.A.}},
\bauthor{\bsnm{Doye}, \binits{J.P.K.}}:
\batitle{{DNA} {Nanotweezers} {Studied} with a {Coarse}-{Grained} {Model} of
  {DNA}}.
\bjtitle{Physical Review Letters}
\bvolume{104}(\bissue{17}),
\bfpage{178101}
(\byear{2010})
\doiurl{10.1103/PhysRevLett.104.178101}
\end{barticle}
\endbibitem

\bibitem[\protect\citeauthoryear{Rovigatti
  et~al.}{2014}]{rovigatti_accurate_2014}
\begin{barticle}
\bauthor{\bsnm{Rovigatti}, \binits{L.}},
\bauthor{\bsnm{Bomboi}, \binits{F.}},
\bauthor{\bsnm{Sciortino}, \binits{F.}}:
\batitle{Accurate phase diagram of tetravalent {DNA} nanostars}.
\bjtitle{The Journal of Chemical Physics}
\bvolume{140}(\bissue{15}),
\bfpage{154903}
(\byear{2014})
\doiurl{10.1063/1.4870467}
\end{barticle}
\endbibitem

\bibitem[\protect\citeauthoryear{Poppleton et~al.}{2023}]{poppleton_oxdna_2023}
\begin{barticle}
\bauthor{\bsnm{Poppleton}, \binits{E.}},
\bauthor{\bsnm{Matthies}, \binits{M.}},
\bauthor{\bsnm{Mandal}, \binits{D.}},
\bauthor{\bsnm{Romano}, \binits{F.}},
\bauthor{\bsnm{Šulc}, \binits{P.}},
\bauthor{\bsnm{Rovigatti}, \binits{L.}}:
\batitle{{oxDNA}: coarse-grained simulations of nucleic acids madesimple}.
\bjtitle{Journal of Open Source Software}
\bvolume{8}(\bissue{81}),
\bfpage{4693}
(\byear{2023})
\doiurl{10.21105/joss.04693}
\end{barticle}
\endbibitem

\bibitem[\protect\citeauthoryear{Marsaglia}{2003}]{marsaglia_xorshift_2003}
\begin{botherref}
\oauthor{\bsnm{Marsaglia}, \binits{G.}}:
Xorshift {RNGs}.
Journal of Statistical Software
\textbf{8}(14)
(2003)
\doiurl{10.18637/jss.v008.i14}
\end{botherref}
\endbibitem

\bibitem[\protect\citeauthoryear{Vigna}{2017}]{vigna_further_2017}
\begin{barticle}
\bauthor{\bsnm{Vigna}, \binits{S.}}:
\batitle{Further scramblings of {Marsaglia}’s xorshift generators}.
\bjtitle{Journal of Computational and Applied Mathematics}
\bvolume{315},
\bfpage{175}--\blpage{181}
(\byear{2017})
\doiurl{10.1016/j.cam.2016.11.006}
\end{barticle}
\endbibitem

\bibitem[\protect\citeauthoryear{Blackman and
  Vigna}{2021}]{blackman_scrambled_2021}
\begin{barticle}
\bauthor{\bsnm{Blackman}, \binits{D.}},
\bauthor{\bsnm{Vigna}, \binits{S.}}:
\batitle{Scrambled {Linear} {Pseudorandom} {Number} {Generators}}.
\bjtitle{ACM Transactions on Mathematical Software}
\bvolume{47}(\bissue{4}),
\bfpage{1}--\blpage{32}
(\byear{2021})
\doiurl{10.1145/3460772}
\end{barticle}
\endbibitem

\bibitem[\protect\citeauthoryear{Savéant}{2006}]{saveant_elements_2006}
\begin{bbook}
\bauthor{\bsnm{Savéant}, \binits{J.M.}}:
\bbtitle{Elements of Molecular and Biomolecular Electrochemistry: an
  Electrochemical Approach to Electron Transfer Chemistry}.
\bsertitle{The {George} {Fisher} {Baker} non-resident lectureship in chemistry
  at {Cornell} {University}}.
\bpublisher{Wiley-Interscience},
\blocation{Hoboken, N.J}
(\byear{2006}).
\bcomment{OCLC: ocm60311799}
\end{bbook}
\endbibitem

\bibitem[\protect\citeauthoryear{Chidsey}{1991}]{chidsey_free_1991}
\begin{barticle}
\bauthor{\bsnm{Chidsey}, \binits{C.E.D.}}:
\batitle{Free {Energy} and {Temperature} {Dependence} of {Electron} {Transfer}
  at the {Metal}-{Electrolyte} {Interface}}.
\bjtitle{Science}
\bvolume{251}(\bissue{4996}),
\bfpage{919}--\blpage{922}
(\byear{1991})
\doiurl{10.1126/science.251.4996.919}
\end{barticle}
\endbibitem

\bibitem[\protect\citeauthoryear{Savéant}{2002}]{saveant_effect_2002}
\begin{barticle}
\bauthor{\bsnm{Savéant}, \binits{J.-M.}}:
\batitle{Effect of the {Electrode} {Continuum} of {States} in {Adiabatic} and
  {Nonadiabatic} {Outer}-{Sphere} and {Dissociative} {Electron} {Transfers}.
  {Use} of {Cyclic} {Voltammetry} for {Investigating} {Nonlinear}
  {Activation}-{Driving} {Force} {Laws}}.
\bjtitle{The Journal of Physical Chemistry B}
\bvolume{106}(\bissue{36}),
\bfpage{9387}--\blpage{9395}
(\byear{2002})
\doiurl{10.1021/jp0258006}
\end{barticle}
\endbibitem

\bibitem[\protect\citeauthoryear{Laviron}{1979}]{laviron_general_1979}
\begin{botherref}
\oauthor{\bsnm{Laviron}, \binits{E.}}:
General expression of the linear potential sweep voltammogram in the case of
  diffusionless electrochemical systems.
Journal of Electroanalytical Chemistry and Interfacial Electrochemistry,
10
(1979)
\doiurl{10.1016/S0022-0728(79)80075-3}
\end{botherref}
\endbibitem

\bibitem[\protect\citeauthoryear{Ren et~al.}{2018}]{ren_single-molecule_2018}
\begin{barticle}
\bauthor{\bsnm{Ren}, \binits{H.}},
\bauthor{\bsnm{Cheyne}, \binits{C.G.}},
\bauthor{\bsnm{Fleming}, \binits{A.M.}},
\bauthor{\bsnm{Burrows}, \binits{C.J.}},
\bauthor{\bsnm{White}, \binits{H.S.}}:
\batitle{Single-{Molecule} {Titration} in a {Protein} {Nanoreactor} {Reveals}
  the {Protonation}/{Deprotonation} {Mechanism} of a {C}:{C} {Mismatch} in
  {DNA}}.
\bjtitle{Journal of the American Chemical Society}
\bvolume{140}(\bissue{15}),
\bfpage{5153}--\blpage{5160}
(\byear{2018})
\doiurl{10.1021/jacs.8b00593}
\end{barticle}
\endbibitem

\bibitem[\protect\citeauthoryear{Tan et~al.}{2016}]{tan_kinetics_2016}
\begin{barticle}
\bauthor{\bsnm{Tan}, \binits{C.S.}},
\bauthor{\bsnm{Riedl}, \binits{J.}},
\bauthor{\bsnm{Fleming}, \binits{A.M.}},
\bauthor{\bsnm{Burrows}, \binits{C.J.}},
\bauthor{\bsnm{White}, \binits{H.S.}}:
\batitle{Kinetics of {T3}-{DNA} {Ligase}-{Catalyzed} {Phosphodiester} {Bond}
  {Formation} {Measured} {Using} the $\alpha$-{Hemolysin} {Nanopore}}.
\bjtitle{ACS Nano}
\bvolume{10}(\bissue{12}),
\bfpage{11127}--\blpage{11135}
(\byear{2016})
\doiurl{10.1021/acsnano.6b05995}
\end{barticle}
\endbibitem

\bibitem[\protect\citeauthoryear{Rosenstein
  et~al.}{2012}]{rosenstein_integrated_2012}
\begin{barticle}
\bauthor{\bsnm{Rosenstein}, \binits{J.K.}},
\bauthor{\bsnm{Wanunu}, \binits{M.}},
\bauthor{\bsnm{Merchant}, \binits{C.A.}},
\bauthor{\bsnm{Drndic}, \binits{M.}},
\bauthor{\bsnm{Shepard}, \binits{K.L.}}:
\batitle{Integrated nanopore sensing platform with sub-microsecond temporal
  resolution}.
\bjtitle{Nature Methods}
\bvolume{9}(\bissue{5}),
\bfpage{487}--\blpage{492}
(\byear{2012})
\doiurl{10.1038/nmeth.1932}
\end{barticle}
\endbibitem

\bibitem[\protect\citeauthoryear{Qing et~al.}{2018}]{qing_directional_2018}
\begin{barticle}
\bauthor{\bsnm{Qing}, \binits{Y.}},
\bauthor{\bsnm{Ionescu}, \binits{S.A.}},
\bauthor{\bsnm{Pulcu}, \binits{G.S.}},
\bauthor{\bsnm{Bayley}, \binits{H.}}:
\batitle{Directional control of a processive molecular hopper}.
\bjtitle{Science}
\bvolume{361}(\bissue{6405}),
\bfpage{908}--\blpage{912}
(\byear{2018})
\doiurl{10.1126/science.aat3872}
\end{barticle}
\endbibitem

\bibitem[\protect\citeauthoryear{Larkin
  et~al.}{2014}]{larkin_high-bandwidth_2014}
\begin{barticle}
\bauthor{\bsnm{Larkin}, \binits{J.}},
\bauthor{\bsnm{Henley}, \binits{R.Y.}},
\bauthor{\bsnm{Muthukumar}, \binits{M.}},
\bauthor{\bsnm{Rosenstein}, \binits{J.}},
\bauthor{\bsnm{Wanunu}, \binits{M.}}:
\batitle{High-{Bandwidth} {Protein} {Analysis} {Using} {Solid}-{State}
  {Nanopores}}.
\bjtitle{Biophysical Journal}
\bvolume{106}(\bissue{3}),
\bfpage{696}--\blpage{704}
(\byear{2014})
\doiurl{10.1016/j.bpj.2013.12.025}
\end{barticle}
\endbibitem

\bibitem[\protect\citeauthoryear{Henley
  et~al.}{2016}]{henley_electrophoretic_2016}
\begin{barticle}
\bauthor{\bsnm{Henley}, \binits{R.Y.}},
\bauthor{\bsnm{Ashcroft}, \binits{B.A.}},
\bauthor{\bsnm{Farrell}, \binits{I.}},
\bauthor{\bsnm{Cooperman}, \binits{B.S.}},
\bauthor{\bsnm{Lindsay}, \binits{S.M.}},
\bauthor{\bsnm{Wanunu}, \binits{M.}}:
\batitle{Electrophoretic {Deformation} of {Individual} {Transfer} {RNA}
  {Molecules} {Reveals} {Their} {Identity}}.
\bjtitle{Nano Letters}
\bvolume{16}(\bissue{1}),
\bfpage{138}--\blpage{144}
(\byear{2016})
\doiurl{10.1021/acs.nanolett.5b03331}
\end{barticle}
\endbibitem

\bibitem[\protect\citeauthoryear{Cao et~al.}{2016}]{cao_discrimination_2016}
\begin{barticle}
\bauthor{\bsnm{Cao}, \binits{C.}},
\bauthor{\bsnm{Ying}, \binits{Y.-L.}},
\bauthor{\bsnm{Hu}, \binits{Z.-L.}},
\bauthor{\bsnm{Liao}, \binits{D.-F.}},
\bauthor{\bsnm{Tian}, \binits{H.}},
\bauthor{\bsnm{Long}, \binits{Y.-T.}}:
\batitle{Discrimination of oligonucleotides of different lengths with a
  wild-type aerolysin nanopore}.
\bjtitle{Nature Nanotechnology}
\bvolume{11}(\bissue{8}),
\bfpage{713}--\blpage{718}
(\byear{2016})
\doiurl{10.1038/nnano.2016.66}
\end{barticle}
\endbibitem

\bibitem[\protect\citeauthoryear{Schneider et~al.}{2010}]{schneider_dna_2010}
\begin{barticle}
\bauthor{\bsnm{Schneider}, \binits{G.F.}},
\bauthor{\bsnm{Kowalczyk}, \binits{S.W.}},
\bauthor{\bsnm{Calado}, \binits{V.E.}},
\bauthor{\bsnm{Pandraud}, \binits{G.}},
\bauthor{\bsnm{Zandbergen}, \binits{H.W.}},
\bauthor{\bsnm{Vandersypen}, \binits{L.M.K.}},
\bauthor{\bsnm{Dekker}, \binits{C.}}:
\batitle{{DNA} {Translocation} through {Graphene} {Nanopores}}.
\bjtitle{Nano Letters}
\bvolume{10}(\bissue{8}),
\bfpage{3163}--\blpage{3167}
(\byear{2010})
\doiurl{10.1021/nl102069z}
\end{barticle}
\endbibitem

\bibitem[\protect\citeauthoryear{Plesa et~al.}{2013}]{plesa_fast_2013}
\begin{barticle}
\bauthor{\bsnm{Plesa}, \binits{C.}},
\bauthor{\bsnm{Kowalczyk}, \binits{S.W.}},
\bauthor{\bsnm{Zinsmeester}, \binits{R.}},
\bauthor{\bsnm{Grosberg}, \binits{A.Y.}},
\bauthor{\bsnm{Rabin}, \binits{Y.}},
\bauthor{\bsnm{Dekker}, \binits{C.}}:
\batitle{Fast {Translocation} of {Proteins} through {Solid} {State}
  {Nanopores}}.
\bjtitle{Nano Letters}
\bvolume{13}(\bissue{2}),
\bfpage{658}--\blpage{663}
(\byear{2013})
\doiurl{10.1021/nl3042678}
\end{barticle}
\endbibitem

\bibitem[\protect\citeauthoryear{Bell and Keyser}{2016}]{bell_digitally_2016}
\begin{barticle}
\bauthor{\bsnm{Bell}, \binits{N.A.W.}},
\bauthor{\bsnm{Keyser}, \binits{U.F.}}:
\batitle{Digitally encoded {DNA} nanostructures for multiplexed,
  single-molecule protein sensing with nanopores}.
\bjtitle{Nature Nanotechnology}
\bvolume{11}(\bissue{7}),
\bfpage{645}--\blpage{651}
(\byear{2016})
\doiurl{10.1038/nnano.2016.50}
\end{barticle}
\endbibitem

\bibitem[\protect\citeauthoryear{Bell and Keyser}{2015}]{bell_specific_2015}
\begin{barticle}
\bauthor{\bsnm{Bell}, \binits{N.A.W.}},
\bauthor{\bsnm{Keyser}, \binits{U.F.}}:
\batitle{Specific {Protein} {Detection} {Using} {Designed} {DNA} {Carriers} and
  {Nanopores}}.
\bjtitle{Journal of the American Chemical Society}
\bvolume{137}(\bissue{5}),
\bfpage{2035}--\blpage{2041}
(\byear{2015})
\doiurl{10.1021/ja512521w}
\end{barticle}
\endbibitem

\bibitem[\protect\citeauthoryear{Derrington
  et~al.}{2015}]{derrington_subangstrom_2015}
\begin{barticle}
\bauthor{\bsnm{Derrington}, \binits{I.M.}},
\bauthor{\bsnm{Craig}, \binits{J.M.}},
\bauthor{\bsnm{Stava}, \binits{E.}},
\bauthor{\bsnm{Laszlo}, \binits{A.H.}},
\bauthor{\bsnm{Ross}, \binits{B.C.}},
\bauthor{\bsnm{Brinkerhoff}, \binits{H.}},
\bauthor{\bsnm{Nova}, \binits{I.C.}},
\bauthor{\bsnm{Doering}, \binits{K.}},
\bauthor{\bsnm{Tickman}, \binits{B.I.}},
\bauthor{\bsnm{Ronaghi}, \binits{M.}},
\bauthor{\bsnm{Mandell}, \binits{J.G.}},
\bauthor{\bsnm{Gunderson}, \binits{K.L.}},
\bauthor{\bsnm{Gundlach}, \binits{J.H.}}:
\batitle{Subangstrom single-molecule measurements of motor proteins using a
  nanopore}.
\bjtitle{Nature Biotechnology}
\bvolume{33}(\bissue{10}),
\bfpage{1073}--\blpage{1075}
(\byear{2015})
\doiurl{10.1038/nbt.3357}
\end{barticle}
\endbibitem

\bibitem[\protect\citeauthoryear{Plesa et~al.}{2016}]{plesa_direct_2016}
\begin{barticle}
\bauthor{\bsnm{Plesa}, \binits{C.}},
\bauthor{\bsnm{Verschueren}, \binits{D.}},
\bauthor{\bsnm{Pud}, \binits{S.}},
\bauthor{\bsnm{Van Der~Torre}, \binits{J.}},
\bauthor{\bsnm{Ruitenberg}, \binits{J.W.}},
\bauthor{\bsnm{Witteveen}, \binits{M.J.}},
\bauthor{\bsnm{Jonsson}, \binits{M.P.}},
\bauthor{\bsnm{Grosberg}, \binits{A.Y.}},
\bauthor{\bsnm{Rabin}, \binits{Y.}},
\bauthor{\bsnm{Dekker}, \binits{C.}}:
\batitle{Direct observation of {DNA} knots using a solid-state nanopore}.
\bjtitle{Nature Nanotechnology}
\bvolume{11}(\bissue{12}),
\bfpage{1093}--\blpage{1097}
(\byear{2016})
\doiurl{10.1038/nnano.2016.153}
\end{barticle}
\endbibitem

\bibitem[\protect\citeauthoryear{Boukhet et~al.}{2016}]{boukhet_probing_2016}
\begin{barticle}
\bauthor{\bsnm{Boukhet}, \binits{M.}},
\bauthor{\bsnm{Piguet}, \binits{F.}},
\bauthor{\bsnm{Ouldali}, \binits{H.}},
\bauthor{\bsnm{Pastoriza-Gallego}, \binits{M.}},
\bauthor{\bsnm{Pelta}, \binits{J.}},
\bauthor{\bsnm{Oukhaled}, \binits{A.}}:
\batitle{Probing driving forces in aerolysin and $\alpha$-hemolysin biological
  nanopores: electrophoresis versus electroosmosis}.
\bjtitle{Nanoscale}
\bvolume{8}(\bissue{43}),
\bfpage{18352}--\blpage{18359}
(\byear{2016})
\doiurl{10.1039/C6NR06936C}
\end{barticle}
\endbibitem

\bibitem[\protect\citeauthoryear{Clarke et~al.}{2009}]{clarke_continuous_2009}
\begin{barticle}
\bauthor{\bsnm{Clarke}, \binits{J.}},
\bauthor{\bsnm{Wu}, \binits{H.-C.}},
\bauthor{\bsnm{Jayasinghe}, \binits{L.}},
\bauthor{\bsnm{Patel}, \binits{A.}},
\bauthor{\bsnm{Reid}, \binits{S.}},
\bauthor{\bsnm{Bayley}, \binits{H.}}:
\batitle{Continuous base identification for single-molecule nanopore {DNA}
  sequencing}.
\bjtitle{Nature Nanotechnology}
\bvolume{4}(\bissue{4}),
\bfpage{265}--\blpage{270}
(\byear{2009})
\doiurl{10.1038/nnano.2009.12}
\end{barticle}
\endbibitem

\bibitem[\protect\citeauthoryear{Uram et~al.}{2008}]{uram_noise_2008}
\begin{barticle}
\bauthor{\bsnm{Uram}, \binits{J.D.}},
\bauthor{\bsnm{Ke}, \binits{K.}},
\bauthor{\bsnm{Mayer}, \binits{M.}}:
\batitle{Noise and {Bandwidth} of {Current} {Recordings} from {Submicrometer}
  {Pores} and {Nanopores}}.
\bjtitle{ACS Nano}
\bvolume{2}(\bissue{5}),
\bfpage{857}--\blpage{872}
(\byear{2008})
\doiurl{10.1021/nn700322m}
\end{barticle}
\endbibitem

\bibitem[\protect\citeauthoryear{Robertson
  et~al.}{2007}]{robertson_single-molecule_2007}
\begin{barticle}
\bauthor{\bsnm{Robertson}, \binits{J.W.F.}},
\bauthor{\bsnm{Rodrigues}, \binits{C.G.}},
\bauthor{\bsnm{Stanford}, \binits{V.M.}},
\bauthor{\bsnm{Rubinson}, \binits{K.A.}},
\bauthor{\bsnm{Krasilnikov}, \binits{O.V.}},
\bauthor{\bsnm{Kasianowicz}, \binits{J.J.}}:
\batitle{Single-molecule mass spectrometry in solution using a solitary
  nanopore}.
\bjtitle{Proceedings of the National Academy of Sciences}
\bvolume{104}(\bissue{20}),
\bfpage{8207}--\blpage{8211}
(\byear{2007})
\doiurl{10.1073/pnas.0611085104}
\end{barticle}
\endbibitem

\bibitem[\protect\citeauthoryear{Gao et~al.}{2020}]{gao_shot_2020}
\begin{barticle}
\bauthor{\bsnm{Gao}, \binits{R.}},
\bauthor{\bsnm{Edwards}, \binits{M.A.}},
\bauthor{\bsnm{Harris}, \binits{J.M.}},
\bauthor{\bsnm{White}, \binits{H.S.}}:
\batitle{Shot noise sets the limit of quantification in electrochemical
  measurements}.
\bjtitle{Current Opinion in Electrochemistry}
\bvolume{22},
\bfpage{170}--\blpage{177}
(\byear{2020})
\doiurl{10.1016/j.coelec.2020.05.010}
\end{barticle}
\endbibitem

\bibitem[\protect\citeauthoryear{Robinson et~al.}{2018}]{robinson_effects_2018}
\begin{barticle}
\bauthor{\bsnm{Robinson}, \binits{D.A.}},
\bauthor{\bsnm{Edwards}, \binits{M.A.}},
\bauthor{\bsnm{Ren}, \binits{H.}},
\bauthor{\bsnm{White}, \binits{H.S.}}:
\batitle{Effects of {Instrumental} {Filters} on {Electrochemical} {Measurement}
  of {Single}‐{Nanoparticle} {Collision} {Dynamics}}.
\bjtitle{ChemElectroChem}
\bvolume{5}(\bissue{20}),
\bfpage{3059}--\blpage{3067}
(\byear{2018})
\doiurl{10.1002/celc.201800696}
\end{barticle}
\endbibitem

\bibitem[\protect\citeauthoryear{Batchelor-McAuley
  et~al.}{2015}]{batchelor-mcauley_situ_2015}
\begin{barticle}
\bauthor{\bsnm{Batchelor-McAuley}, \binits{C.}},
\bauthor{\bsnm{Ellison}, \binits{J.}},
\bauthor{\bsnm{Tschulik}, \binits{K.}},
\bauthor{\bsnm{Hurst}, \binits{P.L.}},
\bauthor{\bsnm{Boldt}, \binits{R.}},
\bauthor{\bsnm{Compton}, \binits{R.G.}}:
\batitle{In situ nanoparticle sizing with zeptomole sensitivity}.
\bjtitle{The Analyst}
\bvolume{140}(\bissue{15}),
\bfpage{5048}--\blpage{5054}
(\byear{2015})
\doiurl{10.1039/C5AN00474H}
\end{barticle}
\endbibitem

\bibitem[\protect\citeauthoryear{Zhou et~al.}{2017}]{zhou_collisions_2017}
\begin{barticle}
\bauthor{\bsnm{Zhou}, \binits{M.}},
\bauthor{\bsnm{Yu}, \binits{Y.}},
\bauthor{\bsnm{Hu}, \binits{K.}},
\bauthor{\bsnm{Xin}, \binits{H.L.}},
\bauthor{\bsnm{Mirkin}, \binits{M.V.}}:
\batitle{Collisions of {Ir} {Oxide} {Nanoparticles} with {Carbon}
  {Nanopipettes}: {Experiments} with {One} {Nanoparticle}}.
\bjtitle{Analytical Chemistry}
\bvolume{89}(\bissue{5}),
\bfpage{2880}--\blpage{2885}
(\byear{2017})
\doiurl{10.1021/acs.analchem.6b04140}
\end{barticle}
\endbibitem

\bibitem[\protect\citeauthoryear{Glasscott and
  Dick}{2019}]{glasscott_fine-tuning_2019}
\begin{barticle}
\bauthor{\bsnm{Glasscott}, \binits{M.W.}},
\bauthor{\bsnm{Dick}, \binits{J.E.}}:
\batitle{Fine-{Tuning} {Porosity} and {Time}-{Resolved} {Observation} of the
  {Nucleation} and {Growth} of {Single} {Platinum} {Nanoparticles}}.
\bjtitle{ACS Nano}
\bvolume{13}(\bissue{4}),
\bfpage{4572}--\blpage{4581}
(\byear{2019})
\doiurl{10.1021/acsnano.9b00546}
\end{barticle}
\endbibitem

\bibitem[\protect\citeauthoryear{Ma et~al.}{2018}]{ma_quantifying_2018}
\begin{barticle}
\bauthor{\bsnm{Ma}, \binits{H.}},
\bauthor{\bsnm{Ma}, \binits{W.}},
\bauthor{\bsnm{Chen}, \binits{J.-F.}},
\bauthor{\bsnm{Liu}, \binits{X.-Y.}},
\bauthor{\bsnm{Peng}, \binits{Y.-Y.}},
\bauthor{\bsnm{Yang}, \binits{Z.-Y.}},
\bauthor{\bsnm{Tian}, \binits{H.}},
\bauthor{\bsnm{Long}, \binits{Y.-T.}}:
\batitle{Quantifying {Visible}-{Light}-{Induced} {Electron} {Transfer}
  {Properties} of {Single} {Dye}-{Sensitized} {ZnO} {Entity} for {Water}
  {Splitting}}.
\bjtitle{Journal of the American Chemical Society}
\bvolume{140}(\bissue{15}),
\bfpage{5272}--\blpage{5279}
(\byear{2018})
\doiurl{10.1021/jacs.8b01623}
\end{barticle}
\endbibitem

\bibitem[\protect\citeauthoryear{Ustarroz et~al.}{2017}]{ustarroz_impact_2017}
\begin{barticle}
\bauthor{\bsnm{Ustarroz}, \binits{J.}},
\bauthor{\bsnm{Kang}, \binits{M.}},
\bauthor{\bsnm{Bullions}, \binits{E.}},
\bauthor{\bsnm{Unwin}, \binits{P.R.}}:
\batitle{Impact and oxidation of single silver nanoparticles at electrode
  surfaces: one shot versus multiple events}.
\bjtitle{Chemical Science}
\bvolume{8}(\bissue{3}),
\bfpage{1841}--\blpage{1853}
(\byear{2017})
\doiurl{10.1039/C6SC04483B}
\end{barticle}
\endbibitem

\bibitem[\protect\citeauthoryear{Gao et~al.}{2018}]{gao_30_2018}
\begin{barticle}
\bauthor{\bsnm{Gao}, \binits{R.}},
\bauthor{\bsnm{Ying}, \binits{Y.}},
\bauthor{\bsnm{Li}, \binits{Y.}},
\bauthor{\bsnm{Hu}, \binits{Y.}},
\bauthor{\bsnm{Yu}, \binits{R.}},
\bauthor{\bsnm{Lin}, \binits{Y.}},
\bauthor{\bsnm{Long}, \binits{Y.}}:
\batitle{A 30 nm {Nanopore} {Electrode}: {Facile} {Fabrication} and {Direct}
  {Insights} into the {Intrinsic} {Feature} of {Single} {Nanoparticle}
  {Collisions}}.
\bjtitle{Angewandte Chemie International Edition}
\bvolume{57}(\bissue{4}),
\bfpage{1011}--\blpage{1015}
(\byear{2018})
\doiurl{10.1002/anie.201710201}
\end{barticle}
\endbibitem

\bibitem[\protect\citeauthoryear{Amatore et~al.}{2001}]{amatore_ultrafast_2001}
\begin{barticle}
\bauthor{\bsnm{Amatore}, \binits{C.}},
\bauthor{\bsnm{Bouret}, \binits{Y.}},
\bauthor{\bsnm{Maisonhaute}, \binits{E.}},
\bauthor{\bsnm{Goldsmith}, \binits{J.I.}},
\bauthor{\bsnm{Abruña}, \binits{H.D.}}:
\batitle{Ultrafast {Voltammetry} of {Adsorbed} {Redox} {Active} {Dendrimers}
  with {Nanometric} {Resolution}: {An} {Electrochemical} {Microtome}}.
\bjtitle{ChemPhysChem}
\bvolume{2}(\bissue{2}),
\bfpage{130}--\blpage{134}
(\byear{2001})
\doiurl{10.1002/1439-7641(20010216)2:2<130::AID-CPHC130>3.0.CO;2-K}
\end{barticle}
\endbibitem

\bibitem[\protect\citeauthoryear{Watkins et~al.}{2003}]{watkins_zeptomole_2003}
\begin{barticle}
\bauthor{\bsnm{Watkins}, \binits{J.J.}},
\bauthor{\bsnm{Chen}, \binits{J.}},
\bauthor{\bsnm{White}, \binits{H.S.}},
\bauthor{\bsnm{Abruña}, \binits{H.D.}},
\bauthor{\bsnm{Maisonhaute}, \binits{E.}},
\bauthor{\bsnm{Amatore}, \binits{C.}}:
\batitle{Zeptomole {Voltammetric} {Detection} and {Electron}-{Transfer} {Rate}
  {Measurements} {Using} {Platinum} {Electrodes} of {Nanometer} {Dimensions}}.
\bjtitle{Analytical Chemistry}
\bvolume{75}(\bissue{16}),
\bfpage{3962}--\blpage{3971}
(\byear{2003})
\doiurl{10.1021/ac0342931}
\end{barticle}
\endbibitem

\bibitem[\protect\citeauthoryear{Güell
  et~al.}{2015}]{guell_redox-dependent_2015}
\begin{barticle}
\bauthor{\bsnm{Güell}, \binits{A.G.}},
\bauthor{\bsnm{Cuharuc}, \binits{A.S.}},
\bauthor{\bsnm{Kim}, \binits{Y.-R.}},
\bauthor{\bsnm{Zhang}, \binits{G.}},
\bauthor{\bsnm{Tan}, \binits{S.-y.}},
\bauthor{\bsnm{Ebejer}, \binits{N.}},
\bauthor{\bsnm{Unwin}, \binits{P.R.}}:
\batitle{Redox-{Dependent} {Spatially} {Resolved} {Electrochemistry} at
  {Graphene} and {Graphite} {Step} {Edges}}.
\bjtitle{ACS Nano}
\bvolume{9}(\bissue{4}),
\bfpage{3558}--\blpage{3571}
(\byear{2015})
\doiurl{10.1021/acsnano.5b00550}
\end{barticle}
\endbibitem

\bibitem[\protect\citeauthoryear{Momotenko
  et~al.}{2015}]{momotenko_high-speed_2015}
\begin{barticle}
\bauthor{\bsnm{Momotenko}, \binits{D.}},
\bauthor{\bsnm{Byers}, \binits{J.C.}},
\bauthor{\bsnm{McKelvey}, \binits{K.}},
\bauthor{\bsnm{Kang}, \binits{M.}},
\bauthor{\bsnm{Unwin}, \binits{P.R.}}:
\batitle{High-{Speed} {Electrochemical} {Imaging}}.
\bjtitle{ACS Nano}
\bvolume{9}(\bissue{9}),
\bfpage{8942}--\blpage{8952}
(\byear{2015})
\doiurl{10.1021/acsnano.5b02792}
\end{barticle}
\endbibitem

\bibitem[\protect\citeauthoryear{Kang et~al.}{2013}]{kang_electrochemical_2013}
\begin{barticle}
\bauthor{\bsnm{Kang}, \binits{S.}},
\bauthor{\bsnm{Nieuwenhuis}, \binits{A.F.}},
\bauthor{\bsnm{Mathwig}, \binits{K.}},
\bauthor{\bsnm{Mampallil}, \binits{D.}},
\bauthor{\bsnm{Lemay}, \binits{S.G.}}:
\batitle{Electrochemical {Single}-{Molecule} {Detection} in {Aqueous}
  {Solution} {Using} {Self}-{Aligned} {Nanogap} {Transducers}}.
\bjtitle{ACS Nano}
\bvolume{7}(\bissue{12}),
\bfpage{10931}--\blpage{10937}
(\byear{2013})
\doiurl{10.1021/nn404440v}
\end{barticle}
\endbibitem

\bibitem[\protect\citeauthoryear{Zevenbergen
  et~al.}{2011}]{zevenbergen_stochastic_2011}
\begin{barticle}
\bauthor{\bsnm{Zevenbergen}, \binits{M.A.G.}},
\bauthor{\bsnm{Singh}, \binits{P.S.}},
\bauthor{\bsnm{Goluch}, \binits{E.D.}},
\bauthor{\bsnm{Wolfrum}, \binits{B.L.}},
\bauthor{\bsnm{Lemay}, \binits{S.G.}}:
\batitle{Stochastic {Sensing} of {Single} {Molecules} in a {Nanofluidic}
  {Electrochemical} {Device}}.
\bjtitle{Nano Letters}
\bvolume{11}(\bissue{7}),
\bfpage{2881}--\blpage{2886}
(\byear{2011})
\doiurl{10.1021/nl2013423}
\end{barticle}
\endbibitem

\bibitem[\protect\citeauthoryear{Bard and
  Fan}{1996}]{bard_electrochemical_1996}
\begin{barticle}
\bauthor{\bsnm{Bard}, \binits{A.J.}},
\bauthor{\bsnm{Fan}, \binits{F.-R.F.}}:
\batitle{Electrochemical {Detection} of {Single} {Molecules}}.
\bjtitle{Accounts of Chemical Research}
\bvolume{29},
\bfpage{572}--\blpage{578}
(\byear{1996})
\doiurl{10.1126/science.267.5199.871}
\end{barticle}
\endbibitem

\bibitem[\protect\citeauthoryear{Byers et~al.}{2015}]{byers_single_2015}
\begin{barticle}
\bauthor{\bsnm{Byers}, \binits{J.C.}},
\bauthor{\bsnm{Paulose~Nadappuram}, \binits{B.}},
\bauthor{\bsnm{Perry}, \binits{D.}},
\bauthor{\bsnm{McKelvey}, \binits{K.}},
\bauthor{\bsnm{Colburn}, \binits{A.W.}},
\bauthor{\bsnm{Unwin}, \binits{P.R.}}:
\batitle{Single {Molecule} {Electrochemical} {Detection} in {Aqueous}
  {Solutions} and {Ionic} {Liquids}}.
\bjtitle{Analytical Chemistry}
\bvolume{87}(\bissue{20}),
\bfpage{10450}--\blpage{10456}
(\byear{2015})
\doiurl{10.1021/acs.analchem.5b02569}
\end{barticle}
\endbibitem

\bibitem[\protect\citeauthoryear{Paiva et~al.}{2022}]{paiva_enzymatic_2022}
\begin{barticle}
\bauthor{\bsnm{Paiva}, \binits{T.O.}},
\bauthor{\bsnm{Schneider}, \binits{A.}},
\bauthor{\bsnm{Bataille}, \binits{L.}},
\bauthor{\bsnm{Chovin}, \binits{A.}},
\bauthor{\bsnm{Anne}, \binits{A.}},
\bauthor{\bsnm{Michon}, \binits{T.}},
\bauthor{\bsnm{Wege}, \binits{C.}},
\bauthor{\bsnm{Demaille}, \binits{C.}}:
\batitle{Enzymatic activity of individual bioelectrocatalytic viral
  nanoparticles: dependence of catalysis on the viral scaffold and its length}.
\bjtitle{Nanoscale}
\bvolume{14}(\bissue{3}),
\bfpage{875}--\blpage{889}
(\byear{2022})
\doiurl{10.1039/D1NR07445H}
\end{barticle}
\endbibitem

\bibitem[\protect\citeauthoryear{German et~al.}{2016}]{german_laplace_2016}
\begin{barticle}
\bauthor{\bsnm{German}, \binits{S.R.}},
\bauthor{\bsnm{Edwards}, \binits{M.A.}},
\bauthor{\bsnm{Chen}, \binits{Q.}},
\bauthor{\bsnm{White}, \binits{H.S.}}:
\batitle{Laplace {Pressure} of {Individual} {H} $_{\textrm{2}}$ {Nanobubbles}
  from {Pressure}–{Addition} {Electrochemistry}}.
\bjtitle{Nano Letters}
\bvolume{16}(\bissue{10}),
\bfpage{6691}--\blpage{6694}
(\byear{2016})
\doiurl{10.1021/acs.nanolett.6b03590}
\end{barticle}
\endbibitem

\bibitem[\protect\citeauthoryear{Hochstetler
  et~al.}{2000}]{hochstetler_real-time_2000}
\begin{barticle}
\bauthor{\bsnm{Hochstetler}, \binits{S.E.}},
\bauthor{\bsnm{Puopolo}, \binits{M.}},
\bauthor{\bsnm{Gustincich}, \binits{S.}},
\bauthor{\bsnm{Raviola}, \binits{E.}},
\bauthor{\bsnm{Wightman}, \binits{R.M.}}:
\batitle{Real-{Time} {Amperometric} {Measurements} of {Zeptomole} {Quantities}
  of {Dopamine} {Released} from {Neurons}}.
\bjtitle{Analytical Chemistry}
\bvolume{72}(\bissue{3}),
\bfpage{489}--\blpage{496}
(\byear{2000})
\doiurl{10.1021/ac991119x}
\end{barticle}
\endbibitem

\bibitem[\protect\citeauthoryear{Gu et~al.}{2019}]{gu_plasticity_2019}
\begin{barticle}
\bauthor{\bsnm{Gu}, \binits{C.}},
\bauthor{\bsnm{Larsson}, \binits{A.}},
\bauthor{\bsnm{Ewing}, \binits{A.G.}}:
\batitle{Plasticity in exocytosis revealed through the effects of repetitive
  stimuli affect the content of nanometer vesicles and the fraction of
  transmitter released}.
\bjtitle{Proceedings of the National Academy of Sciences}
\bvolume{116}(\bissue{43}),
\bfpage{21409}--\blpage{21415}
(\byear{2019})
\doiurl{10.1073/pnas.1910859116}
\end{barticle}
\endbibitem

\bibitem[\protect\citeauthoryear{Li et~al.}{2018}]{li_electrochemical_2018}
\begin{barticle}
\bauthor{\bsnm{Li}, \binits{X.}},
\bauthor{\bsnm{Dunevall}, \binits{J.}},
\bauthor{\bsnm{Ewing}, \binits{A.G.}}:
\batitle{Electrochemical quantification of transmitter concentration in single
  nanoscale vesicles isolated from {PC12} cells}.
\bjtitle{Faraday Discussions}
\bvolume{210},
\bfpage{353}--\blpage{364}
(\byear{2018})
\doiurl{10.1039/C8FD00020D}
\end{barticle}
\endbibitem

\bibitem[\protect\citeauthoryear{Trasobares
  et~al.}{2017}]{trasobares_estimation_2017}
\begin{barticle}
\bauthor{\bsnm{Trasobares}, \binits{J.}},
\bauthor{\bsnm{Rech}, \binits{J.}},
\bauthor{\bsnm{Jonckheere}, \binits{T.}},
\bauthor{\bsnm{Martin}, \binits{T.}},
\bauthor{\bsnm{Aleveque}, \binits{O.}},
\bauthor{\bsnm{Levillain}, \binits{E.}},
\bauthor{\bsnm{Diez-Cabanes}, \binits{V.}},
\bauthor{\bsnm{Olivier}, \binits{Y.}},
\bauthor{\bsnm{Cornil}, \binits{J.}},
\bauthor{\bsnm{Nys}, \binits{J.P.}},
\bauthor{\bsnm{Sivakumarasamy}, \binits{R.}},
\bauthor{\bsnm{Smaali}, \binits{K.}},
\bauthor{\bsnm{Leclere}, \binits{P.}},
\bauthor{\bsnm{Fujiwara}, \binits{A.}},
\bauthor{\bsnm{Théron}, \binits{D.}},
\bauthor{\bsnm{Vuillaume}, \binits{D.}},
\bauthor{\bsnm{Clément}, \binits{N.}}:
\batitle{Estimation of $\pi-\pi$ electronic couplings from current
  measurements}.
\bjtitle{Nano Letters}
\bvolume{17}(\bissue{5}),
\bfpage{3215}--\blpage{3224}
(\byear{2017})
\doiurl{10.1021/acs.nanolett.7b00804}
\end{barticle}
\endbibitem

\bibitem[\protect\citeauthoryear{Zhou
  et~al.}{2017}]{zhou_electrodeposition_2017}
\begin{barticle}
\bauthor{\bsnm{Zhou}, \binits{M.}},
\bauthor{\bsnm{Dick}, \binits{J.E.}},
\bauthor{\bsnm{Bard}, \binits{A.J.}}:
\batitle{Electrodeposition of {Isolated} {Platinum} {Atoms} and {Clusters} on
  {Bismuth}—{Characterization} and {Electrocatalysis}}.
\bjtitle{Journal of the American Chemical Society}
\bvolume{139}(\bissue{48}),
\bfpage{17677}--\blpage{17682}
(\byear{2017})
\doiurl{10.1021/jacs.7b10646}
\end{barticle}
\endbibitem

\bibitem[\protect\citeauthoryear{Zhou et~al.}{2019}]{zhou_probing_2019}
\begin{barticle}
\bauthor{\bsnm{Zhou}, \binits{M.}},
\bauthor{\bsnm{Bao}, \binits{S.}},
\bauthor{\bsnm{Bard}, \binits{A.J.}}:
\batitle{Probing {Size} and {Substrate} {Effects} on the {Hydrogen} {Evolution}
  {Reaction} by {Single} {Isolated} {Pt} {Atoms}, {Atomic} {Clusters}, and
  {Nanoparticles}}.
\bjtitle{Journal of the American Chemical Society}
\bvolume{141}(\bissue{18}),
\bfpage{7327}--\blpage{7332}
(\byear{2019})
\doiurl{10.1021/jacs.8b13366}
\end{barticle}
\endbibitem

\bibitem[\protect\citeauthoryear{Li et~al.}{2019}]{li_transition_2019}
\begin{barticle}
\bauthor{\bsnm{Li}, \binits{Y.}},
\bauthor{\bsnm{Wang}, \binits{H.}},
\bauthor{\bsnm{Wang}, \binits{Z.}},
\bauthor{\bsnm{Qiao}, \binits{Y.}},
\bauthor{\bsnm{Ulstrup}, \binits{J.}},
\bauthor{\bsnm{Chen}, \binits{H.-Y.}},
\bauthor{\bsnm{Zhou}, \binits{G.}},
\bauthor{\bsnm{Tao}, \binits{N.}}:
\batitle{Transition from stochastic events to deterministic ensemble average in
  electron transfer reactions revealed by single-molecule conductance
  measurement}.
\bjtitle{Proceedings of the National Academy of Sciences}
\bvolume{116}(\bissue{9}),
\bfpage{3407}--\blpage{3412}
(\byear{2019})
\doiurl{10.1073/pnas.1814825116}
\end{barticle}
\endbibitem

\bibitem[\protect\citeauthoryear{Grall et~al.}{2021}]{grall_attoampere_2021}
\begin{barticle}
\bauthor{\bsnm{Grall}, \binits{S.}},
\bauthor{\bsnm{Alić}, \binits{I.}},
\bauthor{\bsnm{Pavoni}, \binits{E.}},
\bauthor{\bsnm{Awadein}, \binits{M.}},
\bauthor{\bsnm{Fujii}, \binits{T.}},
\bauthor{\bsnm{Müllegger}, \binits{S.}},
\bauthor{\bsnm{Farina}, \binits{M.}},
\bauthor{\bsnm{Clément}, \binits{N.}},
\bauthor{\bsnm{Gramse}, \binits{G.}}:
\batitle{Attoampere {Nanoelectrochemistry}}.
\bjtitle{Small}
\bvolume{17}(\bissue{29}),
\bfpage{2101253}
(\byear{2021})
\doiurl{10.1002/smll.202101253}
\end{barticle}
\endbibitem

\bibitem[\protect\citeauthoryear{Torbensen
  et~al.}{2019}]{torbensen_immuno-based_2019}
\begin{barticle}
\bauthor{\bsnm{Torbensen}, \binits{K.}},
\bauthor{\bsnm{Patel}, \binits{A.N.}},
\bauthor{\bsnm{Anne}, \binits{A.}},
\bauthor{\bsnm{Chovin}, \binits{A.}},
\bauthor{\bsnm{Demaille}, \binits{C.}},
\bauthor{\bsnm{Bataille}, \binits{L.}},
\bauthor{\bsnm{Michon}, \binits{T.}},
\bauthor{\bsnm{Grelet}, \binits{E.}}:
\batitle{Immuno-{Based} {Molecular} {Scaffolding} of {Glucose} {Dehydrogenase}
  and {Ferrocene} {Mediator} on \textit{fd} {Viral} {Particles} {Yields}
  {Enhanced} {Bioelectrocatalysis}}.
\bjtitle{ACS Catalysis}
\bvolume{9}(\bissue{6}),
\bfpage{5783}--\blpage{5796}
(\byear{2019})
\doiurl{10.1021/acscatal.9b01263}
\end{barticle}
\endbibitem

\bibitem[\protect\citeauthoryear{Nault
  et~al.}{2015}]{nault_electrochemical_2015}
\begin{barticle}
\bauthor{\bsnm{Nault}, \binits{L.}},
\bauthor{\bsnm{Taofifenua}, \binits{C.}},
\bauthor{\bsnm{Anne}, \binits{A.}},
\bauthor{\bsnm{Chovin}, \binits{A.}},
\bauthor{\bsnm{Demaille}, \binits{C.}},
\bauthor{\bsnm{Besong-Ndika}, \binits{J.}},
\bauthor{\bsnm{Cardinale}, \binits{D.}},
\bauthor{\bsnm{Carette}, \binits{N.}},
\bauthor{\bsnm{Michon}, \binits{T.}},
\bauthor{\bsnm{Walter}, \binits{J.}}:
\batitle{Electrochemical {Atomic} {Force} {Microscopy} {Imaging} of
  {Redox}-{Immunomarked} {Proteins} on {Native} {Potyviruses}: {From}
  {Subparticle} to {Single}-{Protein} {Resolution}}.
\bjtitle{ACS Nano}
\bvolume{9}(\bissue{5}),
\bfpage{4911}--\blpage{4924}
(\byear{2015})
\doiurl{10.1021/acsnano.5b00952}
\end{barticle}
\endbibitem

\bibitem[\protect\citeauthoryear{Chennit
  et~al.}{2022}]{chennit_high-density_2022}
\begin{barticle}
\bauthor{\bsnm{Chennit}, \binits{K.}},
\bauthor{\bsnm{Coffinier}, \binits{Y.}},
\bauthor{\bsnm{Li}, \binits{S.}},
\bauthor{\bsnm{Clément}, \binits{N.}},
\bauthor{\bsnm{Anne}, \binits{A.}},
\bauthor{\bsnm{Chovin}, \binits{A.}},
\bauthor{\bsnm{Demaille}, \binits{C.}}:
\batitle{High-density single antibody electrochemical nanoarrays}.
\bjtitle{Nano Research}
(\byear{2022})
\doiurl{10.1007/s12274-022-5137-1}
\end{barticle}
\endbibitem

\end{thebibliography}

\end{document}


\title[QBIOL: A quantum bioelectrochemical software based on point stochastic processes]{
\begin{center}
\textbf{Supplementary information}
\end{center}
QBIOL: A quantum bioelectrochemical software based on point stochastic processes}


\author[1,2]{\fnm{Simon} \sur{Grall}}\email{sgrall@laas.fr}
\author[2]{\fnm{Ignacio} \sur{Madrid}}\email{madrid@sat.t.u-tokyo.ac.jp}
\author[2]{\fnm{Aramis} \sur{Dufour}}\email{aramis.dufour@polytechnique.org}
\author[2]{\fnm{Helen} \sur{Sands}}\email{helen.sands@polytechnique.org}
\author[4]{\fnm{Masaki} \sur{Kato}}\email{kato\_m@eis.hokudai.ac.jp}
\author[3]{\fnm{Akira} \sur{Fujiwara}}\email{akira.fujiwara@ntt.com}
\author[2]{\fnm{Soo Hyeon} \sur{Kim}}\email{shkim@iis.u-tokyo.ac.jp}
\author[5]{\fnm{Arnaud} \sur{Chovin}}\email{arnaud.chovin@u-paris.fr}
\author[5]{\fnm{Christophe} \sur{Demaille}}\email{christophe.demaille@u-paris.fr}
\author*[1,2]{\fnm{Nicolas} \sur{Clement}}\email{nclement@iis.u-tokyo.ac.jp}


\affil[1]{\orgdiv{LAAS}, \orgname{CNRS}, \orgaddress{\street{7, av. du colonel Roche}, \city{Toulouse}, \postcode{31031}, \country{France}}}

\affil[2]{\orgdiv{LIMMS}, \orgname{CNRS}, \orgaddress{\street{4-6-1 Komaba}, \city{Meguro-ku, Tokyo}, \postcode{153-8505}, \country{Japan}}}

\affil[3]{\orgdiv{ Basic Research Laboratories}, \orgname{NTT}, \orgaddress{\street{3-1 Morinosato Wakamiya}, \city{Atsugi-shi, Kanagawa}, \postcode{243-0198}, \country{Japan}}}

\affil[4]{\orgdiv{Graduate School of Chemical Sciences and Engineering}, \orgname{Hokkaido University}, \orgaddress{\street{Kita 13 Nishi 8}, \city{Kita-ku, Sapporo}, \postcode{060-8628}, \country{Japan}}}

\affil[5]{\orgdiv{Laboratoire d'Electrochimie Moléculaire}, \orgname{Université Paris Cité - CNRS}, \orgaddress{\street{15, rue Jean-Antoine de Baïf}, \city{Paris}, \postcode{75013}, \country{France}}}

\keywords{}



\maketitle
\clearpage
\tableofcontents
\clearpage

\section{Rates}\label{rates_section}
QBIOL uses by default Marcus-Hush (MH) rates \cite{marcus_theory_1965,hush_homogeneous_1968} (Eq. \ref{k_marcus}), although the widely used Butler Volmer (BV) model is also available (Eq. \ref{BVox} and \ref{BVred}). The BV model, much simpler, is an asymptotic approximation of the MH approach that cannot account for changes of reorganization energy $\lambda$, eventually failing to reproduce accurately experimental results for bio-molecular systems \cite{zheng_electrochemical_2023, madrid_ballistic_2023}. Unlike the BV model, the MH model accounts for these changes (Eq. \ref{lambda_reorg}), at the cost of a computationally expensive integration.

\begin{align}
k^{MH}_{\text{ox,red}} =& \frac{\rho H^2}{\hbar} e^{-\beta z} \sqrt{\frac{\pi}{k_B T \lambda}} \times \nonumber \\
& \int\limits_{-\infty}^{+\infty} \frac{1}{1 + e^{\frac{x}{k_B T}}} e^{-\frac{(x - \lambda \pm \eta)^2}{4 \lambda k_B T}} \, dx \label{k_marcus} \\
\lambda =& \frac{q^2}{8 \pi \varepsilon_0} \left(\frac{1}{a_0} - \frac{1}{2(z + a_0)}\right) \left(\frac{1}{\varepsilon_{\text{op}}} - \frac{1}{\varepsilon}\right) \label{lambda_reorg}
\end{align}

with $k_{ox}$ the oxidation rate, $k_{red}$ the reduction rate, $\rho$ the density of state in the metallic electrode, $H^2$ the electronic coupling, $\beta$ the tunnel decay ratio, $z$ the distance to the electrode, $\hbar$ the reduced Planck constant, $\lambda$ the reorganization energy, $T$ the temperature, $k_B$ the Boltzmann constant and $\eta=q(E-E^0)$ with $E$ the potential at the electrode, $E^0$ the standard potential of the molecule, $q$ the elementary charge, $\varepsilon_0$ the permittivity of vacuum, $\varepsilon_{op}$ the optical limit for the relative permittivity of water ($\approx 1.78$) and $\varepsilon$ the effective relative permittivity of water. Note that this expression corresponds to the non-adiabatic electron transfer case (weak electronic coupling of the redox molecule with the electrode), typically encountered in bio-electrochemistry. The adiabatic case can be implemented by modifying the pre-integral factor.\\
Due to the computational cost of calculating the MH rates for every $z$ and $E$ over time, rates are pre-computed for efficiency, with a 1 mV resolution in terms of voltage and 0.01 \AA~in terms of space. During the actual simulation, a linear interpolation with the actual positions and voltages is done for precise values of $k_{ox,red}$. No approximation other than the interpolation is made on the calculation of the rates, nor a threshold is used limiting their extension in space as it would prevent from considering intermolecular electron transfer.\\

\begin{figure}
\centering
\includegraphics[clip, trim= 0cm 4.5cm 0.6cm 3cm, width = \textwidth]{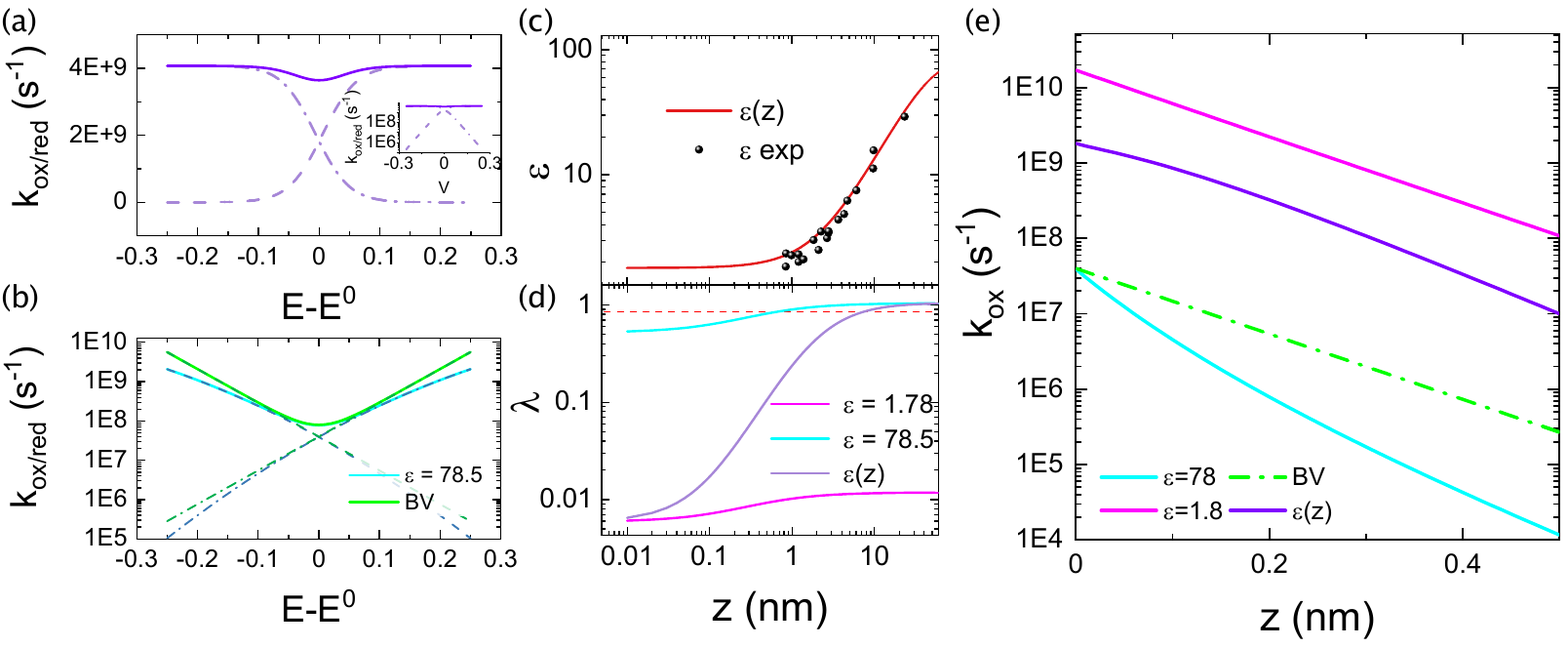}
\caption{\textbf{QBIOL rates.} (a) MH rates with $\varepsilon = 1.8$ (inset shows the logscale in y axis). (b) BV rates and MH rates with $\varepsilon = 78.5$. (c) The permittivity of water has been found to decrease with confinement \cite{fumagalli_anomalously_2018}. We show here the data from the original publication with the arbitrary fit we use to obtain $\varepsilon(z)$. (d) $\lambda(z)$ for different values of $\varepsilon$. The red horizontal dashed line marks $\lambda = 0.85$ eV. (e) Oxidation rates comparison. MH rates calculated with a varying $\lambda$ with $z$ have the $\varepsilon$ used for the calculation indicated in legend.}
\label{ratesfig}
\end{figure}

\begin{figure}
\centering
\includegraphics[clip, trim= 5cm 7cm 5cm 7cm, width = \textwidth]{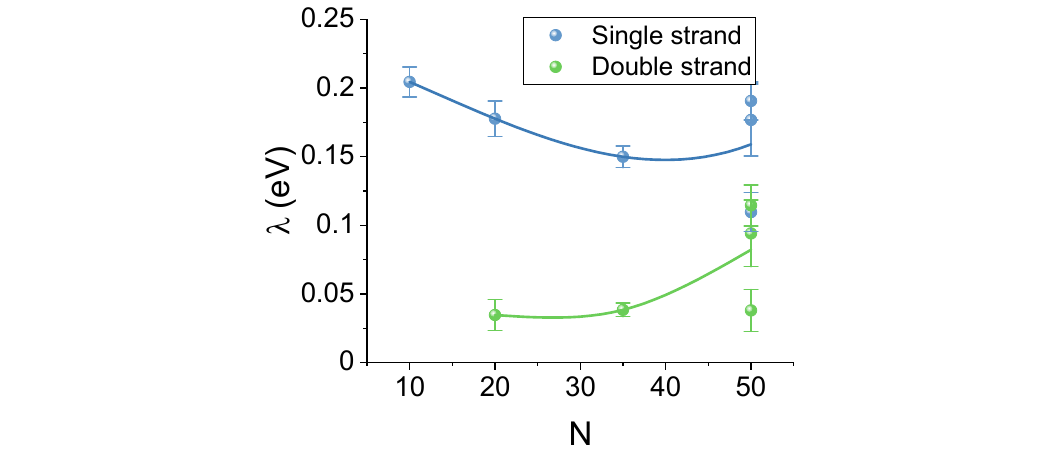}
\caption{\textbf{Dependency of the reorganization energy with the length of DNA.} High scanrate cyclic voltammetry experiments measurements of the reorganization energy $\lambda$ measured on ssDNA and dsDNA with different number of bases $N$. Lines are guides to the eye. Reproduced from \cite{zheng_electrochemical_2023}}
\label{lambda_vs_N}
\end{figure}

In the case of MH rates, we account for the different reorganization energies $\lambda$ seen by the electrochemical center \cite{madrid_ballistic_2023, zheng_activationless_2024} by adjusting the permittivity of water $\varepsilon$. For the DNA end-attached molecules, we consider $\varepsilon = 1.8$ (Fig. \ref{ratesfig} (a)) and for free particles, we consider the macroscopic value $\varepsilon = 78.5$ (Fig. \ref{ratesfig} (b)). Previous work on nanoconfined water \cite{fumagalli_anomalously_2018} showed experimentally very small values of $\varepsilon$ (Fig. \ref{ratesfig} (c)), which arguably could be used to calculate $\lambda(z)$ with a varying $\varepsilon$ (Fig. \ref{ratesfig} (d)). However, confined water molecules within a gap $z_{gap}$ are different from freely moving molecules seen at a distance $\sim z_{gap}$. In practice, all QBIOL simulations with free particles are run with $\varepsilon = 78.5$ and with $\varepsilon = 1.8$ for DNA end-attached molecule, with the oxidation rates obtained when calculating $\lambda(z)$ with $\varepsilon(z)$ shown only for illustrative purposes.\\
On Fig. \ref{ratesfig} (e), we see the different oxidation rates near the interface and in particular, considering $\lambda(z)$ calculated with $\varepsilon = 78.5$ translates into a sharper decrease of $k_{ox}$ at the interface, which could be taken into account in BV formalism with the commonly admitted $\lambda = 0.85$ eV by just increasing slightly the value of the tunnel decay ratio $\beta$.\\
Additional experimental insigths are given in \cite{zheng_electrochemical_2023} where experimental values of $\lambda$ are extracted for different length of single strand (ss) and double strand (ds) DNA molecules through high scanrate cyclic voltammetry (Fig. \ref{lambda_vs_N}), emphasizing that if research is still ongoing for an accurate description of $\lambda$ in biomolecular systems, it does not prevent QBIOL to integrate these findings into its electron transfer rates.
Aside electrochemical rates, QBIOL is fully compatible with any type of rates, such as molecular electronics or the one used for the calculation of the quantum dot statistics in the main paper.

\begin{align}
k_{ox} =& k_0e^{-\beta z}e^{(1-\alpha)\frac{q(V-V^0)}{k_BT}} \label{BVox} \\
k_{red} =& k_0e^{-\beta z}e^{-\alpha\frac{q(V-V^0)}{k_BT}}\label{BVred}
\end{align}

\section{Time constants}

\begin{figure*}
\centering
\includegraphics[clip, trim= 1.5cm 0cm 1.7cm 0cm, width = \textwidth]{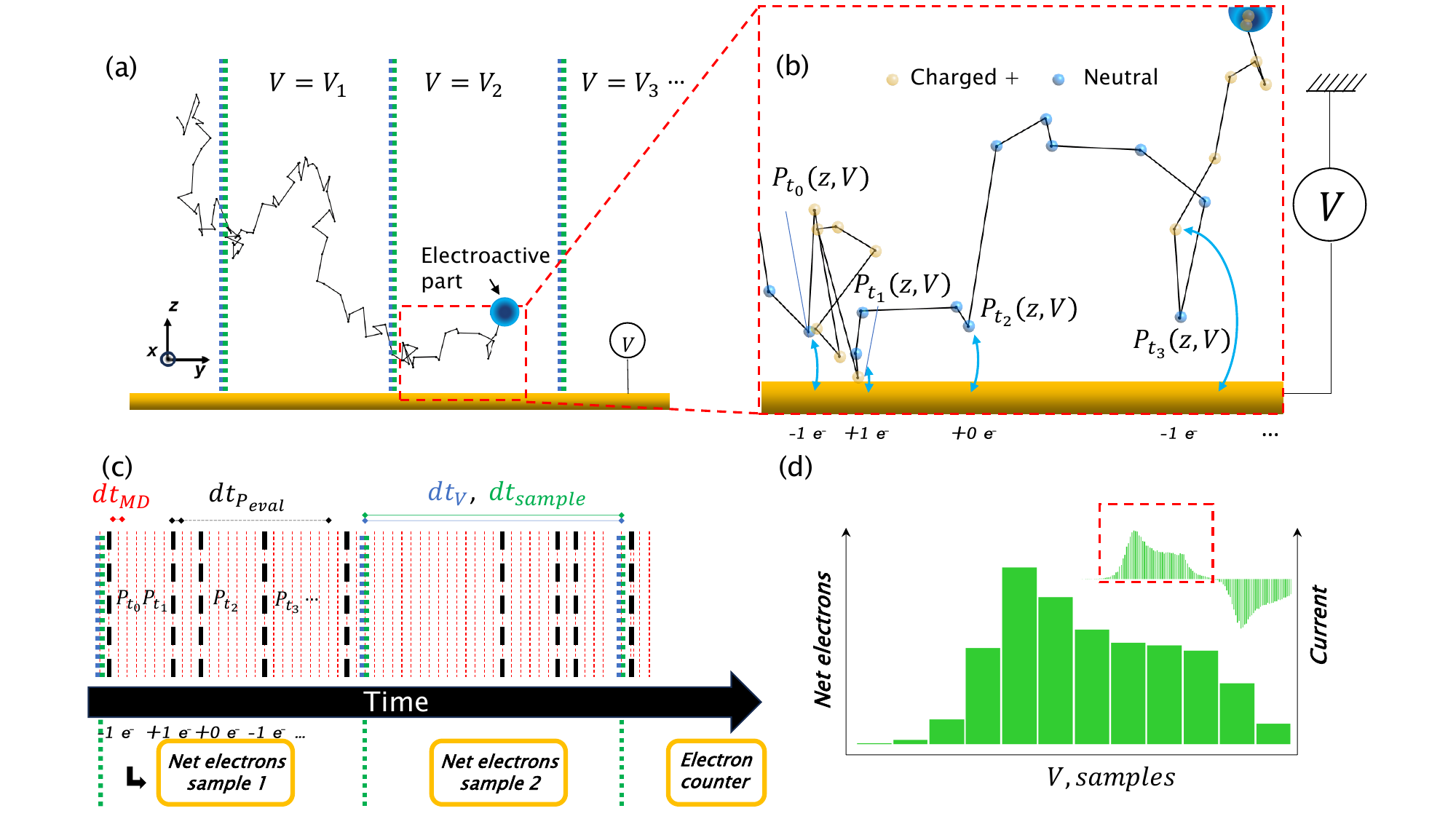}
\caption{\textbf{QBIOL's general workflow.} (a) A trajectory of the electroactive part is obtained from MD. Here it is represented with a single electrode where a varying voltage $E -E^0$ is applied. (b) From the position of the electroactive label and potential $E -E^0$ applied to the electrode, QBIOL can estimate a probability of transferring an electron between the molecule and the electrode. Electron transferred to the electrode are counted here. (c) Different time steps coexist: $dt_{MD}$ is the time step between each simulated MD point, $dt_{P_{eval}}$ is the time step (of variable length) between each probability estimation for electron transfer ($P_{ti}(V,z)$), $dt_{V}$ and $dt_{sample}$, often equal, are the time step between each voltage change and the time step between each sample for net electron transferred estimation, respectively. (d) The simulation is repeated thousands of times, and an average net number of electron transferred is obtained at each sample point. The conversion to current is straightforward using $dt_{sample}$.}
\label{QBIOL_simple}
\end{figure*}

Another strength of QBIOL is its ability to manage time as closely as possible to experimental data acquisition devices. A general description of the time constants is shown Fig. \ref{QBIOL_simple}. The sampling time $dt_{sample}$ is as often as possible the same as the voltage time resolution $dt_{V}$. If $dt_{V} < dt_{sample}$, typically when using high frequency modulations, an interpolation is done to correctly change the voltage during the sample acquisition. The minimum time constant $dt_{MD}$ is defined here as the molecular dynamics time constant, set to $dt_{MD} = 9.09\times 10^{-13}$s. This value is chosen to match the currently used MD library and so that commonly encountered diffusion constants for freely moving particles (for example $D= 6.5 \times 10^{-6}$ cm$^2$/s for Fc in NaClO$_4$ \cite{zevenbergen_fast_2009}) result in reasonably slowly moving particles ($\Delta z \ll 1~\AA$), a good compromise between speed and accuracy for the simulations. Finally, $dt_{P_{eval}}$ corresponds to the time at which the probability to transfer an electron is evaluated, and is defined as $dt_{P_{eval}} = -\frac{1}{\Gamma_{max}}\log(1-u)$, $u \in \mathcal{U}(0,1)$~ with $\Gamma_{max}$ the highest rate encountered during the considered time period and $\mathcal{U}$ the uniform distribution, a consequence of the rejection sampling used in QBIOL (see also section \ref{rejection_sampling}).

\clearpage
\section{Comparison with previous stochastic electrochemical simulations}
\begin{table*}[h!]
\centering
\begin{tabular}{p{4cm}p{2.5cm}lp{2.3cm}c}
\toprule
 \textbf{Implementation}& \textbf{MD} &\textbf{$\tau$ min/max} & \raggedright \textbf{Rates management} &\textbf{Ref.} \\
\toprule
Steady-state CV, concentric nanogaps &Free particle & 0.2 ps / 2 \textmu s& BV& \cite{white_electrochemistry_2008}\\ \midrule
CV, microdisk electrode&Free particle& 2 ms/ 77 min  &\raggedright BV, MH (approx.\cite{oldham_comparison_1988}) & \cite{cutress_how_2011}\\ \midrule
Chronoamperometry, microdisk electrode&Free particle & 1 \textmu s/130 \textmu s &Collision to the electrode & \cite{cutress_electrochemical_2011}\\ \midrule
CV, E-DNA as half-sphere confined springs&Spring attached particle & 0.2 ps/160 \textmu s &BV & \cite{huang_random_2013}\\ \midrule
Redox cycling, confined geometries&Free particle & 2 ns/1 min&\raggedright Threshold near the electrode&\cite{katelhon_noise_2013}\\ \midrule
CV, one dimensional &Free particle& 13.3 ms/ 8 s & BV&\cite{samin_one-dimensional_2016}\\ \midrule
Arbitrary $V$, nanogap &\raggedright Free particle, DNA, RNA,... & 0.909 ps/1 min& BV, MH& This work\\
\botrule
\end{tabular}
\caption{Comparison with previous electrochemical stochastic simulations. $\tau$ min/max indicates the shortest/ longest time resolved. BV stands for Butler-Volmer and MH for Marcus Hush rates.}
\label{stochastic_table}
\end{table*}

\clearpage

\section{Algorithmic considerations}\label{algorithmic_considerations}
\subsection{Rejection sampling}\label{rejection_sampling}
A ``naive'' approach would be to calculate the probability to transfer an electron from a given position of the molecule to the electrode at each $dt_{MD}$. However, considering the timescale of most electrochemical measurements ($> 1$ s), this would involve calculations out of reach for the computing power available at the time we write this article. We use instead rejection sampling, a method of sampling which consists here in estimating the time between two electrochemical events instead of estimating the probability of such events for each $dt_{MD}$ \cite{casella_generalized_2004}. This is especially efficient when the probability for an event to occur at each time step is small, which is the case here. It means in practice that, for example on what is represented Figure \ref{QBIOL_simple}, the probability to transfer an electron is estimated at irregular time intervals following $dt_{P_{eval}} = -\frac{1}{\Gamma_{max}}\log(1-u)$, $u \in \mathcal{U}(0,1)$, with $\Gamma_{max} = \max(k_{ox} + k_{red})$. To give an order of magnitude, the naive simulation of a cylic voltammogram was estimated to take around one year, versus a few minutes with rejection sampling.

\subsection{Electron counting}
The probability of transferring an electron over time is estimated using the rates (section \ref{rates_section}) and rejection sampling. Each time the probability to transfer an electron is realized, the state of the molecule is switched and one electron is added to the counter corresponding to the location (top electrode, bottom electrode,...) and transition (0 to 1 or 1 to 0) of the realized electron jump. The mean waiting time between two transitions $\tau_{01}$ and $\tau_{10}$ are also recorded over the simulation and defined as described by Eq. (\ref{tau01}) and (\ref{tau10}), and allow to calculate further statistics on the current.
\begin{align}
\tau_{01} = \frac{1}{n_{0\rightarrow 1}}\sum_i^{n_{0\rightarrow 1}}{t^i_{01}} \label{tau01}\\
\tau_{10} = \frac{1}{n_{1\rightarrow 0}}\sum_i^{n_{1\rightarrow 0}}{t^i_{10}} \label{tau10}
\end{align} 
with $\tau_{01}$ ($\tau_{10}$) the mean waiting time to transfer an electron from the state 0 to 1 (1 to 0), $t^i_{01}$ ($t^i_{10}$) the time between transitions $i-1$ and $i$ and $n_{0\rightarrow 1}$ ($n_{1\rightarrow 0}$) the total transitions from the state 0 to 1 (1 to 0, respectively) for a single molecule. It also allows to obtain chronoamperometry by binning the transition times $\tau_{01}$ and $\tau_{10}$ and integrating the net number of jump $n_{jp} = n_{0\rightarrow 1} - n_{1\rightarrow 0}$ in each bin (of duration $\Delta t_{bin}$), with the current calculated as $I_{bin} = q \times n_{jp} / \Delta t_{bin}$.
\subsection{``Long'' numerical experiments} \label{long_experiments}
\subsubsection{Algorithms}\label{algorithms}
The computation strategy is different depending on the duration of the experiment. We define $\Delta t_{track}$ the time covered by a MD track, that we assume representative of the total spatial configurations accessible to the molecule. It means for example for a free particle that the MD track covers enough time so that the molecule diffuses in the entire gap defined in the experiment. We also define $dt_{sample}$ as the simulated time for one output in current. For example, for voltage steps $dV = 10$ mV in a cyclic voltammogram at $\nu = 1$ V/s, $dt_{sample} = dV/\nu = 10$ ms per voltage step (here, one current output per voltage step).\\

\begin{figure}[H]%
\centering
\includegraphics[width=1\textwidth,clip, trim = 0cm 3cm 0cm 4cm ]{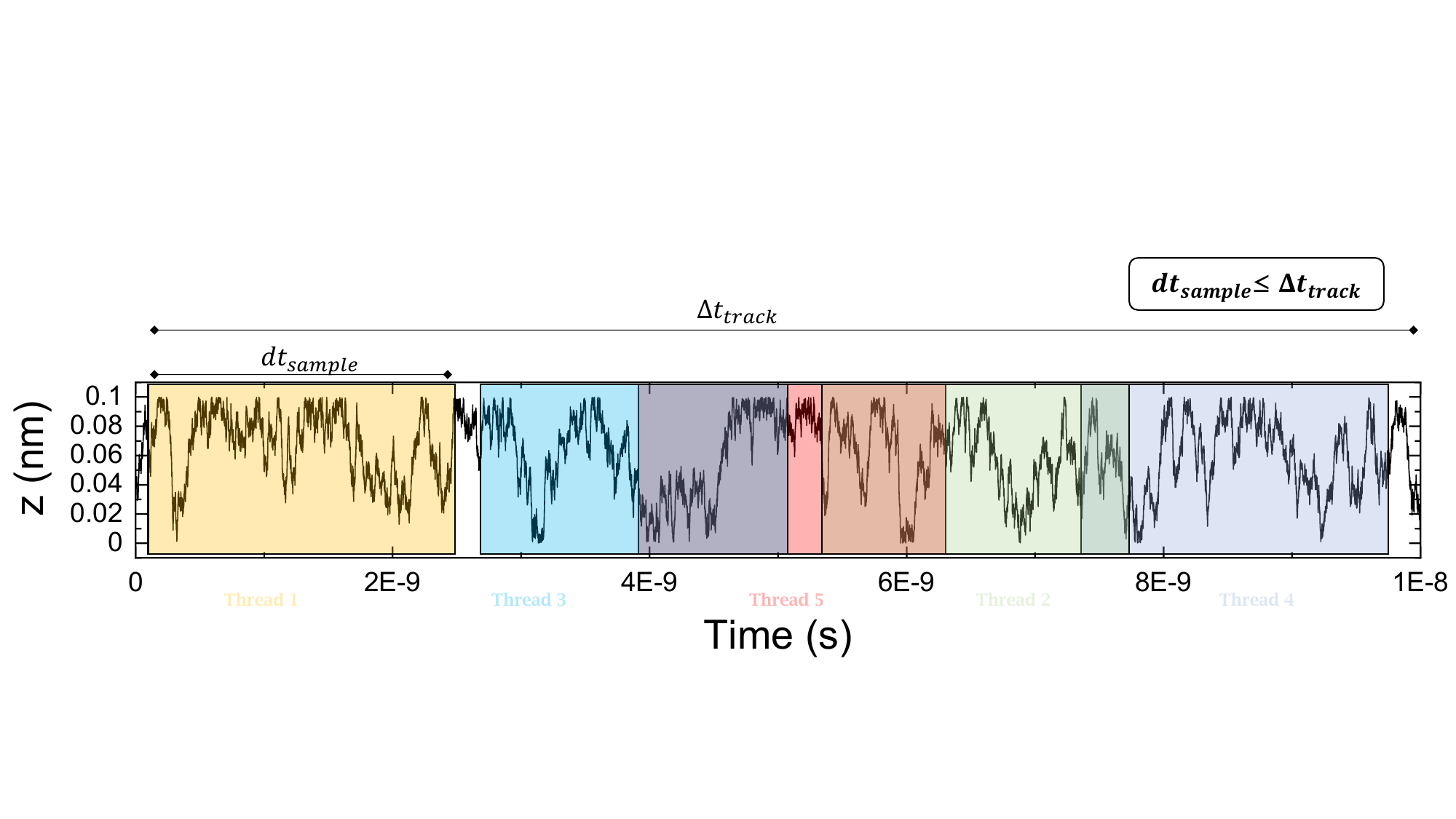}
\caption{\textbf{Chronological algorithm.} Example track (CV, free particle, 0.1 nm gap bounded by two walls of which one or two are electrodes) where each thread covers a different part of the track. The region covered by each thread is assigned randomly, the time window identical in duration, and depends only on the duration of the experiment (i.e. voltage step and sweeprate).}\label{tracks_fast_CV}
\end{figure}

For experiments where $dt_{sample} \leq \Delta t_{track}$, we use the ``chronological'' algorithm described Fig. \ref{tracks_fast_CV}. Each thread goes through a portion of the track that corresponds to $dt_{sample}$ in duration, starting randomly on the track to keep an overall representative track (as each thread only sees a fraction of the whole track). The results are averaged over all threads, each thread representing a single molecule during a duration of $dt_{sample}$ in a different location in space. The history of the states of the molecule is naturally preserved as each thread conducts the whole experiment for one molecule.

\begin{figure}[H]%
\centering
\includegraphics[width=1\textwidth,clip, trim = 0cm 4cm 1cm 4cm ]{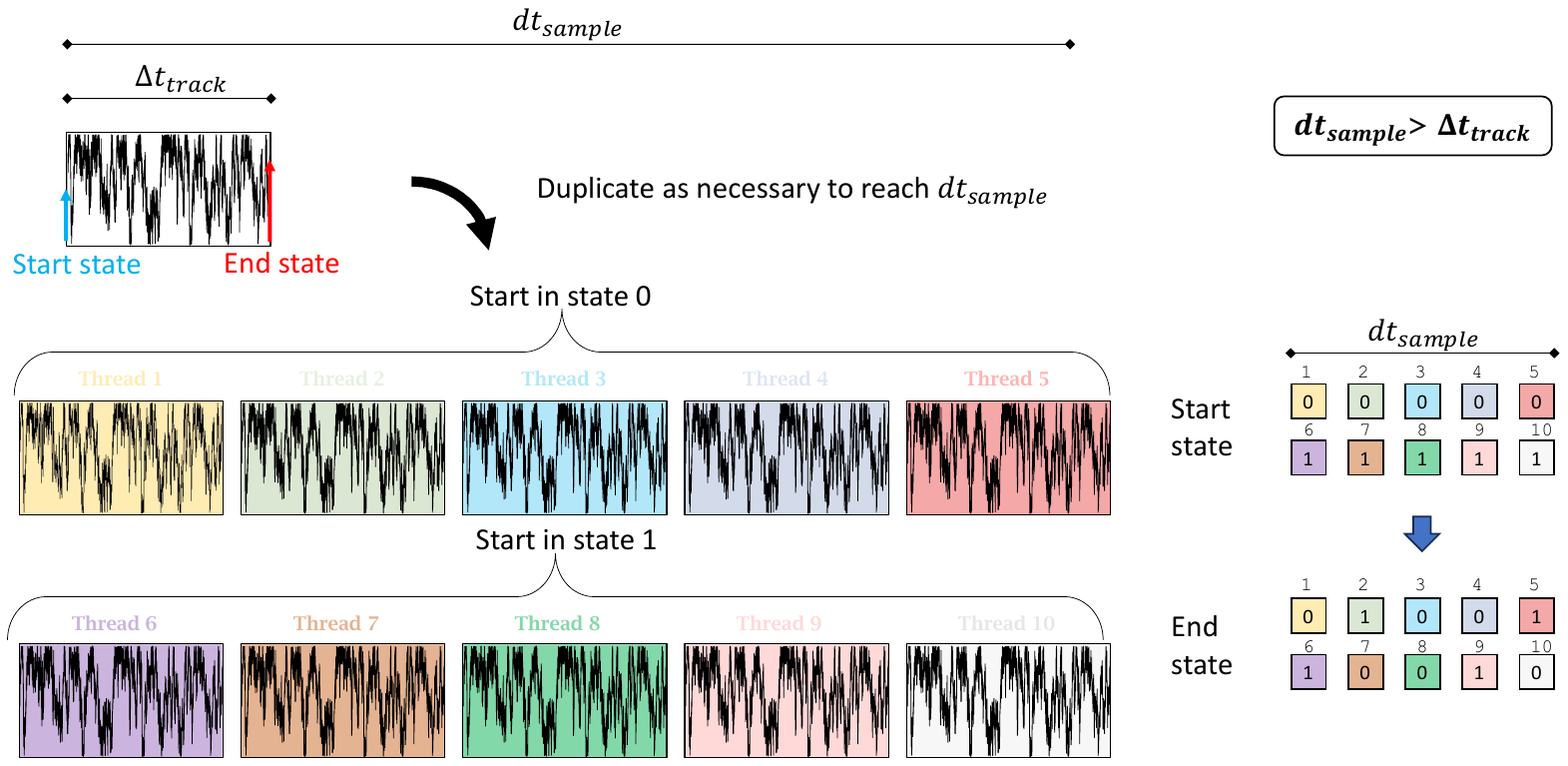}
\caption{\textbf{Parallel algorithm.} When the time to simulate is longer than the available MD data, the MD track is copied until it is long enough to cover the desired duration $dt_{sample}$. Each thread is assigned all the available MD data, and will carry on the simulation starting on 0 for half the threads and 1 for the other half. The end state of each threads is saved for further reconstruction of the state history of the molecule across the whole track.}\label{slow_CV_1}
\end{figure}

\begin{figure}[H]%
\centering
\includegraphics[width=1\textwidth,clip, trim = 0cm 6cm 6cm 6cm ]{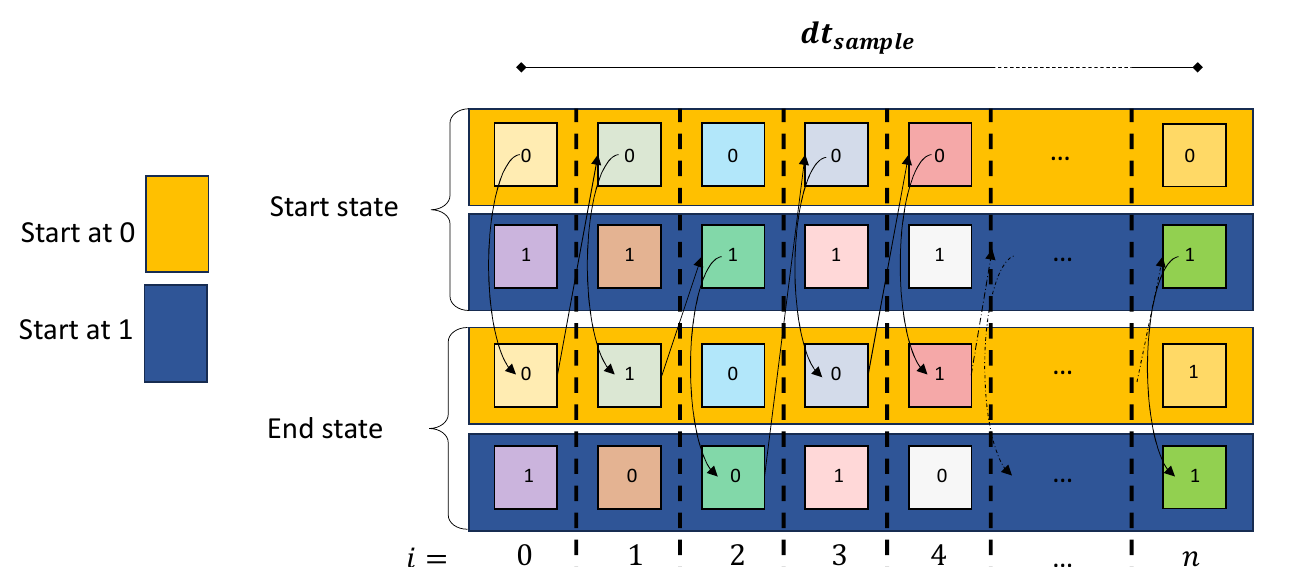}
\caption{\textbf{Parallel algorithm's subroutine for state history reconstruction.} With half the threads starting at 0 and the other half at 1, the actual path of the molecule is recreated by choosing one of the two threads at each step $i$ (with $0\leq i \leq n$) spaced by $\Delta t_{track}$, matching the end state of a thread at $i$ to the start state of a thread at $i+1$.}\label{slow_CV_2}
\end{figure}

Generating MD track with picoseconds resolution over minutes (for longest experiments) is out of the reach of today's computers, and for experiments where $dt_{sample} > \Delta t_{track}$, we use a different ``parallel universe'' algorithm. We here take advantage of the fact that our simulation essentially looks at the state of a molecule that can only have 2 states: 0 or 1 (for oxidation and reduction). The idea consists in slicing the overall $dt_{sample}$ in $n$ slices the size of $\Delta t_{track}$ until $n\times\Delta t_{track} \geq dt_{sample}$. This creates an artificial MD track by just looping over the available MD data, that are again supposed representative of the spatial variations of the molecule. Each thread is assigned a portion of duration $\Delta t_{track}$ (Fig. \ref{slow_CV_1}), with half of them initialized in the state 0, and the other half in the state 1. This step is critical, this is the step that allows for the massive parallelization of the calculations of electrochemical experiments based on molecular dynamics. We now have two ``parallel universes'' where the molecule is always at the state 0 (yellow on Fig. \ref{slow_CV_2}) or 1 (blue) at the beginning of a thread calculation. Looking at the simulation of the first slice, we can use the actual outcome of the experiment, giving the starting state of the next slice. An additional sub-routine is created to recursively reconstruct the actual history of the states of the molecule, allowing to retrieve the current and other fluctuation moments in the same manner than for the chronological algorithm when $dt_{sample} \leq \Delta t_{track}$. In practice, a few thousands of molecules are simulated in parallel this way. A direct consequence is that the number of molecules simulated is in general much higher in the chronological than in the parallel universe algorithm.
 
\section{Simulating electrochemical experiments}
\subsection{Estimation of the current}
\begin{figure}[H]%
\centering
\includegraphics[width=1\textwidth, clip, trim = 0cm 0cm 0cm 0cm]{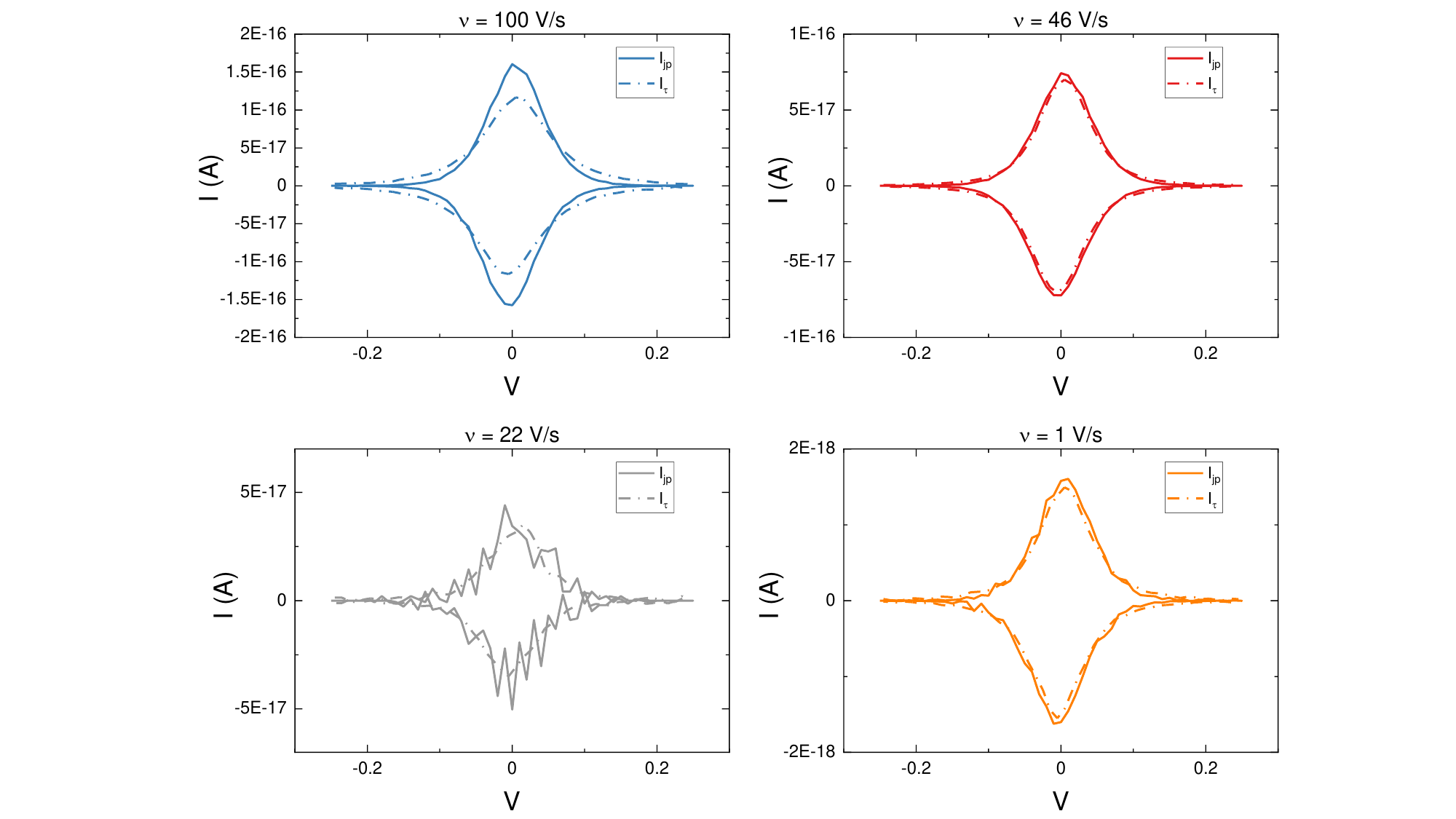}
\caption{\textbf{QBIOL CVs at different rates: comparison of $I_{jp}$ and $I_{\tau}$.} CVs of a single ferrocene particle in a 60 nm gap at decreasing sweep rates. (a) At 100 V/s, the chronological algorithm is used. We see a clear discrepancy between $I_{jp}$ and $I_{\tau}$, due to the small number of event per molecule. (b) At 46 V/s, still using the same algorithm, both currents tend to agree as the number of event per molecule increases. (c) At 22 V/s now using the parallel universe algorithm (because $dt_{sample} > \Delta t_{track}$), a lot less molecules are simulated, which makes $I_{jp}$ very noisy. However, though the number of molecule decreased, the number of events per molecule kept increasing (as longer times are simulated) making $I_{\tau}$ more reliable. (d) At 1 V/s, reliability remain in favor of $I_{\tau}$, though $I_{jp}$ becomes more and more accurate.}\label{Ijp_Itau}
\end{figure}

The current in a one-electrode experiment, such as in CV, can be obtained using different approaches:
\begin{align}
I_{jp} = \frac{1}{N_{mol}}\sum_{i=1}^{N_{mol}}{n_{jp}^i \frac{q\nu}{dV} \label{I_jumps}}\\
I_{\tau} = q\frac{dP_{1}}{dV}\frac{dV}{dt} \label{I_tau}\\
\end{align}
with $n_{jp}^i$ the net number of electron transferred (``jumps'') for the $i^{th}$ molecule, $N_{mol}$ the total number of molecules simulated and $P_{1}=\frac{\tau_{01}}{\tau_{01}+\tau_{10}}$ the probability to be in the state 1.  Fig. \ref{Ijp_Itau} shows a series of CVs taken at decreasing sweep rates around the transition from the chronological algorithm to the parallel universe algorithm, illustrating the reliability advantage of $I_{jp}$ at high sweep rates and of $I_{\tau}$ at low sweep rates.

As shown Fig. \ref{Ijp_Itau}, the reliability of $I_{jp}$ is based on $N_{mol}$ and the net number of jumps \textit{on all threads} and allows to recover a current even at very short time scale using large $N_{mol}$. The reliability of $I_{\tau}$ on the other hand depends on the number of transitions \textit{within each simulated molecule} (i.e. thread) as each time one of these occurs, the accuracy of the estimation of $\tau_{01}$ and $\tau_{10}$ increases. In practice, for short $dt_{sample}$ (i.e. fast sweep rates for example), the chronological algorithm (see section \ref{algorithms}) is used, where a large number of threads can be allocated to simulate individual molecules and give accurate estimations of $I_{jp}$ while typically, $I_{\tau}$ is inaccurate because of the small number of transitions within each simulated molecules. For long experiments the parallel universe algorithm is used and though it has typically a significantly lower number of threads allocated to simulate independent molecules, decreasing the accuracy of $I_{jp}$, the number of transitions within each simulated molecules increases (as there is more time for them to occur) making $I_{\tau}$ more reliable \cite{kutovyi_noise_2020}. More precisely, we empirically estimated the minimum number of transitions within a single simulated molecule for $I_{\tau}$ to be trusted at 2. For CV it means that the values of $I_{\tau}$ near $V^0$ will become reliable before the extrema, as there are the most transitions around this potential.\\
The current in a two-electrode experiment is typically simpler to estimate, as it is stationary, either limited by mass transfer \cite{anderson_thin-layer_1965} or electron transfer rate \cite{madrid_ballistic_2023}. It is measured by counting the electrons at one of the electrodes, using then Eq. \ref{I_jumps} to recover the current.\\
\subsection{Chronoamperometry}\label{chronoamperometry_method}
\begin{figure*}[h]%
\centering
\includegraphics[width=\textwidth,clip, trim = 0cm 4.5cm 0cm 4.5cm ]{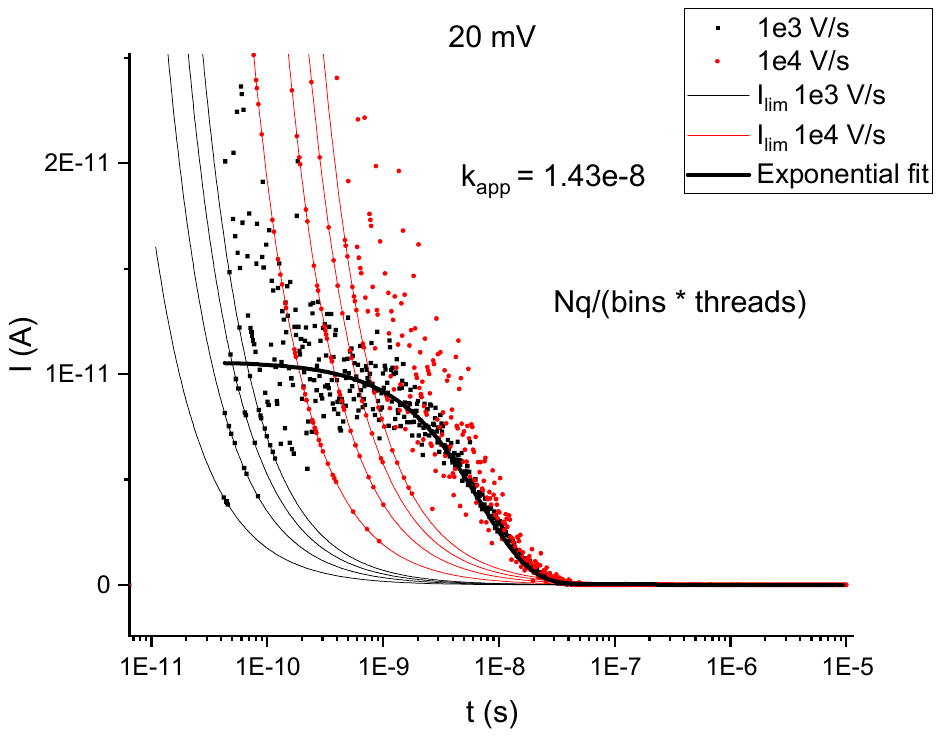}
\caption{\textbf{Chronoamperometry simulation using logarithmically spaced bins with the limits marked for an experiment done at two distinct sweep rates on a dt20ds DNA.} The same current is recovered as it is limited by the DNA movement, but the limit of current detection change due to the different acquisition durations. The limits are found using the Table \ref{Limit_current_QBIOL}, where $I_{lim}= \frac{q}{\Delta tN}$ for $N = 1, 2, 3$ and $4$. Here $N$ represents the number of event from which the current is estimated in this simulation. Simulation parameters: tethered dt20ds-Fc DNA unconfined, E -E$^0$= 0 mV $\rightarrow 20$ mV, $dt_{sample} \approx 1\times 10^{-5}$~s)}
\label{Chronoamp_log}
\end{figure*}

\begin{figure*}[h]%
\centering
\includegraphics[width=\textwidth,clip, trim = 0cm 6cm 0cm 3cm ]{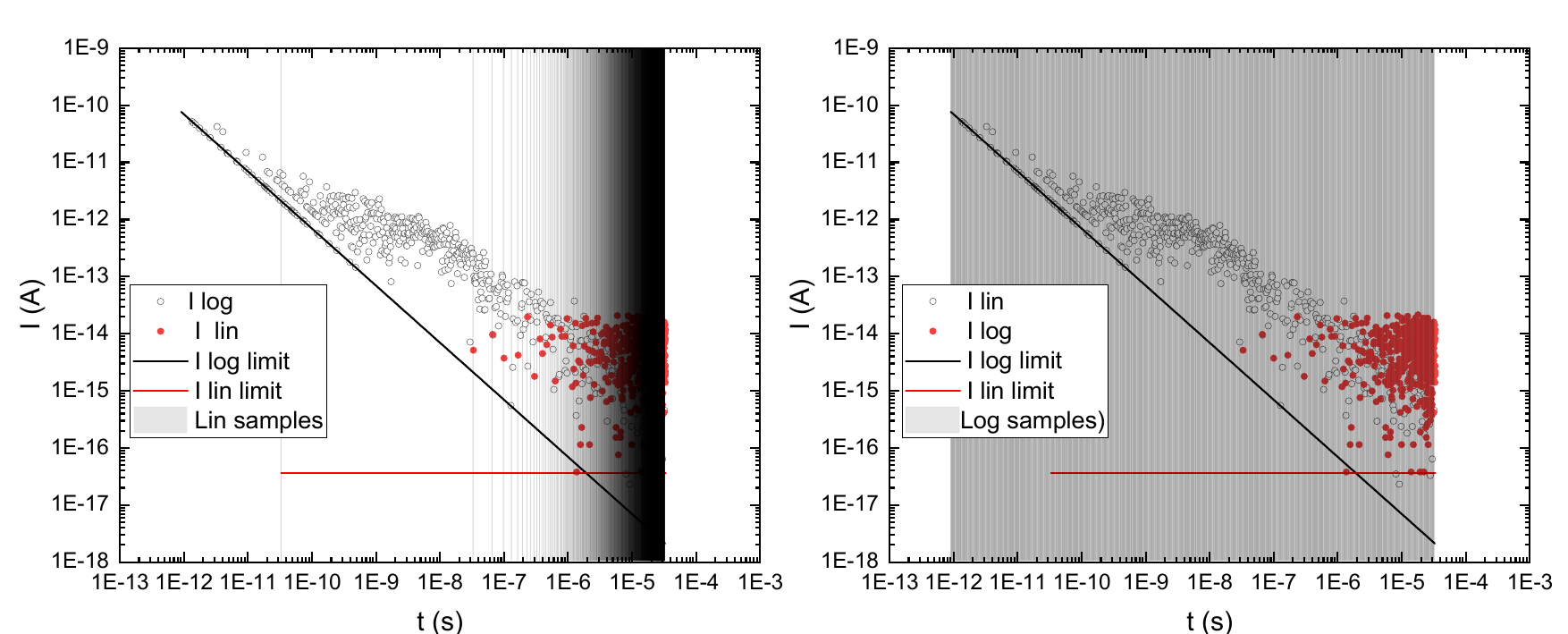}
\caption{\textbf{Chronoamperometry current limits depending on the type of bins.} Chronoamperometry example showing the current traces acquired linearly and logarithmically, with the associated currents limit and samples (left: linear, right logarithmic sampling). Simulation parameters: Fc free particle, 1 nm gap, E -E$^0$= -10 mV $\rightarrow 0$ mV, $dt_{sample} = 3.3\times 10^{-5}$~s)}
\label{Ilim_chronoamp}
\end{figure*}

Chronoamperometry in QBIOL is essentially done in the same way than cyclic voltammetry, except that we record not only the total number of electron jumps but also the individual time at which each jump occurs. Gathering jumps in time bins of width $\tau_{bin}$ spread between the minimum time $dt_{MD}$ and the maximum time of the experiment allows to define a current $i = n_{jp}/\tau_{bin}$ versus time at each voltage. One challenged lies in the timescale at which jumps must be recorded. The smallest time defined in QBIOL is $dt_{MD} = 9.09\times 10^{-13}$~s as a minimum boundary, up until seconds of measurement depending on the experiment. Linearly spaced bins with $\tau_{bin} = dt_{MD}$ would result in $\approx 10^{12}$ elements arrays for each time trace, impractical at best. A reasonable number of bins such as 1000 bins would on the other hand miss almost entirely the exponential decay contribution of the timetrace for most experiments, happening at short time scales compared to one second. A compromise was found using logarithmically spaced bins (Fig. \ref{Chronoamp_log}), with $\tau_{bin}$ growing exponentially larger for longer timescale. This allows to resolve both short and long current contributions with good accuracy.\\
Fig. \ref{Ilim_chronoamp} shows an example showing both linearly and logarithmically spaced bins with the corresponding sampling, currents and current limits. Due to the different nature of the binning, the current limits also differ for both methods, but remain defined as $q/\tau_{bin}$

\subsection{Arbitrary waveform} \label{Arb_section}
Arbitrary voltage waveforms can be used in QBIOL, functioning as a real experimental data acquisition and numerical waveform generator would. The applied voltage is sampled at a given sampling rate, not necessarily matching the output current sampling rate. In such case an interpolation is realized during the experiment to apply continuously the voltage variations. As such, it is possible to simulate any type of signal, such as here a 2 GHz excitation on top of a DC linear sweep at around 1600 V/s, as used to illustrate de EF-SMM simulation in the following figures.

\begin{figure}[H]%
\centering
\includegraphics[clip, trim= 4cm 3cm 2.5cm 3cm, width=\textwidth]{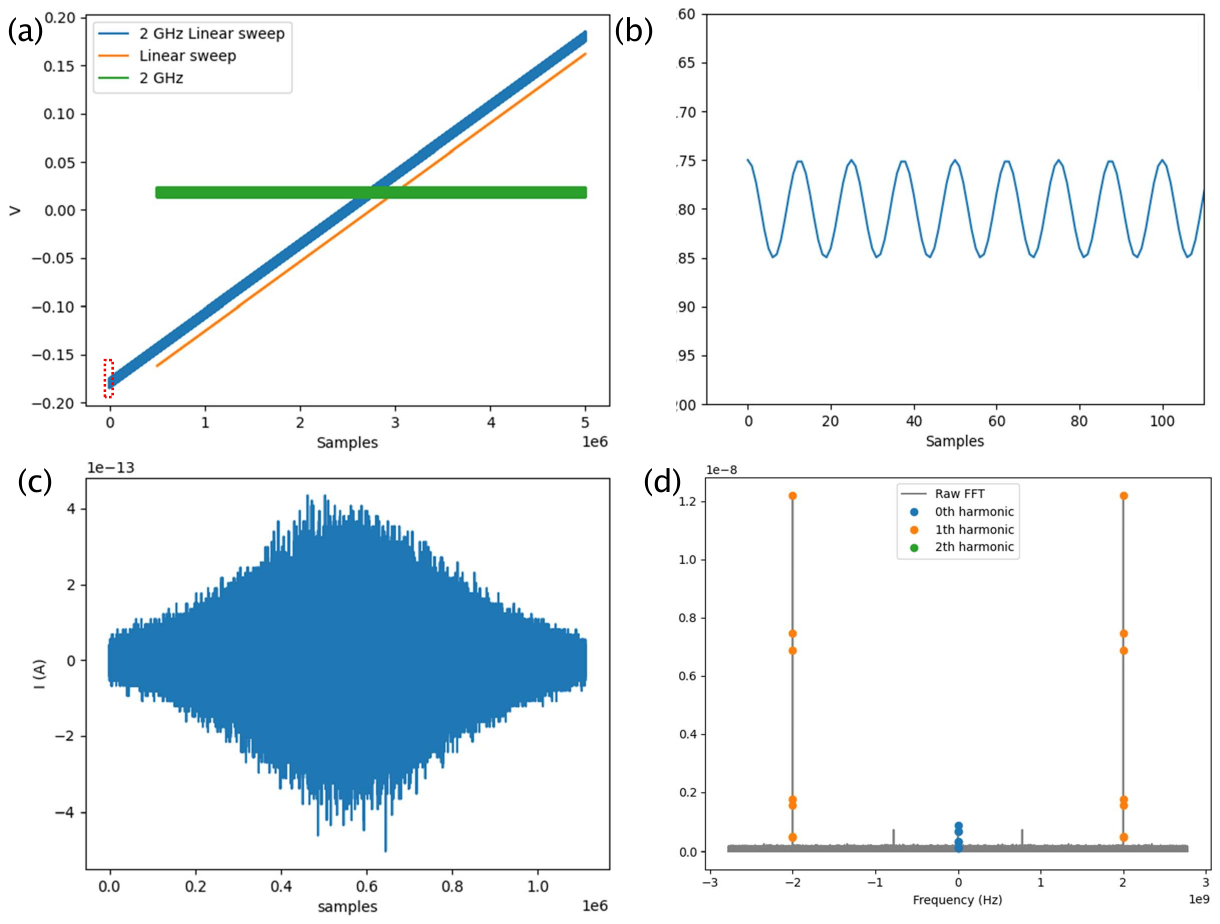}
\caption{\textbf{SMM experiment as implemented in QBIOL.} (a) Voltage signal applied to the electrode with its different components. (2 GHz, 5 mV, 1600 V/s))(b) Magnified voltage as marked by the red dotted rectangle in (a). (c) Resulting current over samples. (d) Fast Fourier Transform of (c). Parameters: Total time: 0.4 ms, time per sample $\approx$ 0.2 ns, output samples = 1111111, voltage samples = 5000000, Marcus-Hush rates.}\label{EF_SMM_QBIOL_1}
\end{figure}

\begin{figure}[H]%
\centering
\includegraphics[clip, trim= 5cm 4cm 4cm 3cm,width=\textwidth]{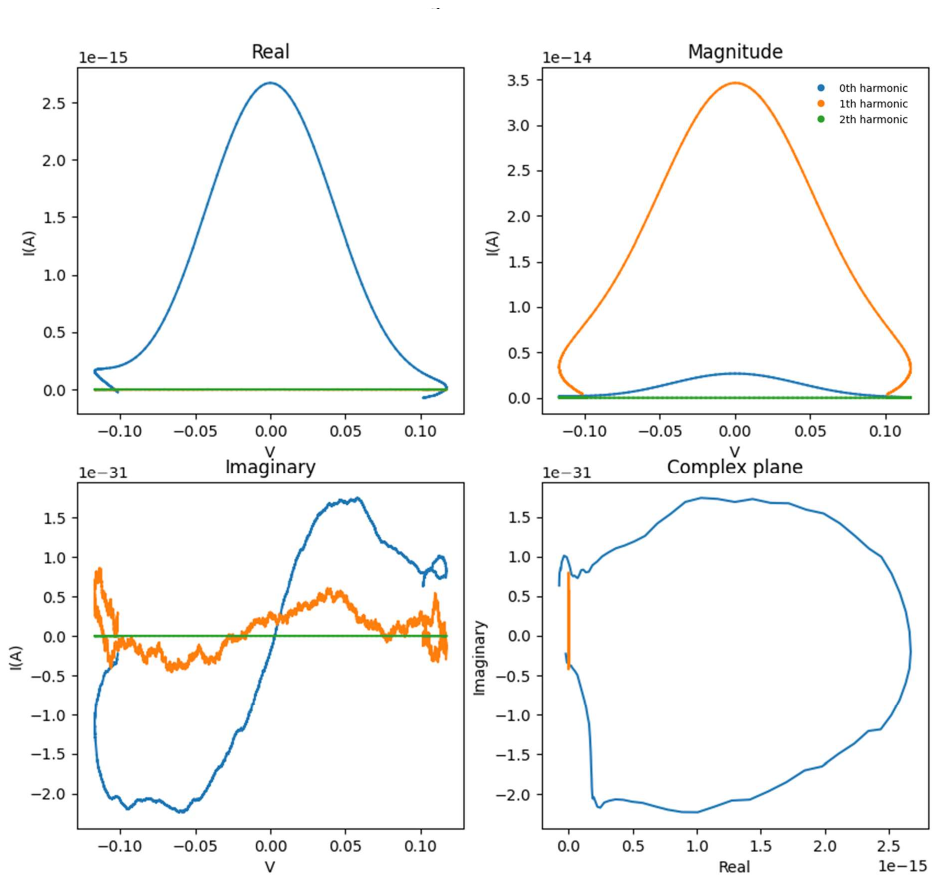}
\caption{\textbf{SMM experiment data treatment}, exhibiting the real and imaginary part of the harmonics of the fundamental at 2 GHz.}\label{EF_SMM_QBIOL_2}
\end{figure}

\begin{figure}[H]%
\centering
\includegraphics[clip, trim= 5cm 6cm 4cm 6cm,width=0.9\textwidth]{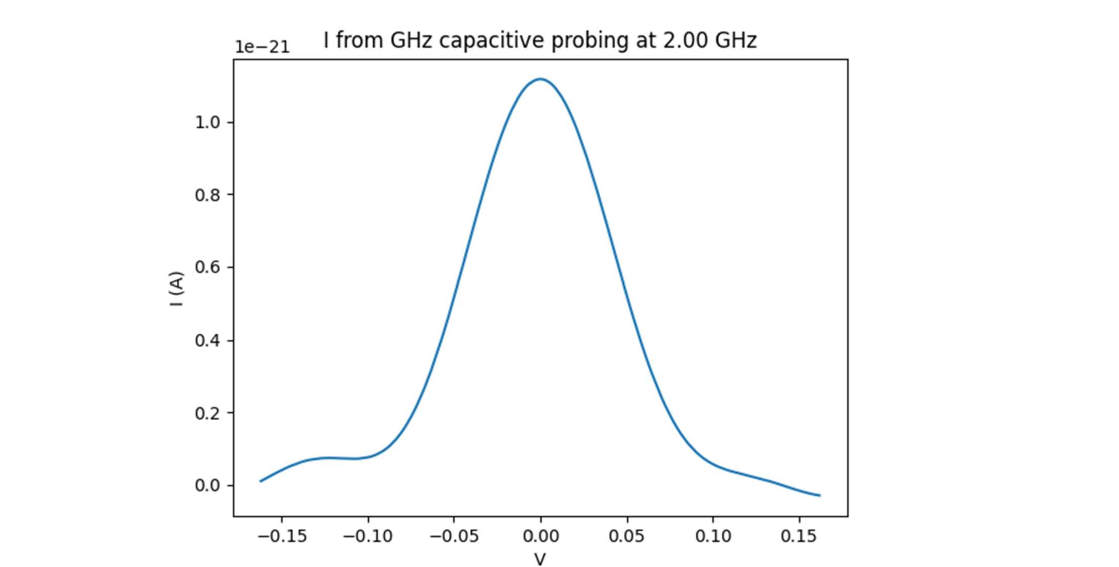}
\caption{\textbf{SMM output current.} Output current as obtained following the procedure described in Eq. 12 in\cite{awadein_nanoscale_2022}: $I = \nu (C_{ac} - G_{ac}/\omega)/2$.}\label{EF_SMM_QBIOL_3}
\end{figure}

\section{Simulation default values}

\begin{table}[h!]
\centering
\begin{tabular}{cp{2cm}p{0.7cm}p{6cm}}
\toprule
 Parameter& Value & Unit & Meaning\\
\toprule
 $k_0$& $4\times 10^7$ & s$^{-1}$& Standard $e^-$ transfer rate BV \\
 $\beta$& $10^{10}$ & m$^{-1}$& Tunneling decay rate\\
 $q$& $1.602\times 10^{-19}$ & C& Elementary charge\\
 $T$& 300 & K& Temperature\\
 $k_B$& $1.38\times 10^{-23}$ & J/K& Boltzmann constant\\
 $E^0$& 0 & V & Standard electrochemical potential\\
 $\alpha$& 0.5 & --& Asymmetry factor BV\\
 $\rho H^2$& $4\times 10^{-6}$ & J & Coupling energy MH\\
 $\hbar$& $1.054\times 10^{-34}$ & J.s& Reduced Planck constant\\
 $\varepsilon_0$& $8.85\times 10^{-12}$ & F/m& Vacuum permittivity\\
 $\varepsilon_{op}$& 1.78 & --& Optical relative permittivity of water\\
 $\varepsilon_{bulk}$& 78.5 & --& Bulk relative permittivity of water\\
 $\varepsilon_{nano}$& 1.8 & --& Nanoscale relative permittivity of water\\
 $a_0$& 3 & $\AA$& Ferrocene radius\\
\botrule
\end{tabular}
\caption{Default values used in QBIOL.}
\label{table_default_values}
\end{table}

\section{Concentration profiles}

\begin{figure}[H]%
\centering
\includegraphics[clip, trim= 4cm 0cm 4cm 0cm, width = 0.6\textwidth]{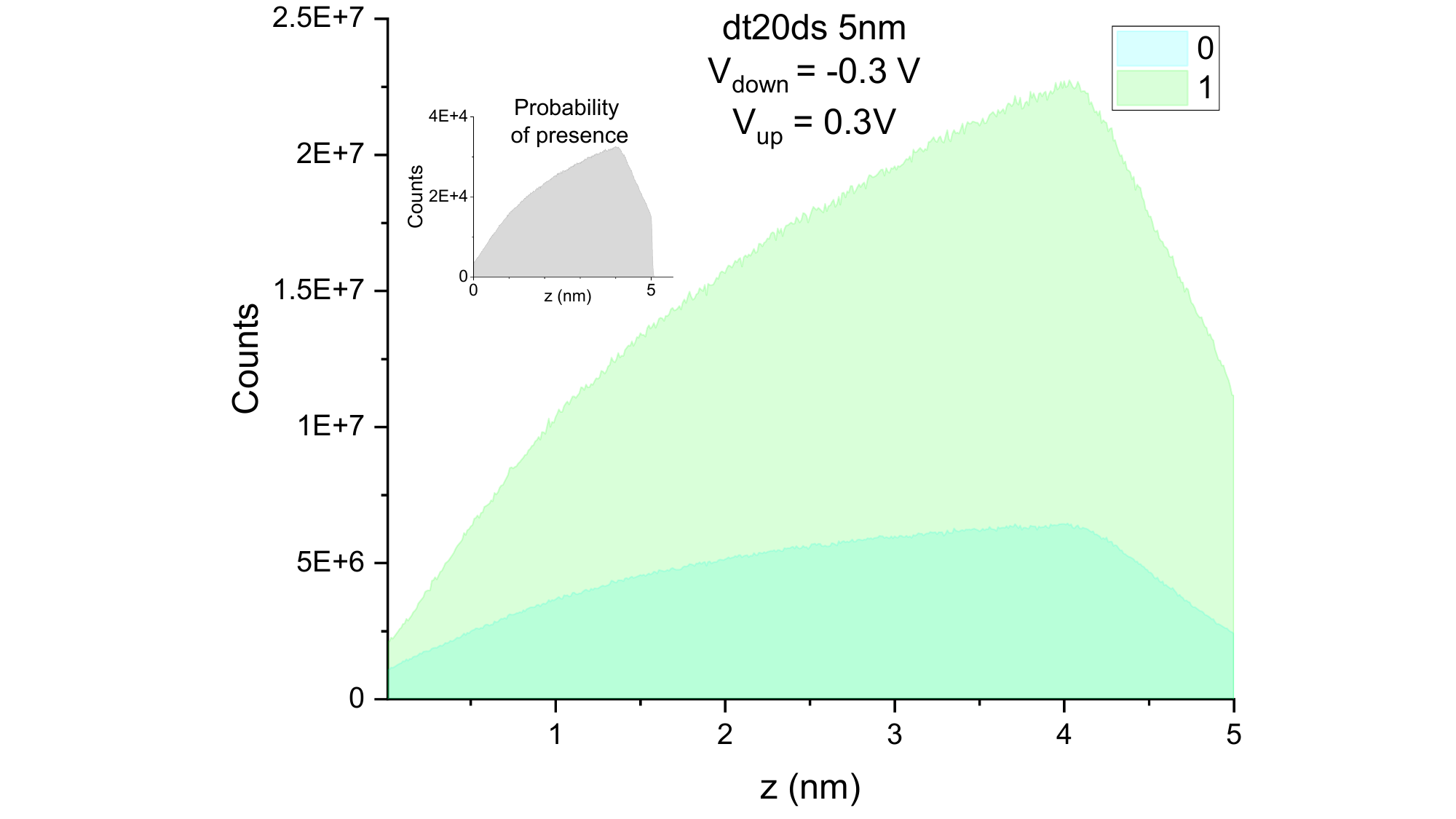}
\caption{\textbf{State distribution of a tethered dt20 DNA double strand} with a Fc head on the 5- end confined in a 5 nm gap. The grey inset shows the total probability of presence of the Fc head of the molecule in the 5 nm gap.}
\label{PoxPred_dt20ds}
\end{figure}

\begin{figure}[H]
\centering
\includegraphics[clip, trim= 1cm 2cm 2cm 4cm, width = \textwidth]{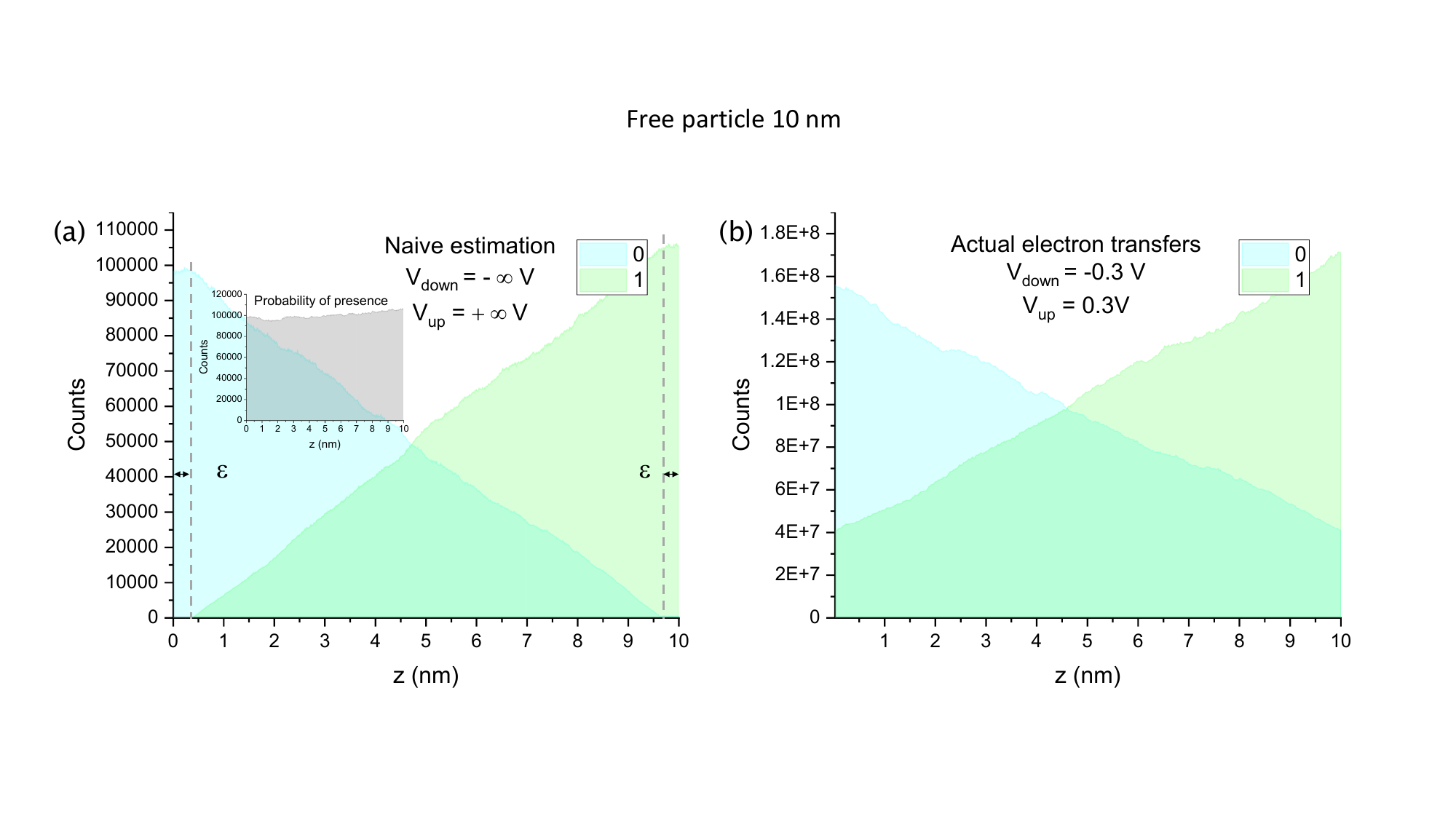}
\caption{\textbf{State distribution of a freely moving Fc molecule} confined in a 10 nm gap. (a) Estimation done considering a threshold on both sides ($\varepsilon = 4 \AA$) where the electron transfer occurs automatically. (b) Estimation done considering actual electron transfer rates on both sides. Grey inset in (a) shows the total probability of presence of the Fc molecule in the 10 nm gap.}
\label{PoxPred_10nm}
\end{figure}

\section{Molecular dynamics (MD)}\label{molecular_dynamics}
QBIOL only needs a position over time to compute a current for a given electroactive molecule. It is thus compatible with any molecular dynamic library able to provide such data for the studied system. In the present configuration, it implements its own random walk generator for freely moving particle, and uses oxDNA \cite{ouldridge_dna_2010, rovigatti_accurate_2014,poppleton_oxdna_2023} for the simulations involving DNA strands.
QBIOL uses MD-generated ``tracks'', files containing a list of time and position of the electroactive molecule, to estimate the probability of electron transfer between the molecule and one (or more) electrode(s). For freely diffusing molecules, QBIOL has its own implementation of random walk diffusion. The number of tracks necessary to correctly model a gap of length $z_{gap}$ increases $\propto z_{gap}^2$, making for now micrometer scales difficult to model in such a way. Other more complex molecules, such as tethered DNA, are simulated with external MD libraries, such as oxDNA (currently used for DNA strands simulations). Unless explicitly mentioned, the minimum time step in QBIOL is $dt_{MD} = 9.09\times 10^{-13}$ s.

\subsection{Random walk and random number generation}
For the simulation of freely moving particle within a nanogap, QBIOL uses a random walk algorithm. The gap is defined in one dimension, between $z = 0$ and $z_{gap}$ with reflective boundaries. At each $dt_{MD}$, a molecule is moved from its previous position of a distance $\Delta z = g\sqrt{2Ddt_{MD}}, g\in \mathcal{N}(0,1)$. The random number $g$ used here and all random number generated for QBIOL are generated using the xoroshiro128+ algorithm on GPU, which guarantees the independence of the number generated as long as a single GPU thread generates less than $2^{64}\approx 10^{19}$ requests (which is by far our case with at worst $\approx 10^9$ calls per GPU thread)\cite{marsaglia_xorshift_2003,vigna_further_2017,blackman_scrambled_2021}.

QBIOL simulates one molecule at a time, and repeat the experiment $n_{threads}$ times, with $n_{threads}$ the total number of GPU threads used to simulate molecules (see Methods for details on the use of GPU threads). This is equivalent to have $n_{threads}$ molecules in the same space with no steric interaction with each other, i.e. two molecules could in principle be at the same place at the same time. Concentrations for a random walk experiments in a 1D confined space can be estimated from the effective probability of presence or directly from $C = \frac{1000}{N_A z_{gap}2Dt}$~M with $t$ the simulated MD time and $N_A$ Avogadro's number. In general QBIOL simulations run with concentrations of electroactive particle on the order of 1 mM.

\subsection{SAM}
\begin{figure}[H]
\centering
\includegraphics[clip, trim= 1.5cm 0cm 0cm 0cm, width = \textwidth]{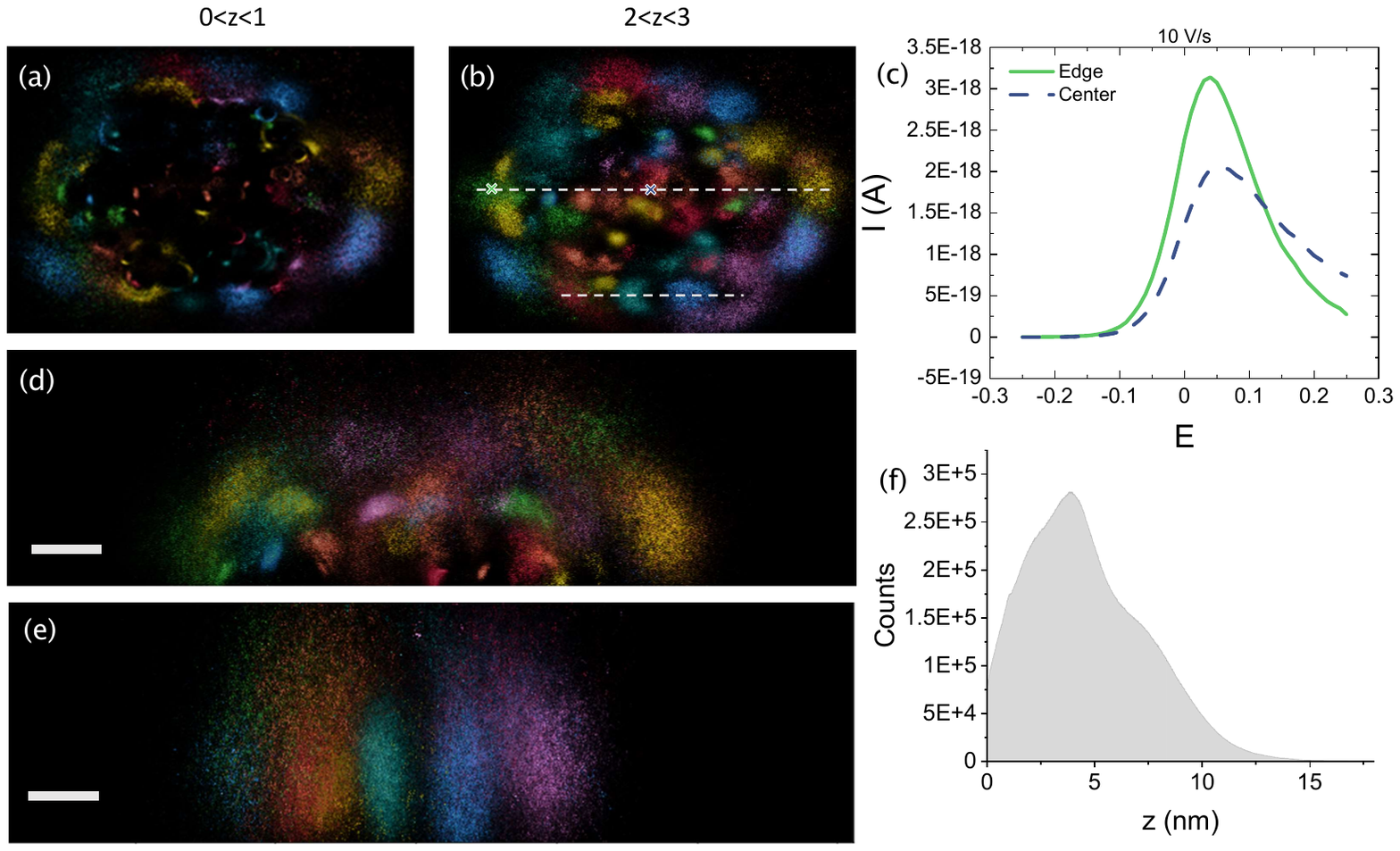}
\caption{\textbf{QBIOL SYL3C SAM simulation with 61 strands.} Top view probability of presence of the extremity of the SYL3C DNA strand for position $z$ within (a) $0<z<1$ nm and (b) $2<z<3$ nm. (c) CVs obtained at the center and the side of the SAM (corresponding positions $\times$ in (b)) (CV parameters: $\nu = 10$ V/s, 10 mV per step, MH rates, $\rho H^2 = 4 \times 10^{-6}\times 10^{-4}$ ). Side view of the same probability along a profile going through (d) the center and (e) the edge of the SAM, as indicated by dashed line in (b). (f) Aggregated probability of presence along the $z$ axis of the extremity of SYL3C for all molecules within the SAM.\\
Intermolecular distance = 5 nm.}
\label{SAM_fig}
\end{figure}

Self-assembled monolayers (SAMs) of  densely packed, laterally interacting, redox-DNA molecules can also be simulated with QBIOL, with the possibility to follow individually each molecules separately and do CVs or other electrochemical experiments on them. The intermolecular electron transfer feature is not the object of the present work and will be presented in a future paper but the effect of lateral interactions between the immobilized DNA molecules on the electron transfer with the electrode is simulated. Fig. \ref{SAM_fig} shows a SYL3C SAM with 61 unconfined molecules, with 5 nm between each molecule. The MD was carried out with oxDNA in a few days on a gaming laptop, i.e. large SAMs with thousands of base pairs are available provided it runs on the appropriate computing platform. However, we believe that a few hundreds of molecules will be enough to avoid edge effects and have representative data for the behavior of macroscopic SAMs. Here the CVs are carried out with $10^4$ reduction of the default electron transfer rates, corresponding to a distancing from the surface of roughly 1 nm. We observe here already a difference between the molecules in the center and at the edge of the SAM. The molecule in the center is more affected by its neighbors, decreasing the probability of presence at the interface.

%

\section{Noise and limit of detection}
The fundamental limit of detection for current is simply of one elementary charge per unit time. In QBIOL, this translates into different actual limits depending on the number of threads used to obtain statistics and the type of measurement.

\begin{table*}[h!]
\centering
\begin{tabular}{|c|p{5cm}|p{5cm}|}
\hline
 \textbf{Exp. type}& \textbf{Resolution} & \textbf{Note} \\
\hline
\textbf{One el. CV} & $\frac{q}{\tau N}$ & Where $\tau$ is the interval between each $dV$ step. \\
\hline
\textbf{Two el. CV} & $\frac{q}{\tau}\sqrt{FF\frac{targetmean}{N}}$ or $q\sqrt{FF\frac{targetmean}{\tau \Delta t_{track}}}$ & Where $\frac{1}{\tau}$ is the sampling frequency. \\
\hline
\textbf{Chronoamp.} & $\frac{q}{\Delta t N}$ & $\Delta t$ is the size of the bin over which $e^-$ jumps are counted.   \\
\hline
\end{tabular}
\caption{Limit current as they are calculated in QBIOL depending on the simulation type.}
\label{Limit_current_QBIOL}
\end{table*}

\section{Cyclic voltammetry response of a Thin Layer Cell} \label{TCL_section}

\begin{figure}[h]
\centering
\includegraphics[clip, trim= 9cm 9cm 10cm 7cm, width = 0.5\textwidth]{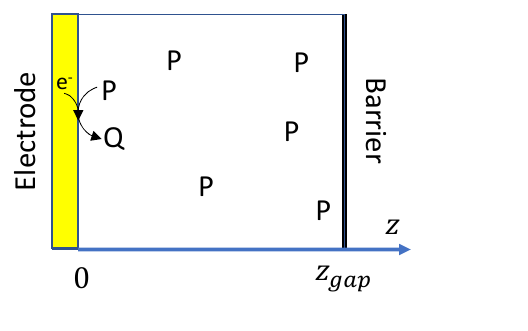}
\caption{\textbf{Depiction of a thin-layer cell (TLC)} delimited by an electrode surface on the left side and by insulating walls on the three other sides. A reference and a counter electrode are also present. A solution of electrolyte containing a redox species (here in its reduced form, $P$) is introduced in the cell. Biasing the electrode at an anodic enough potential (vs. the reference electrode) triggers oxidation of $P$ to its oxidized form $Q$ at the electrode, generating the TLC current. The width of the TLC is $z_{gap}$. }
\label{TCL_1}
\end{figure}

We consider a planar thin layer, delimited by an electrode of surface $S$ and an inert boundary located at a distance $z_{gap}$ (Figure \ref{TCL_1}). A solution of electrolyte containing a redox species (initially in its reduced form, $P$, at an initial concentration of $C^0_p$) is introduced in the cell. The redox species cannot escape the cell sideways, so the problem is one-dimensional, describable using a single spatial coordinate, $x$, ranging from 0 to $z_{gap}$. It is assumed that a reference and a counter-electrode are also in contact with the electrolyte.

Cyclic voltammetry is used to interrogate the cell. The potential of the electrode, $E -E^0$, is scanned linearly with time, from a value sufficiently cathodic versus the standard potential of the redox species, $E^0$, toward a value sufficiently anodic, and back. 
Hence the following electrochemical reaction occurs at the electrode:
\begin{center}
\ce{$P$ - e$^-$ <=>[$V^0$, $k_s$][] $Q$}
\end{center}
$Q$ is the oxidized form of $P$. The current, $i$, is related to the kinetics of the electron transfer (ET) by:
\[
i = FSk_sf(\xi)\{[P]_0-exp(-\xi)[Q]_0\}
\]

Where $[P]_0$ and $[Q]_0$ are the concentrations of $P$ and $Q$ at the electrode surface, respectively. $k_s$ is the standard heterogeneous electron transfer rate constant (i.e. the ET rate constant at zero overpotential, $E = E^0$). It is related to the analogous standard homogenous rate constant, $k_0$, by $k_s = k_0/\beta$ where $\beta$ is the tunneling constant ($\beta \approx$ 1 \AA$^{-1}$, so that $k_s$ (cm/s) $\approx k_0/10^8$ (s$^{-1}$)). $\xi$ is the dimensionless overpotential given by :  
\[
\xi = \frac{F}{RT}(E-E^0) = \frac{q}{k_BT}(E-E^0)
\]

$f(\xi)$ is a function describing the dependence of the rate of electron transfer with overpotential, its functional form depends on the electron transfer kinetic model. For the Butler-Volmer (BV) model: $f(\xi)=e^{\alpha \xi}$ , with $\alpha$ the transfer coefficient (typically of 0.5 \citep{saveant_elements_2006}). For Marcus-Hush-Levisch-Chidsey (MHLC) model \citep{marcus_theory_1965,chidsey_free_1991, saveant_effect_2002}:
\[
f(\xi) = \frac{\int^{+\infty}_{-\infty} \frac{e^{-\frac{(\lambda^* -\xi -x)^2}{4\lambda^*}}}{1+e^x}dx}{\int^{+\infty}_{-\infty} \frac{e^{-\frac{(\lambda^* -x)^2}{4\lambda^*}}}{1+e^x}dx}
\]
With $\lambda^* = \lambda\frac{F}{RT}$ (in eV) being the reorganization energy of the redox species. By virtue of mass conservation and assuming that $P$ and $Q$ have the same diffusion coefficient, $D$, one can write: 
\[
[P]_0 + [Q]_0 = C^0_P
\]
The problem can conveniently be made dimensionless by introducing the following variables and parameters (see Table \ref{dimensionless}). 
\renewcommand{\arraystretch}{1.5}
\begin{table}[ht]
	\centering
		\begin{tabular}{p{0.20\linewidth}  p{0.65\linewidth}}
		\toprule
			\textbf{Dimensionless}		  												& \multirow{2}*{\textbf{Comment}} \\ 
			\textbf{variable}																		&														    \\\midrule
			$p = [P]/ C^0_P$																		& \multirow{2}*{Concentrations are normalized by the initial concentration, $C^0_P$}\\  \cmidrule{1-1}
			
			$q = [Q]/ C^0_P$																		& \\\hline
			$y = z\sqrt{\frac{\nu F}{DRT}}$											& The spatial coordinate $z$ (Figure \ref{TCL_1}) is normalized by the thickness of the transient diffusion layer at the electrode resulting from the potential ramp $\sqrt{\frac{DRT}{\nu F}}$\\\hline
			$\mu = z_{gap}\sqrt{\frac{\nu F}{DRT}}$							& Compares the TLC thickness $z_{gap}$ to that of the transient diffusion layer\\\hline
			$\tau = \frac{t\nu F}{RT}$													& Compares the time to the CV observation time, $\frac{RT}{\nu F}$\\\hline
			$\Lambda = k_s\sqrt{\frac{RT}{\nu DF}}$							& Compares the rate of electron transfer at the electrode to the diffusion rate\\\hline
			$\Psi = \frac{i}{FSC^0_P \sqrt{\frac{\nu DF}{RT}}}$	& Dimensionless current \\\hline
			$\xi =  \frac{q}{k_BT}(E-E^0)$											& Dimensionless potential\\
			\botrule
		\end{tabular}
	\caption{Dimensionless variables and their relationship with usual variables.}
	\label{dimensionless}
\end{table}

The mass conservation equation becomes:
\[
p_0 + q_0 = 1
\]

The current $\Psi$ is related to the dimensionless flux of $P$ at the electrode surface $(\delta P/\delta y_0)$, and to the ET rate via:
\[
\Psi = \frac{\delta  P}{\delta y_0} = \Lambda f(\xi)\left[p_0(1+e^{-\xi})-e^{-\xi}\right]
\]
The concentration profile of $P$ is obtained by solving the $2^{nd}$ Fick's law, describing diffusion in the cell:
\[
\frac{\delta p}{\delta \tau} = \frac{\delta^2 p}{\delta y^2}
\]
with the following initial and boundary conditions:\\
for $\tau = 0$ and $0 \leq y \leq \mu$:      $p=1$;\\
for $\tau \geq 0$ and $y = 0$:               $\frac{\delta p}{\delta y}|_0 = \Lambda f(\xi)\left[p_0(1+e^{-\xi})-e^{-\xi}\right]$\\
for $\tau \geq 0$ and $y = \mu$:             $\frac{\delta p}{\delta y} = 0$.\\

The problem was solved numerically using a home-written FEM solver program which yielded dimensionless CVs, $\Psi$ vs $\xi$, for any value of the $\Lambda$ and $\mu$ parameters, both for BV and MHLC ET kinetics. It appeared that the CV behavior is largely controlled by only two global parameters:
\begin{enumerate}
\item $\mu \Lambda = k_s \frac{z_{gap}}{D}$, which compares the rate of electron transfer to the diffusion rate throughout the TLC
\item $\mu$ which as stated before compares the TLC thickness $z_{gap}$ to that of the diffusion layer.
\end{enumerate}
The parameters $\alpha$ or $\lambda$ have also to be considered, depending on whether BV or MHLC electron transfer kinetics are considered, respectively.

\subsubsection{Limiting cases scenarios}
\begin{figure}[h]
\centering
\includegraphics[clip, trim= 5cm 0cm 4cm 0cm, width = 0.7\textwidth]{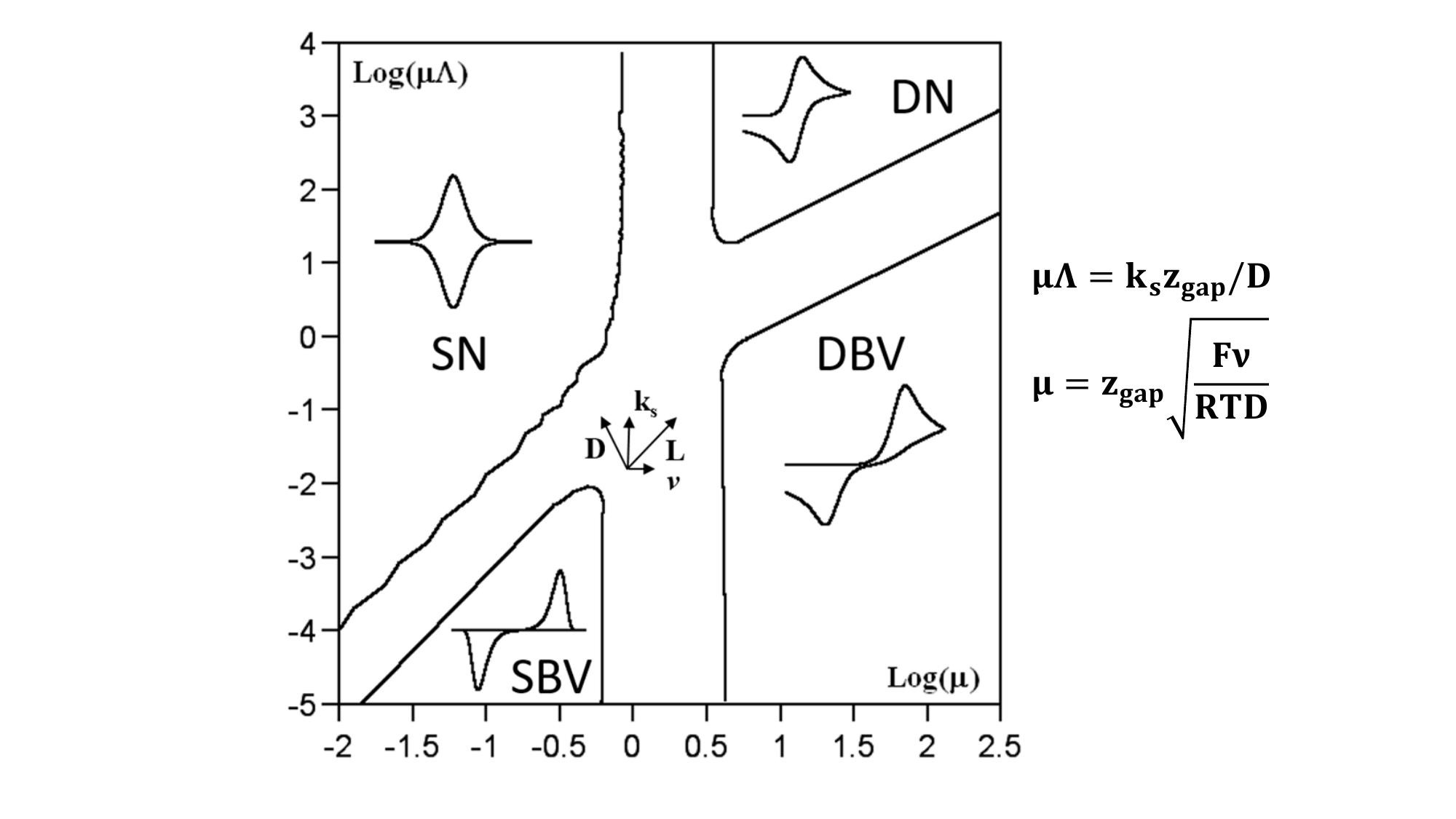}
\caption{\textbf{Cyclic voltammetry interrogation of a thin layer cell (TLC).} Kinetic zone diagram showing the limiting CV responses, reachable for extreme values of the two parameters controlling the system: $\mu$ and $\mu \Lambda$. Butler-Volmer electron transfer kinetics ($\alpha = 0.5$) is assumed. The zones are defined by the peak current being within 10\% (or better) of the analytical values corresponding to each limiting case depicted. The compass shows how altering the values of the parameters $D$, $k_s$, $z_{gap}$ or $\nu$ can modify the CV response of the TLC.}
\label{kinetic_CVs}
\end{figure}

Several important limiting situations, associated with distinct CV shapes, can be identified based on limiting values of the $\mu$ and $\mu \Lambda$ parameters, as shown in the kinetic zone diagram presented in Figure \ref{kinetic_CVs} (calculated for a BV ET model and $\alpha= 0.5$).

\begin{itemize}
\item \textbf{$\mu\rightarrow 0$}: In this case, \emph{always reached at low enough scan rates}, ample time is given to the redox species to diffuse through the cell and reach the electrode so that the signal is identical to that of a surface-confined (paradoxically diffusion-less) species, as described by Laviron\citep{laviron_general_1979}. The intensity of the CV peak current, \emph{$i_p$, is then proportional to $\nu$, i.e. the $i_p/\nu$ ratio is constant}.
 Two subcases ensue, depending on how fast is the electron transfer:\\
	\begin{itemize}
	\item[\textopenbullet] \textbf{$\mu \Lambda \gg 1$}: the electron transfer is fast compared to diffusion through the cell (note that this condition does not depend on the scan rate). The electron transfer is at equilibrium, said to be Nernstian. The shape of the corresponding CV is shown in the upper left corner of Figure \ref{kinetic_CVs}, in the zone labeled ``SN'' (for Surface-Nerstian). This zone is characterized by a constant value of $\Psi_p/\mu = 0.25$, hence the peak current is $i_p= 0.25F^2 \nu SC^0_p\frac{z_{gap}}{RT}$, the peak separation is zero and the peak potential is equal to $V^0$. Note that the $SC^0_pz_{gap}$ term is the amount of redox molecules in the cell.
	\item[\textopenbullet] \textbf{$\mu \Lambda \ll 1$}: the electron transfer is slow compared to diffusion through the cell (note that this condition does not depend on the scan rate). The exact shape of the CV depends on the electron transfer kinetic model. For the BV model the shape of the corresponding CV is shown in the lower left corner of Figure \ref{kinetic_CVs}, zone labeled ``SBV'' (Surface-BV). This zone is characterized by a constant value of $\Psi_p/\mu = 0.184$, the peak current value is $i_p= 0.184F^2 \nu SC^0_p\frac{z_{gap}}{RT}$, the peak positions and separations depend on the exact value of $\mu \Lambda$. The forward (anodic) peak position is : 
\[
E_{pa} = E^0 + 2.3\frac{RT}{\alpha F}log(\frac{RTk_0}{\alpha \nu F})
\]
	\end{itemize}
The peaks shift away from $E^0$ (and from each other) as $\nu$ is increased, e.g. for $\alpha = 0.5$ the peak separation increases by 240 mV per decade of $\nu$ (at 25°C).\\

\item \textbf{$\mu \rightarrow +\infty$}: In this case, always reached at high enough scan rates, the transient diffusion layer developing in the cell is significantly thinner than the cell width $z_{gap}$. In this case planar semi-infinite-diffusion is at play, the CVs present a distinctive diffusional ``tail'', and their peak current is proportional to $\sqrt{\nu}$, i.e the $i_p/\sqrt{\nu}$ ratio is constant. Two subcases ensue, depending on how fast is the electron transfer compared to diffusion:\\
\begin{itemize}
\item[\textopenbullet] \textbf{$\mu \Lambda \gg 1$}: 	the electron transfer is fast compared to diffusion through the cell, the electron transfer is at equilibrium (i.e. ``Nernstian''), case ``DN'' in Figure \ref{kinetic_CVs}. This zone is characterized by a constant value of $\Psi_p = 0.446$. The peak current is expressed by:  $i_p = FSC^0_p \sqrt{\frac{DF\nu}{RT}}$. The forward and return peaks are located at $E^0 \pm 30$ mV, respectively, so that their separation is $\approx 60$ mV (at 25°C).
\item[\textopenbullet] $\mu \Lambda \ll 1$: 	the electron transfer is slow compared to diffusion through the cell. For BV kinetics ($\alpha = 0.5$), case ``DBV'' in Figure \ref{kinetic_CVs}. In this zone $\Psi_p = 0.351$, so that $i_p=0.351FSC_p^0 \sqrt{\frac{DF\nu}{RT}}$.  The positions of the forward and backward peaks depend on the scan rate $\nu$, the higher the scan rate the larger the peak separation. The forward peak potential is: 
\[
E_p=E^0+2\frac{RT}{F} \left[2.3log(k_s \sqrt{\frac{2RT}{DF\nu}}-0.78\right]
\]

The peaks shift away from $E^0$ (and from each other) as $\nu$ is increased, e.g. for $\alpha = 0.5$ the peak separation increases by 120 mV per decade of $\nu$ (at 25°C).

	\end{itemize}
\end{itemize}

The take away message is that it is the $\mu$ parameter which solely decides of whether the CV displays the characteristic of a ``surface'' or `` diffusive'' signal, regardless of the ET rate.  ET kinetics mainly affects the peak position.

\subsubsection{General case scenarios}
\begin{figure}[H]
\centering
\includegraphics[clip, trim= 0cm 0cm 0cm 0cm, width = 0.9\textwidth]{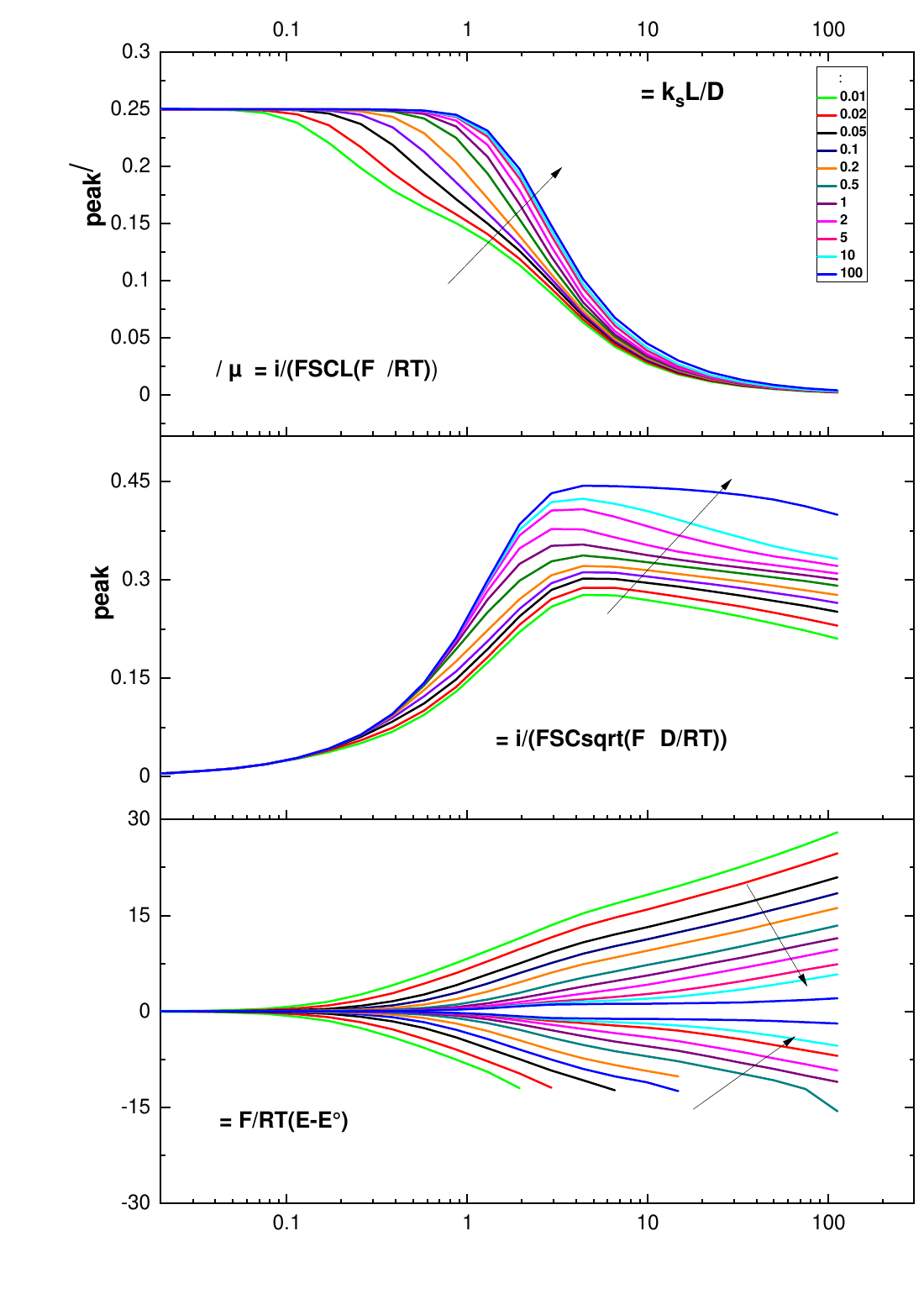}
\caption{\textbf{Cyclic voltammetry interrogation of a Thin Layer Cell.} Variation of the characteristics of the CV as a function of the parameter $\mu$, calculated for various values of the $\mu \Lambda$ parameter (as shown), MHLC ET kinetics ($\lambda = 0.85$ eV). Shown are the variations of : (a) the ratio of the dimensionless anodic peak current, $\Psi_{pa}$, over $\mu$. (b) The dimensionless anodic peak current  $\Psi_{pa}$. (c) The anodic,  $\xi_{pa}$, and cathodic, $\xi_{pc}$ peak overpotentials.}
\label{TLC_3}
\end{figure}

As an illustration of a general case, the variation of the characteristics of the CV wave with $\mu$, calculated for a large panel of $\mu \Lambda$ values and in the case of Marcus kinetics ($\lambda = 0.85$ eV) are shown in Figure \ref{TLC_3}. Since only $\mu$ depends on $\nu$, each curve represents the variation of the CV characteristics with scan rate, for a given $\mu \Lambda= k_sz_{gap}/D$ value, color coded as shown. Scan rate formally increases from left to right.

Part (a) shows the variation of the dimensionless (anodic) peak current value, $\Psi_p$, divided by $\mu$, vs. $\log(\mu)$. This plot is reminiscent of the dimensional $i_p/_\nu$ vs. $\log(\nu)$ graph, typically used as an experimental observable. Part (b) displays the variation of $\Psi_p$ with $\log(\mu)$, which is akin to the experimentally praised $i_p/\sqrt{\nu}$ vs. $\log(\nu)$ plots. Part (c) shows the variation of the dimensionless anodic, $\xi_{pa}$, and cathodic, $\xi_{pc}$, peak potentials vs. $\log(\mu)$ (i.e $\approx$ vs. $\log (\nu)$).
One can notably see from Figure \ref{TLC_3} a that, for $\mu \rightarrow 0$ (i.e. slow scan rates), no matter the $\mu \Lambda$ value, the CV characteristics are those described above for the zone SN in Figure \ref{kinetic_CVs}: a reversible (Nernstian surface CV), featuring a constant $\Psi_p / \mu =$ 0.25 (i.e. $i_p/\nu$ = constant).\\

As $\mu$ is increased one observes in Figure \ref{TLC_3}b that $\Psi_p$ increases and initially tends toward a plateau, equal to 0.446 for high enough $\mu\Lambda$ value, or closer to 0.35 for lower values. This is as predicted in the case of a signal becoming diffusive (zones DN or DBV in Figure \ref{kinetic_CVs}, displaying a constant $i_p/\sqrt{\nu}$ ratio). However, as a result of the ET kinetics now following the MHLC model, the current then slowly decreases at $\nu$ is further raised.\\
Figure \ref{TLC_3}b allows to identify that, for slow enough ET rate (i.e. low $\mu\Lambda$ values) the peaks, and their separation increase with $\mu$ (i.e. $\nu$) roughly following two different slopes, the steepest one corresponding to the lowest µ values. This is reminiscent of the characteristics of zones SBV and DNV in Figure \ref{kinetic_CVs}, where the peak separations increased with v twice as fast for a surface CV than for a diffusive signal, respectively (240 mV vs 120 mV/ decade of $\nu$). Obviously, MHLC kinetics result in the actual variations not being ideally linear, unlike predicted by the BV model.\\

\section{TLC vs QBIOL}

\begin{figure}[H]
\centering
\includegraphics[clip, trim= 2cm 3cm 3cm 3cm, width = \textwidth]{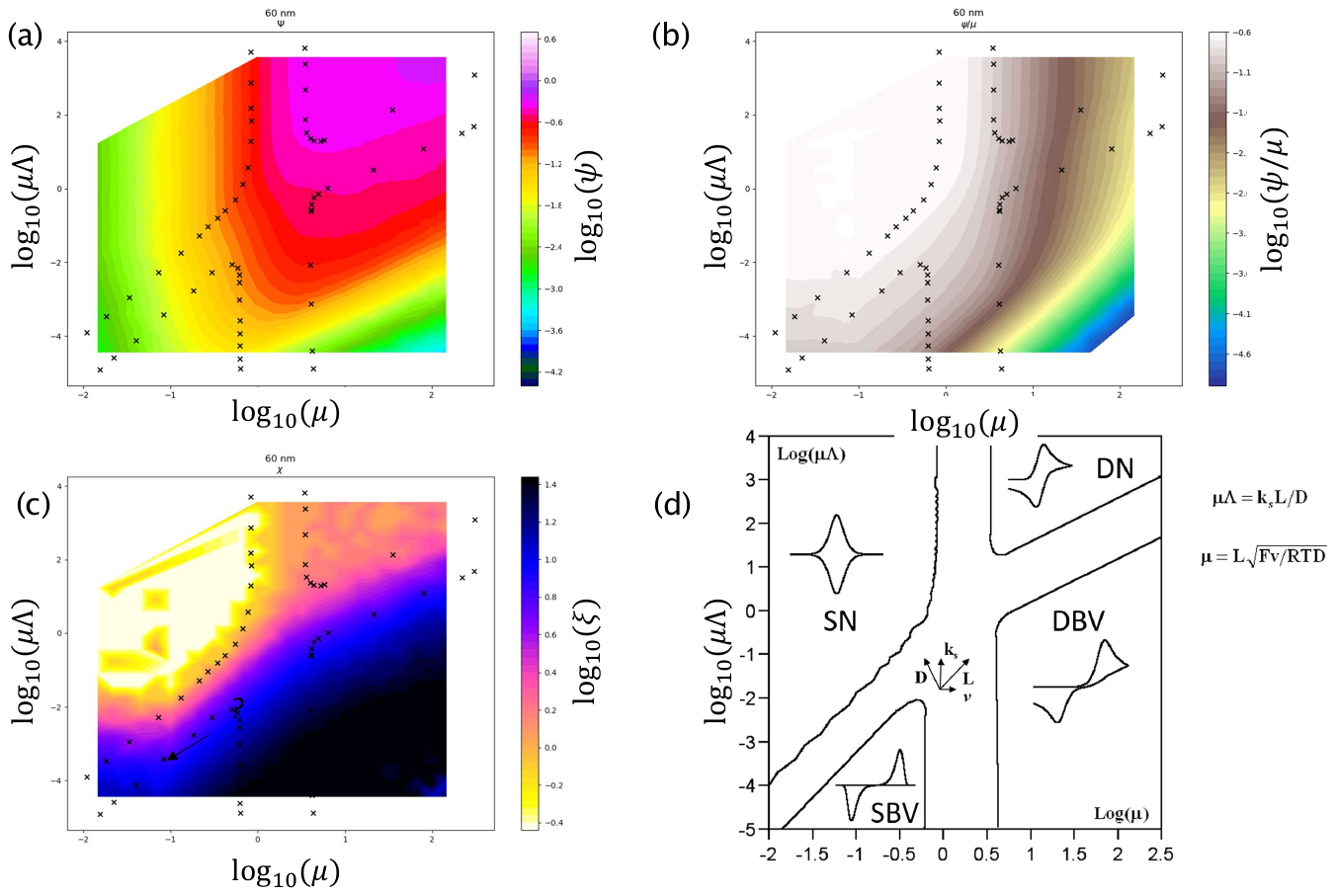}
\caption{\textbf{Comparison TLC model with QBIOL maps.} QBIOL-generated CV oxidation peak's current ($\psi$) (a), current peak to sweeprate ratio ($\psi/\mu$) (b) and voltage ($\chi$) (c) heatmaps. (d) Reprint of the TCL model map (Fig. \ref{kinetic_CVs} for comparison.}
\label{QBIOL_heatmaps_SI}
\end{figure}

\begin{figure}[H]
\centering
\includegraphics[clip, trim= 4cm 2cm 3cm 0cm, width = \textwidth]{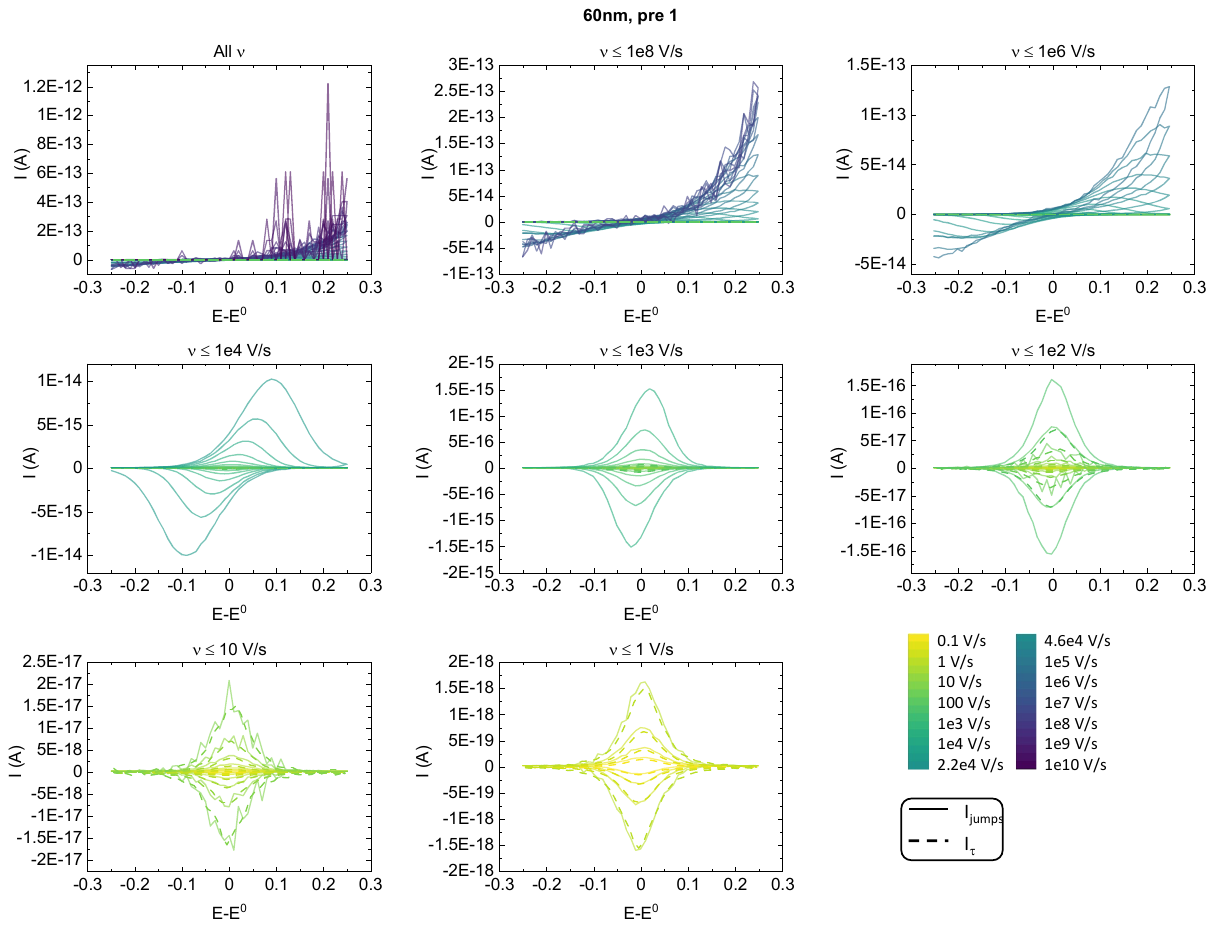}
\caption{\textbf{CVs in QBIOL} QBIOL-generated CV oxidation peak's voltage heatmap.Cyclic voltammograms in a 60 nm gap for Fc free particle as simulated in QBIOL, showing both current calculations  methods $I_{jumps}$ and $I_{t}$ when available.}
\label{CVs_60nm}
\end{figure}

\begin{figure}[H]
\centering
\includegraphics[clip, trim= 3cm 3cm 4cm 3cm, width = \textwidth]{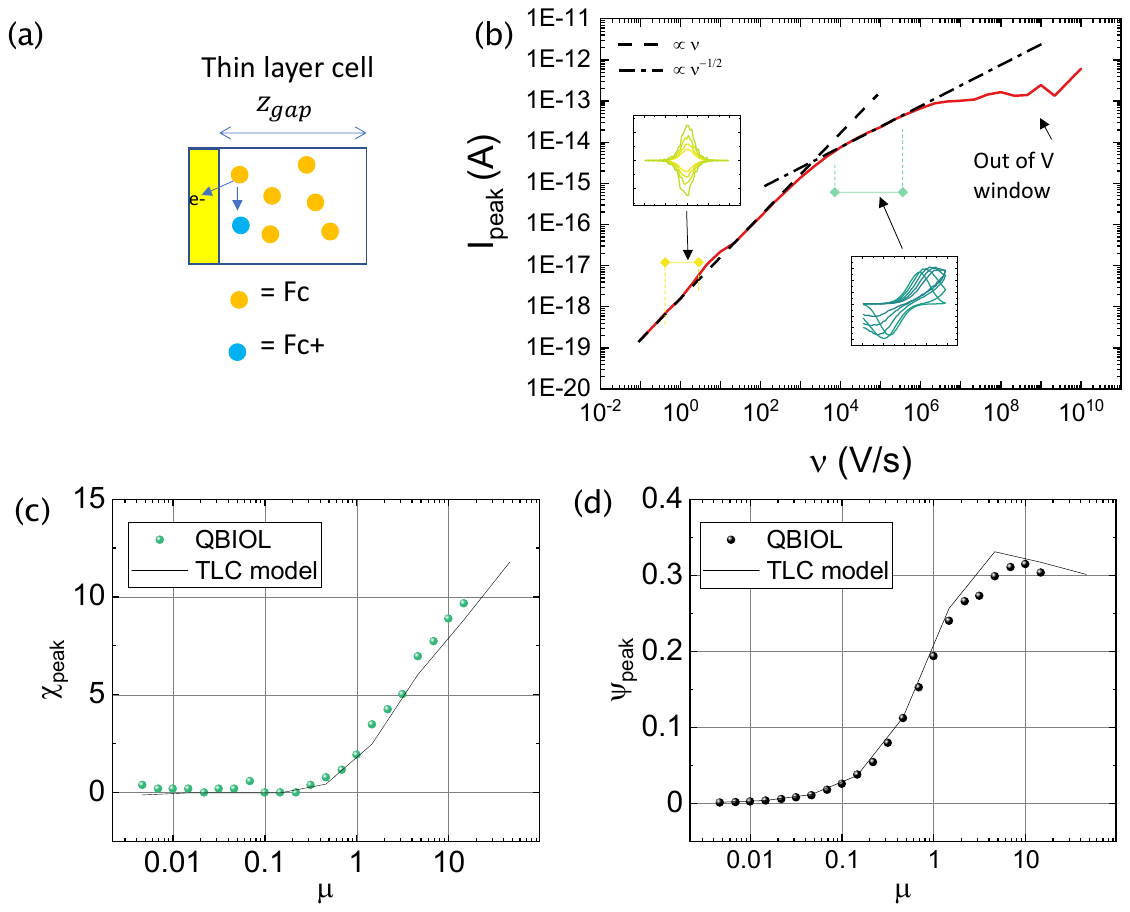}
\caption{\textbf{Comparison TLC model with QBIOL.} (a) Schematics of the model. (b) QBIOL oxidative current peak versus sweeprate $\nu$, with corresponding CVs in inset.  Normalized ocidative peak potential $\chi_{peak}$ (c) and current $\psi_{peak}$ versus normalized sweeprate $\mu$ for both QBIOL and the TLC model. The slight discrepancy between QBIOL and the TLC model observed at high sweeprate can come from uncertainty on the peak as we reach the limit of the potantial window explored, and to the fact that QBIOL takes into account $\lambda(z)$ whereas TLC uses $\lambda=0.85$~eV. QBIOL simulations done in 60 nm gap.}
\label{TCL_vs_QBIOL}
\end{figure}

\subsection{Calculation efficiency and convergence of moments}\label{convergence_moments}
\begin{figure}[H]%
\centering
\includegraphics[width=1.1\textwidth]{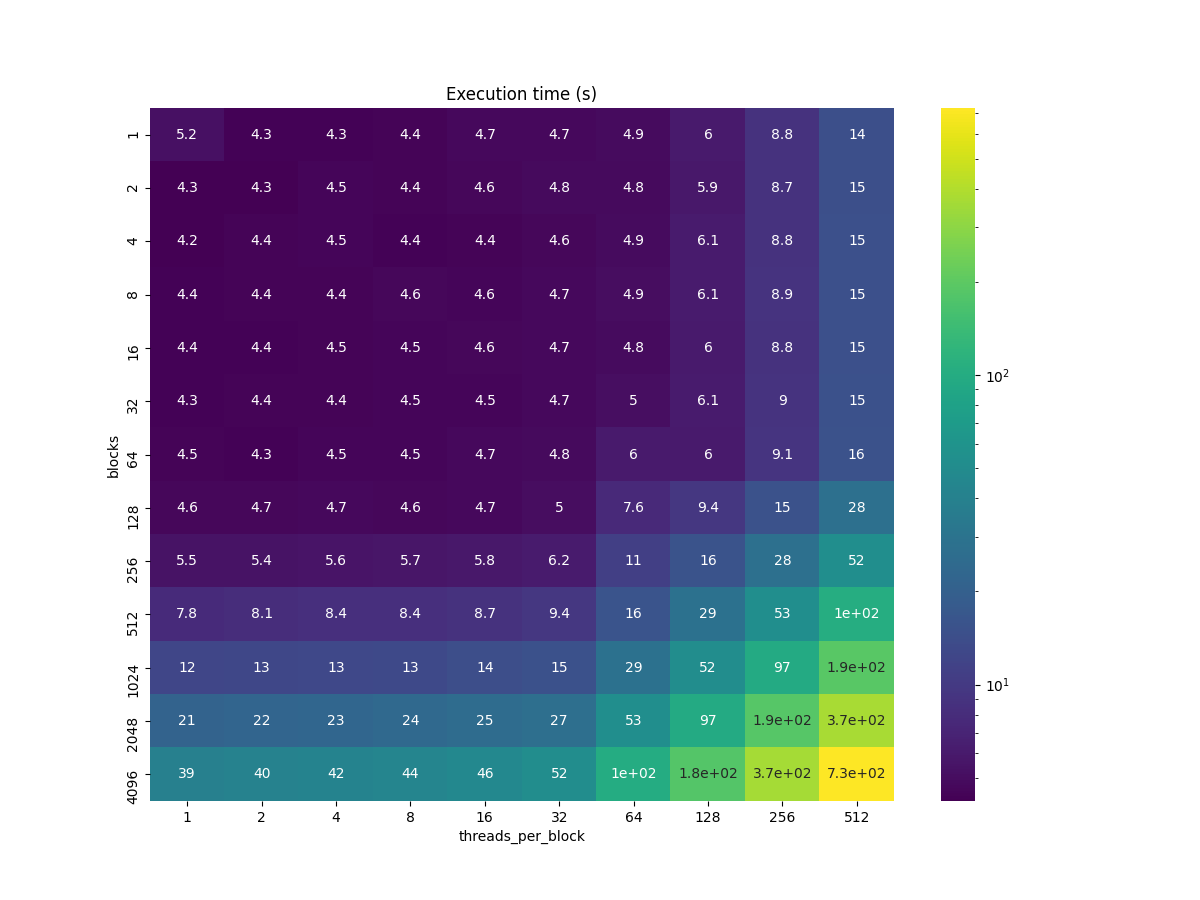}
\caption{\textbf{QBIOL execution time vs total number of GPU threads.} Time necessary to finish the simulation (in s). The total number of threads is blocks $\times$ threads\_per\_block.}\label{GPU_time}
\end{figure}

\begin{figure}[H]%
\centering
\includegraphics[width=1.1\textwidth]{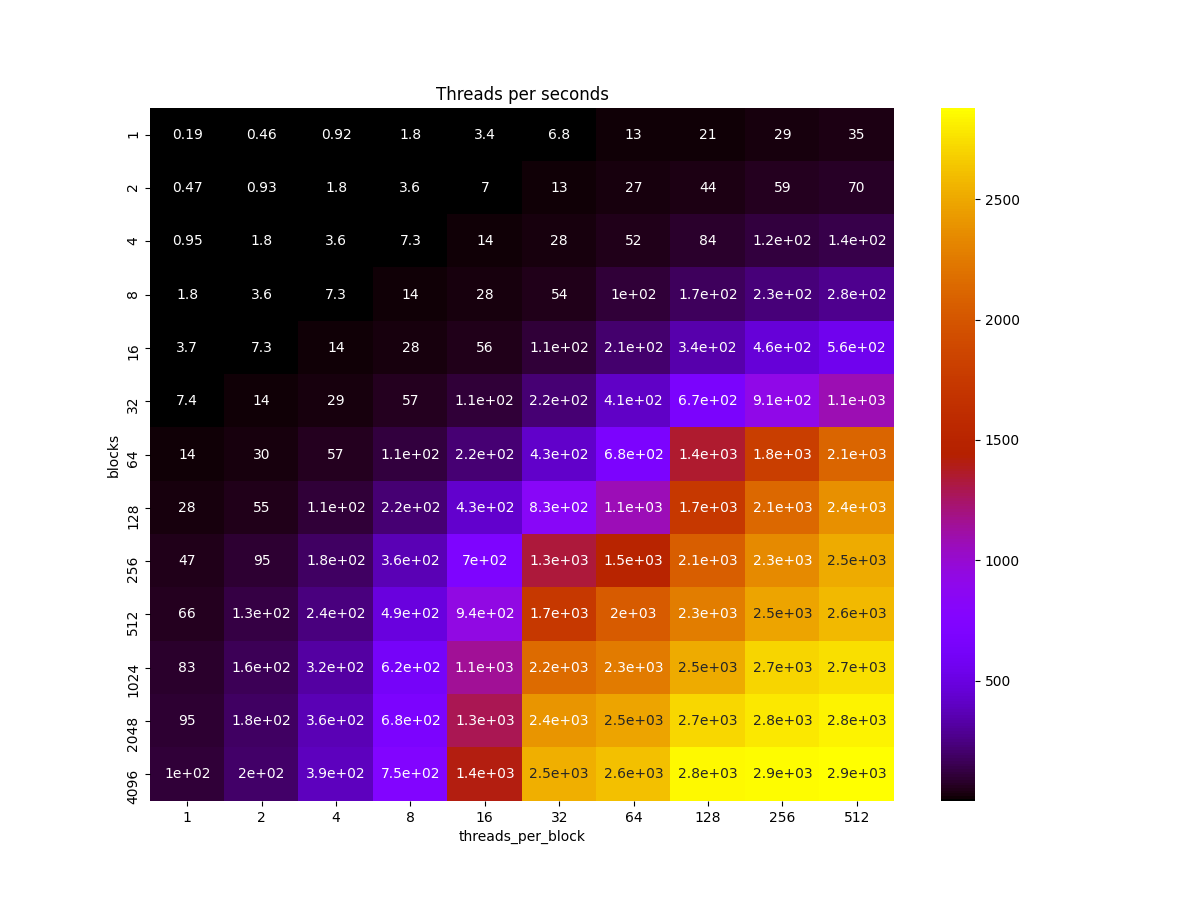}
\caption{\textbf{Number of threads finishing the simulation per second vs total number of GPU threads.} Number of threads finishing the simulation per second. The total number of threads is blocks $\times$ threads\_per\_block.}\label{GPU_threads_per_s}
\end{figure}

\begin{figure}[H]%
\centering
\includegraphics[width=1.1\textwidth]{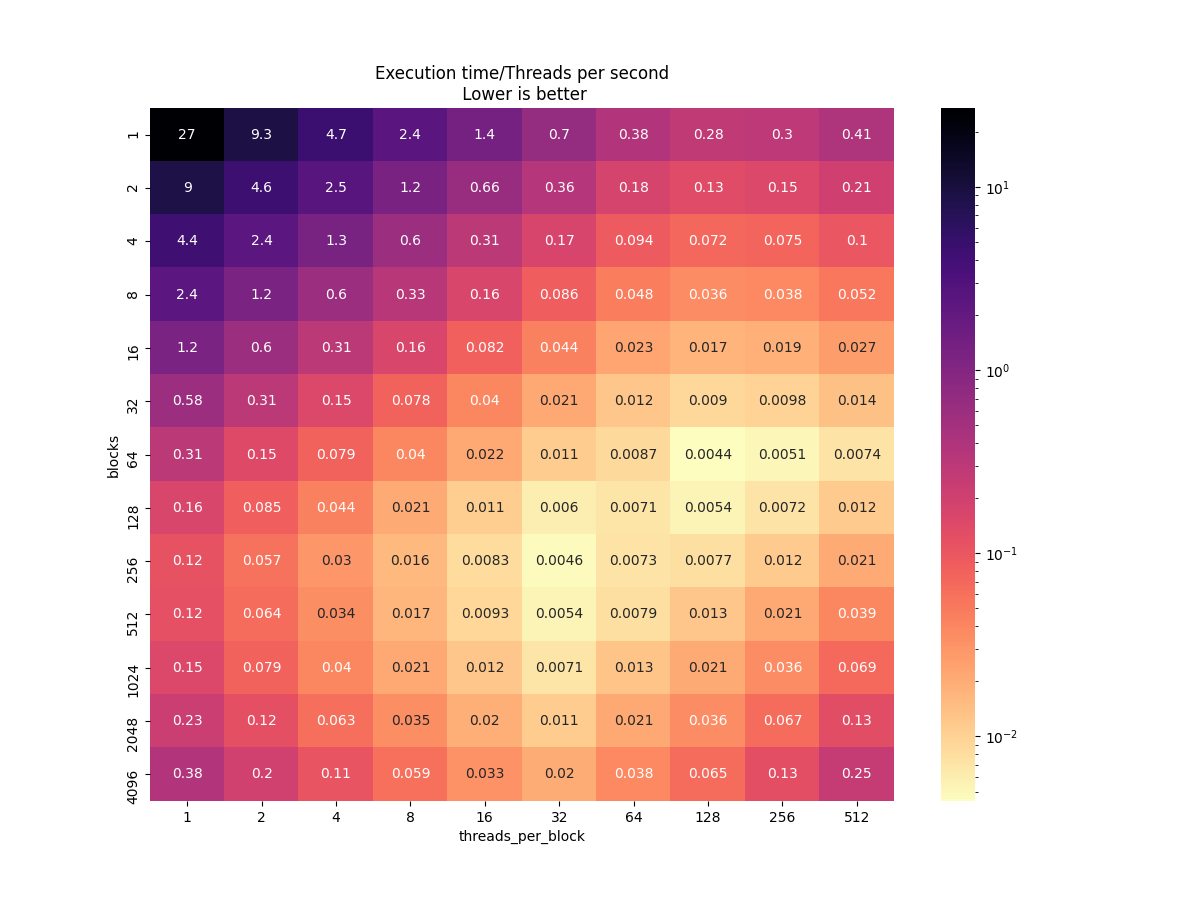}
\caption{\textbf{How many GPU threads for best efficiency ?} Threads/s/execution time, i.e. the previous results Fig.\ref{GPU_threads_per_s} / Fig. \ref{GPU_time}. The lower the value, the faster we get the most simulations done per second. There is a local optimum around 64-128 blocks and 128-256 threads per block, where one can obtain the most simulations done in the least amount of time spent per simulation.}\label{GPU_time_threads_per_s}
\end{figure}

\begin{figure}[H]%
\centering
\includegraphics[width=\textwidth]{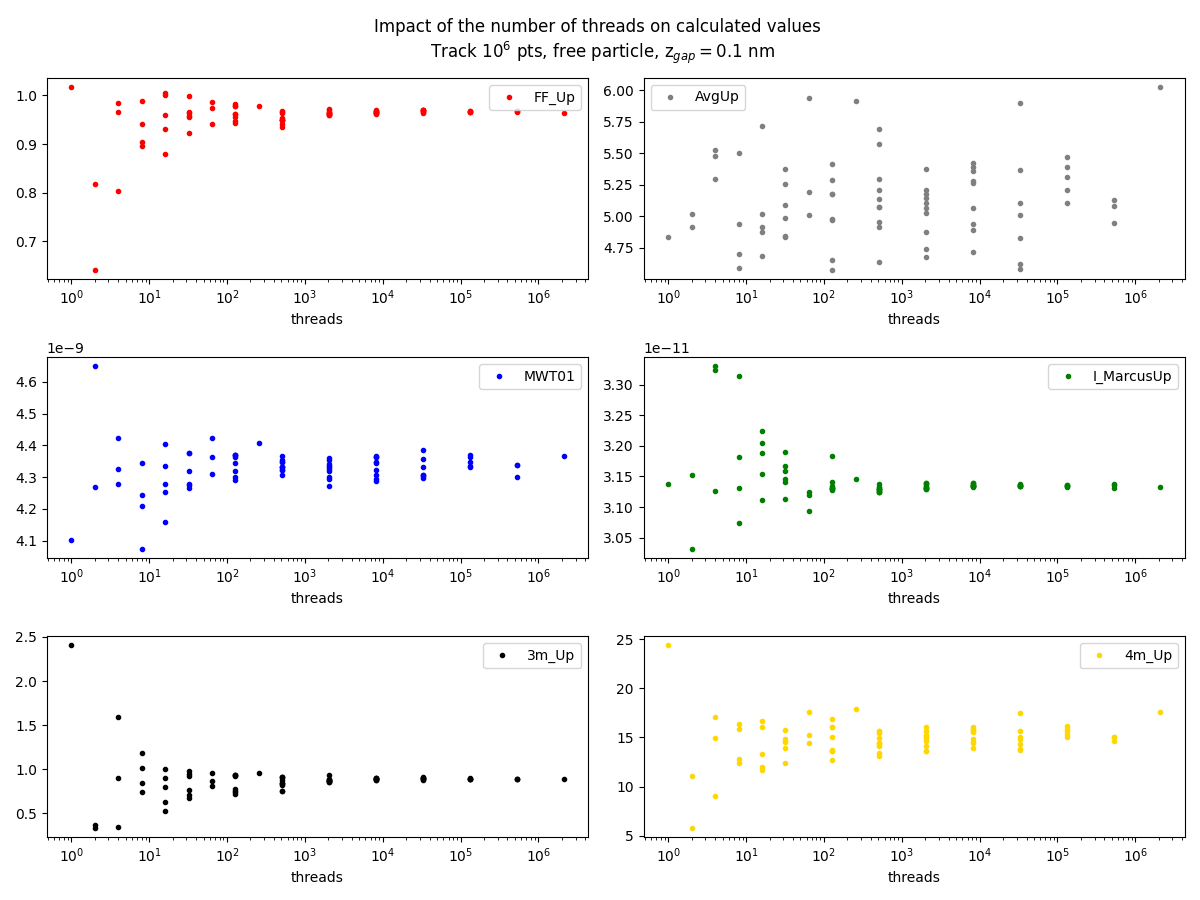}
\caption{\textbf{Convergence of statistical moments vs total number of GPU threads.} Convergence of the fano factor (FF\_Up), the average number of jumps per voltage step and per thread (AvgUp), the mean waiting time to go from 0 to 1 states (MWT01), the current (I\_MarcusUp), skewness (3m\_Up) and the kurtosis (4m\_Up). All moments are evaluated at the top electrode.}\label{moments_convergence}
\end{figure}

One adjustable variable in QBIOL is the number of molecule simulated, which, in general, corresponds to the number of threads used in the GPU calculations. Along with an issue of execution time, comes an issue of reliability: how many molecules are enough to measure the current for a reliable cyclic voltammogram \cite{cutress_how_2011}? How about the higher moments? We did a systematic study of the impact of the total number of threads, spread as blocks and threads per blocks in the GPU, with the following conditions: a ferrocene free particle in a 1 nm gap, using a track of $9.09 \times 10^{-7}$ s and doing a cyclic voltammetry with the bottom electrode kept at -0.3 V and the top electrode varying from -0.3 V to 0.3 V. Results are shown in Fig. \ref{GPU_time} to \ref{GPU_time_threads_per_s}. Fig. \ref{moments_convergence} shows the convergence of different variables with the number of threads. Since it seems difficult to reach full convergence for the 4$^{th}$ moment, most experiments are carried out with at least $128 \times 256 = 32768$ threads, which allows to have convergence of the 3$^{rd}$ moment while remaining near the optimal configuration determined in Fig. \ref{GPU_time_threads_per_s}.


\clearpage

\section{Pushing electrochemistry at the limit: currents tables}
\begin{table}[h]
\centering
\makebox[\textwidth]{
\begin{tabular}{ccccccp{5cm}}
\multicolumn{7}{c}{\textbf{Nanopore i-t recordings}} \\
\toprule
\textbf{$1/(\Delta t)$} & \textbf{Ionic flow}  & \multirow{2}{*}{\textbf{$N$}}        & \textbf{$\Delta i$} & \textbf{$\Delta i\sqrt{N}$} & \multirow{2}{*}{\textbf{Ref}} & \multirow{2}{*}{\textbf{Note}} \\
\textbf{(s$^{-1}$)}    & \textbf{(C/s)}        &                                      & \textbf{(pA)}       & \textbf{(pA)}               &                               &  \\
\toprule
2000                    & $2.30\times 10^{-12}$& 7179    & 1     & 84.73              & \cite{ren_single-molecule_2018}                   & Open current from Fig. 1 \\
$\times 10^{4}$         & $1.7\times 10^{-11}$ & 10612   & 4.5   & 464                & \cite{tan_kinetics_2016}                          & Current baseline in Fig. 2 \\
$2\times 10^{5}$        & $5\times 10^{-9}$    & 156055  & 300   & $1.12\times 10^{5}$& \multirow{2}{*}{\cite{rosenstein_integrated_2012}}& \multirow{2}{*}{Current baseline in Fig. 4} \\
$10^{6}$                & $5\times 10^{-9}$    & 31211   & 760   & $1.34\times 10^{5}$&                                                   & \\
$10^{4}$                & $10^{-10}$           & 62422   & 7     & 1748.9             & \cite{qing_directional_2018}                      & Fig. 3B open pore current \\
$2\times 10^{5}$        & $2\times 10^{-9}$    & 62422   & 1000  & $2.5 \times 10^{5}$& \cite{larkin_high-bandwidth_2014}                 & Current baseline in Fig. 1B \\
$2\times 10^{5}$        & $2\times 10^{-9}$    & 62422   & 700   & $1.75\times 10^{5}$& \cite{henley_electrophoretic_2016}                & Largest delta I seen in Fig. 1A \\
$10^{4}$                & $4.7\times 10^{-11}$ & 29338   & 10    & 1713               & \cite{cao_discrimination_2016}                    & Open pore current from Fig. 1C \\
$2\times 10^{5}$        & $9\times 10^{-10}$   & 28090   & 770   & $1.29\times 10^{5}$& \cite{schneider_dna_2010}                         & Conductance taken from the largest delta G visible in Fig 4a, multiplied by the applied voltage \\
$1.04\times 10^{5}$     & $4.60\times 10^{-10}$& 27610   & 1050  & $1.75\times 10^{5}$& \cite{plesa_fast_2013}                            & Largest conductance delta taken from Fig. 1b, applied voltage from Table S1 (50 mV) \\
$10^{5}$                & $3\times 10^{-10}$   & 18727   & 50    & 6842               & \cite{bell_digitally_2016}                        & Largest delta I taken from Fig. 1 \\
$10^{5}$                & $3.75\times 10^{-10}$& 23408   & 45    & 6885               & \cite{bell_specific_2015}                         & Largest delta I taken from Fig. 2 \\
$10^{4}$                & $7\times 10^{-11}$   & 43695   & 5     & 1045               & \cite{derrington_subangstrom_2015}                & Only blockade currents available, largest one taken \\
$6\times 10^{4}$        & $2\times 10^{-9}$    & 208073  & 560   & $2.55\times 10^{5}$& \cite{plesa_direct_2016}                          & Largest delta I taken from Fig. 2 \\
$2\times 10^{4}$        & $1.4\times 10^{-10}$ & 43695   & 22    & 4599               & \cite{boukhet_probing_2016}                       & Current of KCl at 90 mV in Fig. 1, corresponding to measurements in Fig. 2 \\
2000                    & 1.40E-10             & 436954  & 5     & 3305               & \cite{clarke_continuous_2009}                     & Open current from Fig. 1a \\
$8\times 10^{4}$        & $10^{-11}$           & 780     & 1000  & $2.79\times 10^{4}$& \cite{uram_noise_2008}                            & Baseline current in Fig. 1 \\
$2\times 10^{4}$        & $1.5\times 10^{-10}$ & 46816   & 15    & 3246               & \cite{robertson_single-molecule_2007}             & Open current from Fig. 1 \\
\botrule
\end{tabular}
}
\caption{Method: used the open pore current to estimate the ion flux. The number of molecule is estimated assuming one electron per ion and a time integration over the time found in Gao et. al. \cite{gao_shot_2020}}
\label{ci_1}
\end{table}

\begin{table}[h]
\centering
\makebox[\textwidth]{
\begin{tabular}{cccccp{5cm}}
\multicolumn{6}{c}{\textbf{Single-Particle Electrochemistry}} \\
\toprule
\textbf{$1/(\Delta t)$} &  \multirow{2}{*}{\textbf{$N$}} & \textbf{$\Delta i$} & \textbf{$\Delta i\sqrt{N}$} & \multirow{2}{*}{\textbf{Ref}} & \multirow{2}{*}{\textbf{Note}} \\
\textbf{(s$^{-1}$)}     &                                & \textbf{(pA)}       & \textbf{(pA)}               &                               &  \\
\toprule
$2\times 10^{5}$ & $3.12\times 10^{6}$ & 4000  & $7.07\times 10^{6}$ & \multirow{5}{*}{\cite{robinson_effects_2018}} & \multirow{5}{*}{Charge value from Table S2} \\
$6\times 10^{4}$ & $2.5\times 10^{6}$ & 3000  & $4.74\times 10^{6}$ & & \\
$2\times 10^{4}$ & $2.5\times 10^{6}$ & 2000  & $3.16\times 10^{6}$ & & \\
2000     & $3.12\times 10^{6}$ & 300   & $5.3\times 10^{5}$ & & \\
200      & $2.5\times 10^{6}$ & 90    & $1.42\times 10^{5}$ & & \\
500      & $3.99\times 10^{6}$  & 0.23  & 145.37 & \cite{batchelor-mcauley_situ_2015} & Integration of the peak in Fig. 2 inset gave 6.4e-14 C. \\
$2\times 10^{4}$ & $1.31\times 10^{5}$  & 15    & 5430.87 & \cite{zhou_collisions_2017} & Integration of the peak in figure 5 C gave 2.1e-14 C. \\
$2\times 10^{5}$ & $4.24\times 10^{8}$ & 55  & $1.13\times 10^{6}$  & \cite{glasscott_fine-tuning_2019} & Integration of the peak in Fig. 4B gave 6.8e-11 C. \\
$10^{4}$ & 8583    & 44    & 4076.36 & \cite{ma_quantifying_2018} & Based on the integration of the peak Fig.2a (blue star) and the Q statistics Fig. 2e, a charge of 5.5 fC is used, with 4 electrons per events counted \\
2000     & $1.56\times 10^{5}$  & 12    & 4740.46 & \multirow{2}{*}{\cite{ustarroz_impact_2017}} & Integration of the peak in Fig. 3a v gave 2.5e-14 C, in accordance with estimations in the text \\
$2\times 10^{4}$ & 624  & 2     & 50 & & Data taken from a histogram, from the first bin. C estimated around 0.1 fC (rectangle integration). \\
$10^{4}$ & 612     & 5     & 127 & \cite{gao_30_2018} & Integration of the peak in Fig 1 gave 9.8e-17 C. \\
\botrule
\end{tabular}
}
\caption{Method: find the charge associated with the current peak, assume 1 electron per molecule if the information is not known, and divide by the elementary charge for the number of molecule.}
\label{ci_2}
\end{table}

\begin{table}[h]
\centering
\makebox[\textwidth]{
\begin{tabular}{cccccp{5cm}}
\multicolumn{6}{c}{\textbf{Fast-scan voltammetry}} \\
\toprule
\textbf{$1/(\Delta t)$} & \multirow{2}{*}{\textbf{$N$}} & \textbf{$\Delta i$} & \textbf{$\Delta i\sqrt{N}$} & \multirow{2}{*}{\textbf{Ref}} & \multirow{2}{*}{\textbf{Note}} \\
\textbf{(s$^{-1}$)}     &                               & \textbf{(pA)}       & \textbf{(pA)}               &                               &  \\
\toprule
$2.88\times 10^{6}$ & $9.36\times 10^{7}$ & $1.56\times 10^{6}$ & $1.51\times 10^{10}$ & \multirow{6}{*}{\cite{amatore_ultrafast_2001}}& \multirow{6}{*}{Integration of the peaks in Fig. 3.} \\
$1.12\times 10^{7}$ & $6.96\times 10^{7}$ & $6.30\times 10^{6}$ & $5.25\times 10^{10}$ & & \\
$4.32\times 10^{7}$ & $5.43\times 10^{7}$ & $1.75\times 10^{7}$ & $1.29\times 10^{11}$ & & \\
$8.64\times 10^{7}$ & $5.35\times 10^{7}$ & $2.57\times 10^{7}$ & $1.88\times 10^{11}$ & & \\
$1.44\times 10^{8}$ & $3.27\times 10^{7}$ & $3.57\times 10^{7}$ & $2.04\times 10^{11}$ & & \\
$2.00\times 10^{8}$ & $3.81\times 10^{7}$ & $5.00\times 10^{7}$ & $3.09\times 10^{11}$ & & \\
$8.00\times 10^{4}$ & $1.62\times 10^{9}$ & 20 & $8.06\times 10^{5}$ &  \cite{watkins_zeptomole_2003} & Integration of the peak in Figure 4c \\
\botrule
\end{tabular}
}
\caption{Method: integration the voltammetry peak.}
\label{ci_3}
\end{table}

\begin{table}[h]
\centering
\makebox[\textwidth]{
\begin{tabular}{ccccccp{5cm}}
\multicolumn{7}{c}{\textbf{Nanoscale Electrochemical Imaging}} \\
\toprule
\textbf{$1/(\Delta t)$} & \textbf{Ionic flow}  & \multirow{2}{*}{\textbf{$N$}}        & \textbf{$\Delta i$} & \textbf{$\Delta i\sqrt{N}$} & \multirow{2}{*}{\textbf{Ref}} & \multirow{2}{*}{\textbf{Note}} \\
\textbf{(s$^{-1}$)}    & \textbf{(C/s)}        &                                      & \textbf{(pA)}       & \textbf{(pA)}               &                               &  \\
\toprule
10       & $5.3\times 10^{-12}$& $3.31\times 10^{6}$  & 0.15 & 273  & \cite{guell_redox-dependent_2015} & SECM. $i=4FCnDA$. Used the baseline current of Figure 6a to determine the number of molecules involved \\
125      & $2.2\times 10^{-12}$ & $1.10\times 10^{5}$  & 2    & 663   & \multirow{2}{*}{\cite{momotenko_high-speed_2015}} & SECM. ~ 2.2 pA on Figure 3c. \\
$8\times 10^{-4}$ & $2.2\times 10^{-12}$  & $1.72\times 10^{10}$ & 0.2  & $2.62\times 10^{4}$ & & Assumed same noise than data from the same group. \\
20       & -        & 1000    & 0.2  & 6    & \cite{ren_single-molecule_2018} & Actually redox cycling. N estimated from the area and the density of PEG announced in the paper \\
\hline
\end{tabular}
}
\caption{Method: for SECM, use the baseline current and assume a diffusion limited current to get the number of ions involved in the detection process, similarly to nanopores.}
\label{ci_4}
\end{table}

\begin{table}[h]
\centering
\makebox[\textwidth]{
\begin{tabular}{cccccp{5cm}}
\multicolumn{6}{c}{\textbf{Single-Molecule Redox Cycling}} \\
\toprule
\textbf{$1/(\Delta t)$} & \multirow{2}{*}{\textbf{$N$}} & \textbf{$\Delta i$} & \textbf{$\Delta i\sqrt{N}$} & \multirow{2}{*}{\textbf{Ref}}                    & \multirow{2}{*}{\textbf{Note}} \\
\textbf{(s$^{-1}$)}     &                               & \textbf{(pA)}       & \textbf{(pA)}               &                                                  &  \\
\toprule
2                       & 1                             & $7\times 10^{-3}$   & $7\times 10^{-3}$           & \cite{kang_electrochemical_2013}                 & Single molecule. \\
100                     & 1                             & 0.02                & 0.02                        & \cite{zevenbergen_stochastic_2011}               & $N=CVNa$, with $C = 120$ pM and $V = 60$ \textmu m $\times$ 70 nm $\times$ 1.5 \textmu m $\ll 1 \rightarrow$  1 molecule. \\
0.8                     & 1                             & 0.75                & 0.75                        & \multirow{3}{*}{\cite{bard_electrochemical_1996}}& \multirow{3}{5cm}{$N=CVNa$ gives $N \ll 1 \rightarrow 1$ molecule (assuming $d \sim 1$ nm for $i(d)$, as stated for Fig. 6)} \\
0.8                     & 1                             & 1.5                 & 1.5                         &                                                  & \\
0.8                     & 1                             & 3                   & 3                           &                                                  &  \\
20                      & 1                             & $3.6\times 10^{-3}$ & $3.6\times 10^{-3}$         & \cite{byers_single_2015}                         & Single molecule. \\
0.18                    & 20                            & $4\times 10^{-4}$   & $1.78\times 10^{-3}$       & \cite{paiva_enzymatic_2022}                      & Given in the text. \\
\botrule
\end{tabular}
}
\caption{Method: information from the text for molecule number.}
\label{ci_5}
\end{table}

\begin{table}[h]
\centering
\makebox[\textwidth]{
\begin{tabular}{cccccp{5cm}}
\multicolumn{6}{c}{\textbf{Nanobubble cavity}} \\
\toprule
\textbf{$1/(\Delta t)$} & \multirow{2}{*}{\textbf{$N$}} & \textbf{$\Delta i$} & \textbf{$\Delta i\sqrt{N}$} & \multirow{2}{*}{\textbf{Ref}}                    & \multirow{2}{*}{\textbf{Note}} \\
\textbf{(s$^{-1}$)}     &                               & \textbf{(pA)}       & \textbf{(pA)}               &                                                  &  \\
\toprule

2000 & $1.25\times 10^{5}$ & 100 & $3.53\times 10^{4}$ & \cite{german_laplace_2016} & Faradaic current. \\
\hline
\end{tabular}
}
\caption{Method: Faradaic current estimation.}
\label{ci_6}
\end{table}

\begin{table}[h]
\centering
\makebox[\textwidth]{
\begin{tabular}{cccccp{5cm}}
\multicolumn{6}{c}{\textbf{Neurotransmitter Release}} \\
\toprule
\textbf{$1/(\Delta t)$} & \multirow{2}{*}{\textbf{$N$}} & \textbf{$\Delta i$} & \textbf{$\Delta i\sqrt{N}$} & \multirow{2}{*}{\textbf{Ref}}                    & \multirow{2}{*}{\textbf{Note}} \\
\textbf{(s$^{-1}$)}     &                               & \textbf{(pA)}       & \textbf{(pA)}               &                                                  &  \\
\toprule
$2.00\times 10^{4}$ & $6.87\times 10^{4}$ & 31 & 8123 & \multirow{4}{*}{\cite{hochstetler_real-time_2000}} & \multirow{4}{5cm}{From the integration of the I vs t peak, assuming 1 e- per ion giving ~ 1.1e-14 C.} \\
2000 & $6.87\times 10^{4}$ & 17 & 4455 &  &  \\
400  & $6.87\times 10^{4}$ & 6.7 & 1756 &  &  \\
80   & $6.87\times 10^{4}$ & 1.5 & 393 &  &  \\
2000 & $1.99\times 10^{5}$ & 20 & 8918 & \cite{gu_plasticity_2019} & From peak and half max value, created a corresponding Gaussian peak and integrated it, yielding $\approx$3.185e-14 C \\
4000 & $3.70\times 10^{5}$ & 41 & $2.47\times 10^{4}$ & \cite{li_electrochemical_2018} & Peak integration in the inset of Fig. 1 \\

\hline
\end{tabular}
}
\caption{Method: integration of I vs time peak, assuming 1 $e^-$/molecule.
}
\label{ci_7}
\end{table}

\begin{table}[h]
\centering
\makebox[\textwidth]{
\begin{tabular}{cccccp{5cm}}
\multicolumn{6}{c}{\textbf{Square Wave Voltammetry}} \\
\toprule
\textbf{$1/(\Delta t)$} & \multirow{2}{*}{\textbf{$N$}} & \textbf{$\Delta i$} & \textbf{$\Delta i\sqrt{N}$} & \multirow{2}{*}{\textbf{Ref}}                    & \multirow{2}{*}{\textbf{Note}} \\
\textbf{(s$^{-1}$)}     &                               & \textbf{(pA)}       & \textbf{(pA)}               &                                                  &  \\
\toprule
100	&	10&	$10^{-5}$&	$3.16\times 10^{-5}$&\cite{trasobares_estimation_2017}	 	&Number of molecule provided in the paper (obtained through cyclic voltammetry peak integration)\\
\\
\hline
\end{tabular}
}
\caption{Method: peak integration through the SWV/CV correspondence provided by the authors.}
\label{ci_8}
\end{table}

\begin{table}[h]

\centering
\makebox[\textwidth]{
\begin{tabular}{cccccp{5cm}}
\multicolumn{6}{c}{\textbf{Single-Atom Electrochemistry}} \\
\toprule
\textbf{$1/(\Delta t)$} & \multirow{2}{*}{\textbf{$N$}} & \textbf{$\Delta i$} & \textbf{$\Delta i\sqrt{N}$} & \multirow{2}{*}{\textbf{Ref}}                    & \multirow{2}{*}{\textbf{Note}} \\
\textbf{(s$^{-1}$)}     &                               & \textbf{(pA)}       & \textbf{(pA)}               &                                                  &  \\
\toprule
7.78  & $1.36\times 10^{9}$  & 56  & $2.07\times 10^{6}$ & \multirow{6}{*}{\cite{zhou_electrodeposition_2017}} & \multirow{7}{5cm}{Obtained from the integration of current over time, assuming 2 $e^-$ transferred per molecule (2H$^+$ → H$_2$). Notice, the value retained by Gao et al. is more a maximum current than a delta I.} \\
7.78  & $2.21\times 10^{9}$  & 91  & $4.28\times 10^{6}$  &  &  \\
7.78  & $2.65\times 10^{9}$  & 109 & $5.61\times 10^{6}$  &  &  \\
7.78  & $2.94\times 10^{9}$  & 121 & $6.56\times 10^{6}$  &  &  \\
7.78  & $3.50\times 10^{9}$  & 144 & $8.52\times 10^{6}$  &  &  \\
7.78  & $4.10\times 10^{9}$  & 169 & $1.08\times 10^{7}$  &  &  \\
4     & $1.87\times 10^{9}$  & 150 & $6.49\times 10^{6}$ & \cite{zhou_probing_2019} &  \\
\\
\hline
\end{tabular}
}
\caption{Method: see notes.}
\label{ci_9}
\end{table}

\begin{table}[h]
\centering
\makebox[\textwidth]{
\begin{tabular}{cccccp{5cm}}
\multicolumn{6}{c}{\textbf{Single-molecule Electrochemical STM}} \\
\toprule
\textbf{$1/(\Delta t)$} & \multirow{2}{*}{\textbf{$N$}} & \textbf{$\Delta i$} & \textbf{$\Delta i\sqrt{N}$} & \multirow{2}{*}{\textbf{Ref}}   & \multirow{2}{*}{\textbf{Note}} \\
\textbf{(s$^{-1}$)}     &                               & \textbf{(pA)}       & \textbf{(pA)}               &                                 &  \\
\toprule
100                     & 1                             & $1.2\times 10^{5}$  & $1.2\times 10^{5}$          & \cite{li_transition_2019}       & Conductance of a single Fc molecule. \\
\botrule
\end{tabular}
}
\caption{Method: see notes.}
\label{ci_10}
\end{table}

\begin{table}[h]
\centering
\makebox[\textwidth]{
\begin{tabular}{cccccp{5cm}}
\multicolumn{6}{c}{\textbf{Electrochemical high Frequency STM (EF-STM)}} \\
\toprule
\textbf{$1/(\Delta t)$} & \multirow{2}{*}{\textbf{$N$}} & \textbf{$\Delta i$} & \textbf{$\Delta i\sqrt{N}$} & \multirow{2}{*}{\textbf{Ref}}   & \multirow{2}{*}{\textbf{Note}} \\
\textbf{(s$^{-1}$)}     &                               & \textbf{(pA)}       & \textbf{(pA)}               &                                 &  \\
\toprule
2                       & 200                           & $2.00\times 10^{-8}$&$2.83\times 10^{-7}$         & \cite{grall_attoampere_2021}    &I from $\sigma$ in Fig. 2b. $\Delta t$ from the Fig. 1b: assuming the data is obtained at the highest possible sampling rate, it gives around 2pts/s. N molecules given in the paper.\\

2                       & 1600                          & $2.00\times 10^{-7}$&$8.00\times 10^{-6}$         &\cite{awadein_nanoscale_2022}    & 666 nm$^2$ $\times$ 2.4 molecule/nm$^2 \approx 1600$ molecules probed. $\Delta i$ taken from the uncertainty on current given in the text. With Fig. 3c, a similar consideration than for \cite{grall_attoampere_2021} gives also around 2pts/s.\\ 

\botrule
\end{tabular}
}
\caption{Method: see notes.}
\label{ci_11}
\end{table}

\begin{table}[h!]
\centering
\makebox[\textwidth]{
\begin{tabular}{cccccp{5cm}}
\multicolumn{6}{c}{\textbf{AFM-SECM}} \\
\toprule
\textbf{$1/(\Delta t)$} & \multirow{2}{*}{\textbf{$N$}} & \textbf{$\Delta i$} & \textbf{$\Delta i\sqrt{N}$} & \multirow{2}{*}{\textbf{Ref}}   & \multirow{2}{*}{\textbf{Note}} \\
\textbf{(s$^{-1}$)}     &                               & \textbf{(pA)}       & \textbf{(pA)}               &                                 &  \\
\toprule
10		                  &20	                            &$4.00\times 10^{-4}$ &$1.79\times 10^{-3}$         &	\cite{paiva_enzymatic_2022}     &Molecules in the text, 10Hz lowpass filter used and noise taken from Fig. 10a.\\	
10		                  &4000	                          &$2.00\times 10^{-3}$	&0.13                         &	\cite{torbensen_immuno-based_2019}&Molecules estimated from Fig.2, lower AFM-SECM image. We estimated 4000 Fc (text) on the fd, and obtained a standard deviation of 2 fA on the image when excluding saturated values.\\
10                      &20	                            &0.01	                &$4.5\times 10^{-2}$          &\cite{nault_electrochemical_2015}&Sensitivity in  the text, taken the average number of Fc head/virus as a representative value of the number of Fc molecule.\\	
20		                  &140	                            &0.2	                &2.28                       &\cite{chennit_high-density_2022} & Current bumps from Fig.4 and 5 $\approx 20$ fA. 5 IgG-PEG-Fc per dot, 28 Fc per Ig $\approx 140$ Fc/dot.\\ 		
\botrule
\end{tabular}
}
\caption{Method: most experiments are done on SAMs, with no diffusion-limited current. See notes for details.}
\label{ci_12}
\end{table}

\begin{table}[h!]
\centering
\makebox[\textwidth]{
\begin{tabular}{cccccp{5cm}}
\multicolumn{6}{c}{\textbf{QBIOL}} \\
\toprule
\textbf{$1/(\Delta t)$} & \multirow{2}{*}{\textbf{$N$}} & \textbf{$\Delta i$} & \textbf{$\Delta i\sqrt{N}$} & \multirow{2}{*}{\textbf{Ref}}   & \multirow{2}{5cm}{\textbf{Note}} \\
\textbf{(s$^{-1}$)}     &       & \textbf{(pA)}         & \textbf{(pA)}               &                                 &  \\
\toprule
200	                    &1024	                          &$6.44\times 10^{-7}$	&2.06$\times 10^{-5}$ 				&This work (voltnoisogram)  		& $\sqrt{\text{PSD}}$~at f = 200 Hz and $E = E^0$.\\
200	                    &1024	  												&$1.96\times 10^{-7}$	&6.29$\times 10^{-6}$ 				&This work (CV)							    & Std of $i$ over 0.005 s at $E=E^0$.\\
$5\times 10^{8}$        &65536													&$2.00\times 10^{-12}$&5.12$\times 10^{-10}$ 				&This work (SMM)						    & Std around $E=E^0$, $\Delta t$~= integration time.\\

\botrule
\end{tabular}
}
\caption{Method: CVs and chronoamperometry are done at different voltage sweep to get different time constants. Redox cycling experiments were done on nanogaps of 1 to 10 nm.}
\label{ci_13}
\end{table}

\clearpage

\bibliography{sn-bibliography}